\documentclass[a4paper,fleqn]{cas-sc}
\usepackage[numbers]{natbib}
\usepackage{algorithm, algorithmic}
\usepackage{multirow}
\usepackage{diagbox}
\usepackage{subcaption}
\usepackage{graphicx}

\def\tsc#1{\csdef{#1}{\textsc{\lowercase{#1}}\xspace}}
\tsc{WGM}
\tsc{QE}
\tsc{EP}
\tsc{PMS}
\tsc{BEC}
\tsc{DE}
\begin{document}
\let\WriteBookmarks\relax
\def\floatpagepagefraction{1}
\def\textpagefraction{.001}

\shorttitle{PhysioFormer}    

\shortauthors{Wang et al.}  

\title [mode = title]{PhysioFormer: Integrating Multimodal Physiological Signals and Symbolic Regression for Explainable Affective State Prediction}  

\tnotemark[1]

\tnotetext[1]{This work was supported by Self-determined Research Funds of CCNU from the Colleges’ Basic Research and Operation of MOE (No. CCNU24JC033).}

\author[1]{Zhifeng Wang}[orcid=0000-0001-6960-509X]
\affiliation[1]{organization={Department of Digital Media Technology, Central China Normal University},
	city={Wuhan},
	postcode={430079}, 
	country={China}}
\cormark[1]
\ead{zfwang@ccnu.edu.cn}

\author[1]{Wanxuan Wu}
\ead{wuwx0428@163.com}

\author[2]{Chunyan Zeng}
\affiliation[2]{organization={High-Efficiency Utilization of Solar Energy and Operation Control of Energy Storage System, Hubei University of Technology},
	city={Wuhan},
	postcode={430068}, 
	country={China}}
\cormark[1]
\ead{cyzeng@hbut.edu.cn}

\cortext[cor1]{Corresponding authors}

\begin{abstract}
  As affective computing becomes increasingly crucial in health monitoring and psychological intervention, accurately identifying affective states is a key challenge. While traditional machine learning models have achieved some success in affective computation, their ability to handle complex, multimodal physiological signals is limited. Most affective computing tasks still rely heavily on traditional methods, with few deep learning models applied, particularly in multimodal signal processing. Given the importance of stress monitoring for mental health, developing a highly reliable and accurate affective computing model is essential. In this context, we propose a novel model—PhysioFormer, for affective state prediction using physiological signals. PhysioFormer model integrates individual attributes and multimodal physiological data to address inter-individual variability, enhancing its reliability and generalization across different individuals. By incorporating feature embedding and affective representation modules, PhysioFormer model captures dynamic changes in time-series data and multimodal signal features, significantly improving accuracy. The model also includes an explainability model that uses symbolic regression to extract laws linking physiological signals to affective states, increasing transparency and explainability. Experiments conducted on the Wrist and Chest subsets of the WESAD dataset confirmed the model's superior performance, achieving over 99\% accuracy, outperforming existing SOTA models. Sensitivity and ablation experiments further demonstrated PhysioFormer’s reliability, validating the contribution of its individual components. The integration of symbolic regression not only enhanced model explainability but also highlighted the complex relationships between physiological signals and affective states. Future work will focus on optimizing the model for larger datasets and real-time applications, particularly in more complex environments. Additionally, further exploration of physiological signals and environmental factors will help build a more comprehensive affective computing system, advancing its use in health monitoring and psychological intervention.

\end{abstract}

% \includegraphics{figs/cas-grabs.pdf}

% \begin{highlights}
% \item Research highlights item 1
% \item Research highlights item 2
% \item Research highlights item 3
% \end{highlights}


\maketitle

\section{Introduction}

\subsection{Background}

With the rapid development of society and the increasing pace of life, the importance of emotions for individual physical and mental health has become increasingly evident \cite{Chen2023b}. Emotions are spontaneously generated from daily experiences, stimulating the body to produce hormones and affecting various aspects, including bodily movements, facial expressions, and physiological characteristics \cite{Wang2024b}. The accumulation of negative emotions can lead to depression. It is estimated that about one in five men and one in three women worldwide will experience major depression during their lifetime. Although other mental illnesses, such as schizophrenia and bipolar disorder, are less common, they still have a significant impact on individuals' lives \citep{Saloni2023}. The contemporary focus on mental and physical stress is growing, as the effects of stress on the human body are both widespread and profound, particularly in its impact on the brain, cardiovascular system, immune system, and metabolism. Chronic stress not only leads to mental health issues but may also alter gene expression through epigenetic mechanisms, further affecting physiological functions \citep{McEwen2012}.

In existing research, physiological signals have been applied across various fields. First, in the domain of healthcare, monitoring physiological signals enables early warnings of chronic diseases and health management \citep{Ertin2011}. Second, in affective computation and mental health, analyzing signals such as electrodermal activity and electrocardiograms can identify individuals' affective states, providing effective tools for emotion management. Additionally, in intelligent human-computer interaction, the application of physiological signals can optimize user experience by allowing devices to respond according to users' physiological and affective states \citep{Nadai2016}. Given the significant impact of negative emotions on daily life, monitoring affective states through physiological signals is particularly important. Moreover, the current global healthcare trend is shifting toward preventing chronic diseases and reducing treatment costs, with wearable devices playing a crucial role in this development \citep{Hosseini2023}. These devices can monitor users' physiological signals in real time, providing personalized health recommendations and helping individuals detect potential health issues early, thus promoting a shift in healthcare systems from reactive treatment to proactive prevention.

Traditional affective computation has largely relied on questionnaires \citep{Andreou2011}, which are often highly subjective. The reliability of the results is closely tied to the respondent's attentiveness and the seriousness with which they approach the questionnaire. Moreover, individuals are often unable to accurately perceive their own affective states. For instance, people who are under chronic stress may not be fully aware of the extent of the stress they are experiencing. Therefore, using questionnaires to identify affective states is undoubtedly time-consuming and inefficient. As a result, subsequent research has shifted towards utilizing machine learning and deep learning methods to predict affective states through physiological signals.

\subsection{Gap}

Numerous studies have demonstrated that physiological signals such as electrocardiograms (ECG), electrodermal activity (EDA), and electroencephalograms (EEG) play a crucial role in affective computation. For example, Han et al. \citep{Han2017} proposed a method that combines graph signal processing (GSP) with convolutional neural networks (CNNs), utilizing GSP to process EEG data in order to better capture the spatial features within EEG signals, for EEG-based affective classification. Song et al. \citep{Song2020} introduced a dynamic graph convolutional neural network (DGCNN) for EEG affective recognition. The core of this approach involves constructing a graph structure based on EEG signals and using graph convolutional networks (GCNs) to capture the spatiotemporal dependencies between different brain regions, thereby improving the accuracy of affective computation. Ashwin et al. \citep{Ashwin2022} assessed individual stress states by monitoring multiple physiological indicators such as heart rate (HR), EDA, and body temperature, and applying machine learning algorithms (e.g., k-nearest neighbors (KNN), support vector machine (SVM)) to classify the extracted features.

Despite significant progress in the field of affective recognition, several key challenges and research gaps remain. First, in terms of reliability, while the complexity and diversity of physiological signals are widely acknowledged, many studies still inadequately collect and utilize multimodal physiological data. This shortfall not only limits the model’s generalization capability under multimodal data conditions but also affects its stability in cross-individual affective recognition tasks. Due to the high variability of individual physiological signals, the generalizability and reliability of existing models in practical applications are weakened, and the lack of data support further restricts their applicability in diverse scenarios. Second, although models exhibit high accuracy in certain tasks, affective states are inherently dynamic processes. Existing methods significantly lack the ability to capture the continuity and temporal features of emotional changes, especially when dealing with complex physiological signals. This limitation reduces the model's ability to accurately recognize dynamic affective states. Finally, the issue of model explainability remains prominent. Most deep learning models still function as “black boxes,” making it difficult to clearly explain their decision-making processes and feature selection \cite{Zeng2023b}. This lack of transparency not only reduces users' trust in model predictions but also limits the model's explainability and transparency in real-world applications. Therefore, developing an affective recognition model that emphasizes the integration and utilization of multimodal physiological data, enhances the ability to capture dynamic processes, and improves decision-making transparency could not only advance the accuracy and practicality of affective recognition but also drive progress in stress monitoring and mental health interventions.

\subsection{Summary of Contribution}

To address the aforementioned issues, we developed the PhysioFormer model, which is capable of processing high-dimensional, complex, and sequential physiological signal data in parallel, further enhancing the ability to capture the timeliness and continuity of emotional changes. This section will introduce the contributions of the PhysioFormer model to the field of affective computation as below.

\begin{enumerate}
  \item \textbf{Handling high-dimensional, complex temporal data and introducing individual attributes features.} Due to the high-dimensional, complex, and temporally continuous nature of physiological signal data, traditional machine learning models and existing deep learning models tend to exhibit instability. Therefore, in designing the PhysioFormer model, we incorporated three submodules: \textit{ContribNet}, \textit{AffectNet}, and \textit{AffectAnalyser}, which not only process physiological signal data in parallel but also effectively capture the temporal dynamics and complexity of physiological signals. This enhances the model's ability to detect the timeliness and continuity of emotional changes. Additionally, we introduced an upper triangular matrix encoding in the \textit{ContribNet} module, allowing the model to flexibly focus on important information at different time points while also sensitizing it to the temporal dynamics of physiological signals. This addresses the limitations of traditional machine learning models and existing deep learning networks in effectively capturing the intrinsic relationships within temporally continuous data. Furthermore, to account for individual baseline physiological state differences, we incorporated features describing individual attributes into the dataset, such as age, gender, height, weight, smoking habits, and whether the individual exercised today. These individual attributes features provide rich contextual information, helping the model generalize better across different individuals and improving the accuracy and reliability of affective state prediction. The introduction of individual attributes features allows the model to better understand and explain physiological signals, thereby enhancing its adaptability.
  \item \textbf{Improving reliability in cross-subject affective computation.} To enhance the reliability of cross-subject affective computation, we employed a cross-validation training approach. In each epoch, the model sequentially uses data from each individual for training, explicitly learning and adapting to the physiological signal characteristics of different individuals. This training method effectively reduces the risk of overfitting to specific individuals, thereby improving the model's generalization ability and making it more robust in cross-subject affective computation.
  \item \textbf{Development of a high-precision affective computation model.} The PhysioFormer model has demonstrated exceptional ability in capturing the intrinsic relationships within physiological signal data, significantly improving the accuracy of predicting individuals' affective states. Through rigorous experiments and in-depth analysis on the WEASD dataset, we validated the effectiveness of the PhysioFormer model. On both the Wrist and Chest sub-datasets, the PhysioFormer model achieved over 99\% accuracy, far surpassing the performance of current state-of-the-art (SOTA) models. These results not only highlight the potential of PhysioFormer in the field of affective computation but also demonstrate its reliability and reliability in handling complex physiological signals, providing strong support for precise prediction of individual affective states.
  \item \textbf{Enhancing model explainability.} To improve the model's explainability, we integrated an Explanation Model using symbolic regression to generate mathematical formulas that describe the relationship between input variables and outputs \citep{Udrescu2020}. In our research, this model analyzed the influence of each physiological indicator on affective states, generating formulas that reveal the impact of signals like heart rate variability (HRV), EDA, and ECG on affective prediction. These formulas not only enhance transparency but also help users better understand the model’s decision-making process, increasing both explainability and trust in the model's predictions.
\end{enumerate}

The structure of this paper is arranged as follows. Section 2 provides a detailed overview of the related work in the field. Section 3 defines the mathematical notation and model architecture used in this study. In section 4, we propose and describe the PhysioFormer model and the symbolic regression task in detail. Section 5 covers the dataset, baseline models, experimental setup, and evaluation metrics. Finally, section 6 summarizes the study and discusses future research directions.

\section{Related Work}

\subsection{Affective Computation}
Affective computing aims to analyze physiological signals, facial expressions, vocal patterns, and other behavioral data using sensors, algorithms, and machine learning techniques to predict or identify an individual's affective state \cite{Garc2024}. In recent years, the research focus of affective computing has shifted from single-mode affective classification to the fusion of multimodal data. By integrating multiple sources of signals—such as physiological signals (e.g., heart rate, electrodermal activity), facial expressions, vocal features, and environmental factors—researchers can more accurately capture and recognize an individual's affective state. This multimodal approach not only improves the accuracy of affective recognition but also enhances the timeliness and continuity of tracking affective state changes.

Bernhard et al. \citep{Bernhard2018} proposed a deep learning-based method for text-based affective recognition, with key innovations including bidirectional processing using a Bidirectional Long Short-Term Memory (BiLSTM) network, combined with Dropout regularization and a weighted loss function to address small datasets and class imbalance. Additionally, the study introduced a transfer learning method called Sent2Affect, where the network is pre-trained on affective analysis tasks and then fine-tuned for affective recognition by adjusting the output layer. This approach improved the model’s performance on small-scale datasets. Experiments conducted on six benchmark datasets demonstrated that this method significantly outperformed traditional machine learning approaches, achieving a 23.2\% increase in F1 scores for classification tasks and an 11.6\% reduction in mean squared error for regression tasks.

Ashwin et al. \citep{Ashwin2022} induced stress states in participants through acute stress manipulation tasks (such as the Maastricht Acute Stress Task \citep{Quaedflieg2013} and the Montreal Imaging Stress Test \citep{Dedovic2005}) and recorded physiological signals, including HR, HRV, EDA, and respiratory rate, using wearable sensors. The resulting stress detection models achieved accuracy rates of 97\% in controlled environments and 93\% in everyday settings. This study confirmed the effectiveness of multimodal physiological signal fusion in stress detection and demonstrated the feasibility of wearable devices for stress monitoring across different environments. Similarly, Sarkar et al. \citep{Sarkar2022} proposed an affective recognition method based on ECG signals using a self-supervised deep multi-task learning framework. By pre-training the network on signal transformation tasks and then transferring it to affective classification, the model showed significant performance improvements on the AMIGOS, DREAMER, WESAD, and SWELL datasets, outperforming traditional fully supervised methods. This validates the efficacy of multi-task learning in ECG-based affective recognition. Akre et al. \citep{Akre2023} introduced a depression symptom detection framework using data collected from iPhones and Apple Watches. A gradient boosting classifier processed health data, including vital signs, activity levels, and sleep patterns. The model exhibited moderate predictive accuracy, with ROC AUC values ranging from 0.63 to 0.72, demonstrating the potential of personalized sensor data for depression detection. These studies collectively highlight the significant potential of multimodal physiological signals and personalized health data in the detection and recognition of stress, emotions, and depression symptoms.

Koldijk et al. \citep{Koldijk2014} collected a multimodal dataset specifically designed for research on stress and user modeling, incorporating physiological signals (such as ECG, EMG, EEG), facial expressions, and behavioral data from mouse and keyboard usage. In a controlled experimental environment, participants performed various tasks (e.g., writing, reading, and meetings) to simulate real-world work scenarios. Stress levels were validated using a combination of questionnaires, interviews, and objective measurements. The experimental results indicated a significant correlation between induced stress conditions and both physiological and behavioral data, demonstrating the value of the SWELL dataset for developing robust stress detection models.

Siddharth et al. \citep{Siddharth2019} explored affective computing using multimodal data, leveraging biosignals (EEG, ECG, GSR, HRV) and visual data (facial video) in experiments conducted on four publicly available multimodal affective datasets: DEAP, AMIGOS, MAHNOB-HCI, and DREAMER. These datasets include a range of physiological signals and visual information, with participants self-reporting affective states after watching videos. The results showed that combining biosignals with visual data significantly improved affective classification performance, with the model outperforming previous research particularly on the DEAP and MAHNOB-HCI datasets. Additionally, the study addressed issues such as discrepancies in sampling rates, sensor positions, and the number of signal channels through dataset fusion and transfer learning. Both studies emphasize the potential of multimodal data in stress and affective computing, offering valuable resources for the development of personalized stress management and affective recognition systems.

Inspired by the success of deep learning in processing multimodal information \cite{Zeng2024e,Wang2023a,Zeng2023c,Li2023h,Zeng2022,Zheng2024,Zeng2024g,Wang2023f,Zeng2024b,Wang2011a,Zeng2024f,Wang2011,Zeng2024c,Zhu2013,Zeng2024d,Wang2015b,Zeng2024,Wang2018a,Zeng2024a,Wang2020h,Zeng2023a,Zeng2023,Zeng2018}, this study focuses exclusively on using physiological signals for affective computing, aiming to accurately identify participants' affective states through these signal features. In contrast to multimodal data fusion approaches, this research emphasizes optimizing the processing and classification of single physiological signals, exploring their potential in affective recognition. This approach offers the possibility of simplifying affective computing systems while reducing the complexity of data collection and processing.

\subsection{Explainability Method}

Symbolic regression is defined as the process of discovering symbolic expressions that fit data for an unknown function. Although this problem is theoretically considered NP-hard, in practice, many functions exhibit simplifying properties such as symmetry, separability, and compositionality, making the task feasible \cite{Udrescu2020}. Symbolic regression can be applied across various fields, including economics, psychology, and biomedical engineering. It aids researchers in uncovering hidden mathematical models from experimental data, thus revealing the underlying dynamics of complex systems.

The SINDy (Sparse Identification of Nonlinear Dynamics) method proposed by Brunton et al. \citep{Brunton2016} combines symbolic regression with sparse regression to overcome the limitations of traditional symbolic regression in handling complex nonlinear equations, such as high computational cost and overfitting. SINDy reduces the candidate function space and uses convex optimization to generate concise and explainable equations, ensuring both efficiency and explainability in large-scale systems. This method has been successfully applied to complex systems such as fluid vortices and chaotic Lorenz systems, demonstrating broad applicability. Rogers et al. \citep{Rogers2024} further expanded the application of symbolic regression by integrating it with model-based design of experiments (MBDoE). Symbolic regression generates explainable expressions, while MBDoE optimizes experimental conditions, allowing for rapid differentiation between model candidates and significantly improving process optimization efficiency. This approach has performed exceptionally well in industrial processes like multiphase product synthesis. By incorporating physical knowledge constraints, the complexity of symbolic regression is controlled, enhancing both the explainability and practical value of the models. This combined method shows significant potential for applications in digital manufacturing, process engineering, and new product development, making it suitable for both laboratory research and industrial production optimization.

The expanded application of symbolic regression in psychology and other fields demonstrates its significant interdisciplinary potential. Masato et al. \citep{Masato2023} were the first to apply the symbolic regression tool AI-Feynman to intertemporal choice experiments in psychology, successfully generating seven candidate discount function models, some of which outperformed existing hyperbolic discount models. This study shows that symbolic regression can not only automatically uncover hidden patterns in psychology but also significantly improve the automation and precision of research, shifting away from traditional approaches reliant on human intuition and experience. The introduction of symbolic regression offers psychologists a novel data analysis method, enhancing the accuracy of theoretical modeling and the reproducibility of studies, thereby showcasing the unique value of applying AI technology to the social sciences. Liu et al. \citep{Liu2024} further expanded the application of symbolic regression by integrating it with deep learning in the field of knowledge tracing. Their method automatically extracts algebraic expressions of learners’ cognitive states, revealing the underlying patterns of skill acquisition. In experiments on the large-scale Lumosity training dataset, symbolic regression not only improved the model’s fit but also discovered entirely new patterns of skill acquisition, verifying some existing theoretical findings. This approach provides theoretical support for analyzing large-scale behavioral data, particularly excelling in dynamic learning processes and naturally generated data, addressing challenges like the unobservability of cognitive states and the expanding search space of symbolic models. The application of symbolic regression in psychology, education, and related fields opens new avenues for research on skill acquisition, cognitive diagnostics, and personalized learning path modeling, demonstrating its broad potential in automated pattern discovery and model building.

Inspired by the high performance of deep learning modeling \cite{Zeng2022a,Wang2022t,Zeng2021a,Wang2021m,Zeng2021b,Wang2020h,Zeng2020,Wang2021,Zeng2021c,Tian2018,Wang2015a,Zeng2020a,Min2018,Wang2017}, this paper introduces symbolic regression into affective computing and combines it with deep learning models, offering a new approach for automated affective state inference. Traditional affective computing relies on black-box models, which, while accurate, lack explainability. Symbolic regression addresses this by extracting explainable algebraic expressions from neural networks, enabling the analysis of complex relationships between physiological signals and behavioral features. This improves model transparency and provides a theoretical basis for the quantitative analysis of affective states.

\subsection{Sequential Model}

Sequential models play a crucial role in affective computing, particularly when handling time-series data, where they have demonstrated exceptional performance. Traditional sequence models such as Hidden Markov Models (HMM) \citep{Rabiner1989} and Conditional Random Fields (CRF) \citep{Lafferty2001} have been widely applied in fields like natural language processing. However, with the rise of deep learning, more advanced models like Recurrent Neural Networks (RNN) \citep{Mikolov2010} and Long Short-Term Memory Networks (LSTM) \citep{Frinken2012} have been introduced into affective computing to better capture temporal dependencies.

RNN are among the earliest deep learning models used for processing sequential data. They capture contextual information in time series through recurrently connected neurons and are suitable for various sequence tasks, such as speech recognition and machine translation. However, RNNs face challenges when dealing with long-term dependencies due to the vanishing and exploding gradient problems, making it difficult to retain distant dependencies \citep{Mikolov2010}. To address these limitations, LSTM networks were developed. LSTMs use unique memory cells and gating mechanisms to effectively overcome the information loss problem in traditional RNNs when handling long sequences. Their memory cells retain important information over time, controlling what to keep and what to forget \citep{Gers1999}. LSTM networks excel at tasks involving long-term dependencies, particularly in affective computing, such as speech affective recognition \citep{Dhavale2022} and affective state prediction based on physiological signals \citep{Shyam2024}. The memory cells in LSTM allow the model to capture and retain affective state information over extended periods, improving the accuracy of affective recognition. The combination of LSTM with CNNs further enhances the performance of multimodal affective recognition. CNNs are adept at processing spatial features, such as facial expressions in images or other static physiological signals, while LSTMs handle temporal features, such as dynamic changes in speech and heart rate. This combination allows multimodal affective computing to process both complex time-series data and integrate information from different modalities, leading to more accurate affective state predictions.

Recent studies have further advanced the application of Transformer-based sequence models in affective computing. Unlike LSTM, Transformer models process sequential data in parallel through the self-attention mechanism, overcoming the efficiency bottlenecks that RNNs and LSTMs face when handling long sequences. This makes Transformers particularly suitable for processing longer time-series data \citep{Samira2024}. The self-attention mechanism flexibly captures global contextual information within a sequence without relying on sequential order, making Transformer models more effective in feature extraction and temporal modeling. Mittal et al. \citep{Mittal2020} proposed an M3ER model based on multimodal data, integrating facial expressions, text, and speech modalities. The model incorporates multiplicative fusion techniques to weigh the reliability of each modality, allowing it to automatically emphasize more reliable modalities while suppressing those with higher noise levels. Additionally, M3ER applies canonical correlation analysis (CCA) to filter out irrelevant signals from modalities and generate proxy features, enhancing the model’s robustness against noise and missing data. Experimental results on the IEMOCAP and CMU-MOSEI benchmark datasets show that this method improved accuracy in affective computation tasks by approximately 5\% compared to previous models.

Recent researchers have applied the aforementioned approaches to affective computation tasks. Kumar et al. \citep{Kumar2023} proposed two deep learning-based methods for speech affective recognition: CNN-LSTM and Vision Transformer (ViT). The study experimentally compared the performance of these two models in handling affective recognition tasks, focusing on the advantages of CNN-LSTM in audio feature extraction and affective classification, as well as the potential of ViT for processing speech signals through Mel-spectrograms. The CNN-LSTM model achieved an accuracy of 88.50\% on the EMO-DB dataset, while the Vision Transformer model reached 85.36\%. This work highlights the potential of deep learning technologies, particularly attention mechanisms and image processing techniques, in speech affective recognition, demonstrating the effectiveness of combining CNN and LSTM to extract affective features from speech signals.

Inspired by the powerful information extraction capabilities of deep learning \cite{Wang2025,Liao2024,Wang2024m,Dong2024,Wang2023g,Wang2023j,Wang2023w,Wang2023l,Wang2023d,Ma2023b,Li2023i,Li2023g,Li2023f,Wang2022at,Wang2022as,Lyu2022,Min2019}, this paper focuses on physiological signal sequence models, aiming to explore how to better utilize physiological signals for affective state prediction and enhance the accuracy and generalization of affective computing through multimodal approaches.

\section{Preliminary}

\subsection{Notations and Definitions}

In this section, we formally define the computational modules involved in the affective computation tasks of this study and provide a set of mathematical notations used throughout the paper, as presented in the Table \ref{table1_notations}.

\begin{table}[htbp]
  \centering
  \caption{Definitions of mathematical notation used in this paper.}
  \label{table1_notations}
  \begin{tabular}{c|c}
  \toprule
  Notations & Description \\ \hline
  $\mathcal{P}$ & A set containing $N$ participants is denoted as $\mathcal{P}={p_1,p_2,\ldots,p_N}$ \\ \hline
  $\mathcal{A}$ & A set of $k$ individual attributes is denoted as $\mathcal{A} = \{a_1, a_2, \ldots, a_k\}$, each element represents a feature. \\ \hline
  $\mathcal{B}$ & A set of $M$ physiological indicators obtained from a monitoring device is denoted as $\mathcal{B} = \{b_1, b_2, \ldots, b_M\}$. \\ \hline
  $D$ & Physiological monitoring data of $N$ participants with $M$ indicators. \\ \hline
  $e$ & The affective state of the participants at the current moment. \\ \hline
  $\mathcal{X},X$ & The set of all windows after window segmentation, and the data contained within a specific window. \\ \hline
  $\xi$ & The number of window segments. \\ \hline
  $PF$ & Integrated features of $A$ and $B$, denoted as $PF=A \oplus B$. \\ \hline
  $\alpha^{b_j}$ & The contribution level of physiological indicator $b_j$  \\ \hline
  $\beta^{b_j}$ & The encoded physiological indicator $b_j$. \\ \hline
  $\gamma^{b_j}$ & Integrated features of $A$ and $\beta^{b_j}$, denoted as $\gamma^{b_j}=A \oplus \beta^{b_j}$. \\ \hline
  $\theta^{b_j}$ & The affective state level reflected by the physiological indicator $b_j$.  \\ \hline
  $u$ & Initial affective state level. \\ \hline
  $\Theta$ & The affective state level reflected by all physiological indicators. \\ \hline
  $\Phi$ & Affective state level from all indicators, adjusted by initial state. \\ \bottomrule
  \end{tabular}
  \end{table}

\textbf{Definition 1 (\textit{ContribNet}): }\textit{ContribNet} is a neural network model used to compute the contribution level of physiological indicator $b_j$, consisting of a batch normalization layer and two linear transformation layers. Specifically, the input data is first processed through batch normalization, followed by a linear transformation through the first weight matrix $\mathcal{W}_1^{b_j}$ with an added bias $\mathscr{b}_1^{b_j}$, then activated by a nonlinear activation function $\sigma$. The activated output is further linearly transformed by the second weight matrix $\mathcal{W}_2^{b_j}$, with an added bias $\mathscr{b}_2^{b_j}$, to obtain the final output, representing the contribution level of physiological indicator $b_j$ to the prediction of individual $p_i$'s affective state. This process can be formally represented as:
\begin{equation}
  ContribNet(\cdot)=\mathcal{W}_2^{b_j}\left(\mathrm{\sigma}\left(\mathcal{W}_1^{b_j}\cdot BN \left(\cdot \right)+\mathscr{b}_1^{b_j}\right)\right)+\mathscr{b}_2^{b_j}\ 
\end{equation}
Where: 
\begin{itemize}
  \item $\mathcal{W}_1 \in \mathbb{R}^{H \times (k+m)}$ and $\mathcal{W}_2 \in \mathbb{R}^{1 \times H}$ represent the weight matrices for each layer;
  \item $\mathscr{b}_1 \in \mathbb{R}^H$ and $\mathscr{b}_2 \in \mathbb{R}$ represent the biases;
  \item $H$ denotes the hidden layer dimension;
  \item $\sigma$ denotes the activation function ReLU;
  \item $BN$ represents the batch normalization operation.
\end{itemize}

\textbf{Definition 2 (\textit{AffectNet}): }\textit{AffectNet} is used to compute the affective state level reflected by physiological indicator $b_j$. It extracts and transforms input features progressively through a combination of linear transformations and nonlinear activation functions across multiple hidden layers, ultimately generating the final affective state prediction. Specifically, after the input features are processed through batch normalization, they first undergo a linear transformation by the first layer weights $\mathfrak{W}_1^{b_j}$, with an added bias $\mathfrak{b}_1^{b_j}$, followed by nonlinear activation through the activation function $\sigma$. This process continues in subsequent layers until the final layer $l$, where the weights $\mathfrak{W}_l$ produce the output. This output represents the affective state level reflected by physiological indicator $b_i$ for participant $p_i$. The deep structure of the network allows it to capture complex feature interactions through layer-by-layer learning, enabling the estimation of the participant's affective state. This model can be formally expressed as:
\begin{equation}
  AffectNet\left(\cdot\right)=\left(\mathfrak{W}_l^{b_j}\left(\sigma\left(\mathfrak{W}_{l-1}^{b_j}\left(\cdots\sigma\left(\mathfrak{W}_1^{b_j}\cdot BN\left(\cdot\right)+\mathfrak{b}_1^{b_j}\right)\cdots\right)+\mathfrak{b}_{l-1}^{b_j}\right)\right)+\mathfrak{b}_l^{b_j}\right)
\end{equation}
Where:
\begin{itemize}
  \item $\mathfrak{W}_1^{b_j}, \mathfrak{W}_2^{b_j}, \ldots, \mathfrak{W}_l^{b_j}$ represent the weight matrices for each layer;
  \item $\mathfrak{b}_1^{b_j}, \mathfrak{b}_2^{b_j}, \ldots, \mathfrak{b}_l^{b_j}$ represent the bias vectors for each layer;  
  \item $\sigma$ denotes the activation function ReLU;
  \item $BN$ represents the batch normalization operation.
\end{itemize}

\textbf{Definition 3 (\textit{AffectAnalyser}): }\textit{AffectAnalyser} is a network model used to calculate an individual's affective state at the current moment. Each layer performs a linear transformation on the input features using weight matrices and bias vectors, progressively deriving the individual's affective prediction. Specifically, the input features undergo a linear transformation in the first layer, where a linear operation is performed using the weight matrix $\mathscr{g}_1$ and a bias $c_1$ is added to obtain an intermediate result. This is then followed by a second linear transformation, processed by the weight matrix $\mathscr{g}_2$ and bias vector $c_2$, ultimately generating the affective state estimate for individual $p_i$. This model can be formally represented as:
\begin{equation}
  AffectAnalyser(\cdot)=\mathscr{g}_2\left(\mathscr{g}_1\cdot+c_1\right)+c_2
\end{equation}
Where:
\begin{itemize}
  \item $\mathscr{g}_1$ and $\mathscr{g}_2$ represent the weight matrices of the first and second layers, respectively;  
  \item $c_1$ and $c_2$ represent the bias vectors of the first and second layers, respectively.
\end{itemize}

\subsection{Problem Formulation}

The objective of this study is to identify an individual's affective state using data obtained from human monitoring devices. The task is defined as a set $\mathcal{P} = \{p_1, p_2, \ldots, p_N\}$ consisting of $N$ individuals, where each individual has $k$ corresponding features $\mathcal{A} = \{a_1, a_2, \ldots, a_k\}$. It is assumed that a monitoring device can capture $M$ types of physiological indicators, denoted as $\mathcal{B} = \{b_1, b_2, \ldots, b_M\}$, where $\mathcal{B} \subseteq \{\text{ACC}, \text{BVP}, \text{ECG}, \text{EDA}, \text{EMG}, \text{RESP}, \text{TEMP}\}$. The physiological monitoring data of a set of participants can be represented as $D = \{d_{p_i, b_j} \mid p_i \in \mathcal{P}, b_j \in \mathcal{B}\}$. Based on these definitions, the task of this study can be defined as $e_{p_i} = \text{PhysioFormer}(A_{p_i}||d_{p_i, b_j})$, where $e_{p_i} \in \{0,1,2\}$, with $0$ representing a normal state, $1$ representing an excited state, and $2$ representing a stressed state.

\section{Research Method}

The research task of this study can be divided into two parts. First, a PhysioFormer model is constructed to predict an individual's affective state at a given moment. Through the collaborative functioning of three submodules: feature embedding, affective representation, and affective state prediction. The model effectively transforms physiological data into predicted affective states. Second, symbolic regression is applied to fit various monitoring indicators, enabling a more precise capture and understanding of the relationships between these indicators and affective states. This approach not only enhances the accuracy of affective state predictions but also provides a deeper analysis and explanations of the physiological indicators.

\subsection{The PhysioFormer Model}

\subsubsection{Model Overview}

\begin{figure}[htbp]
	\centering
	\includegraphics[width=.8\textwidth]{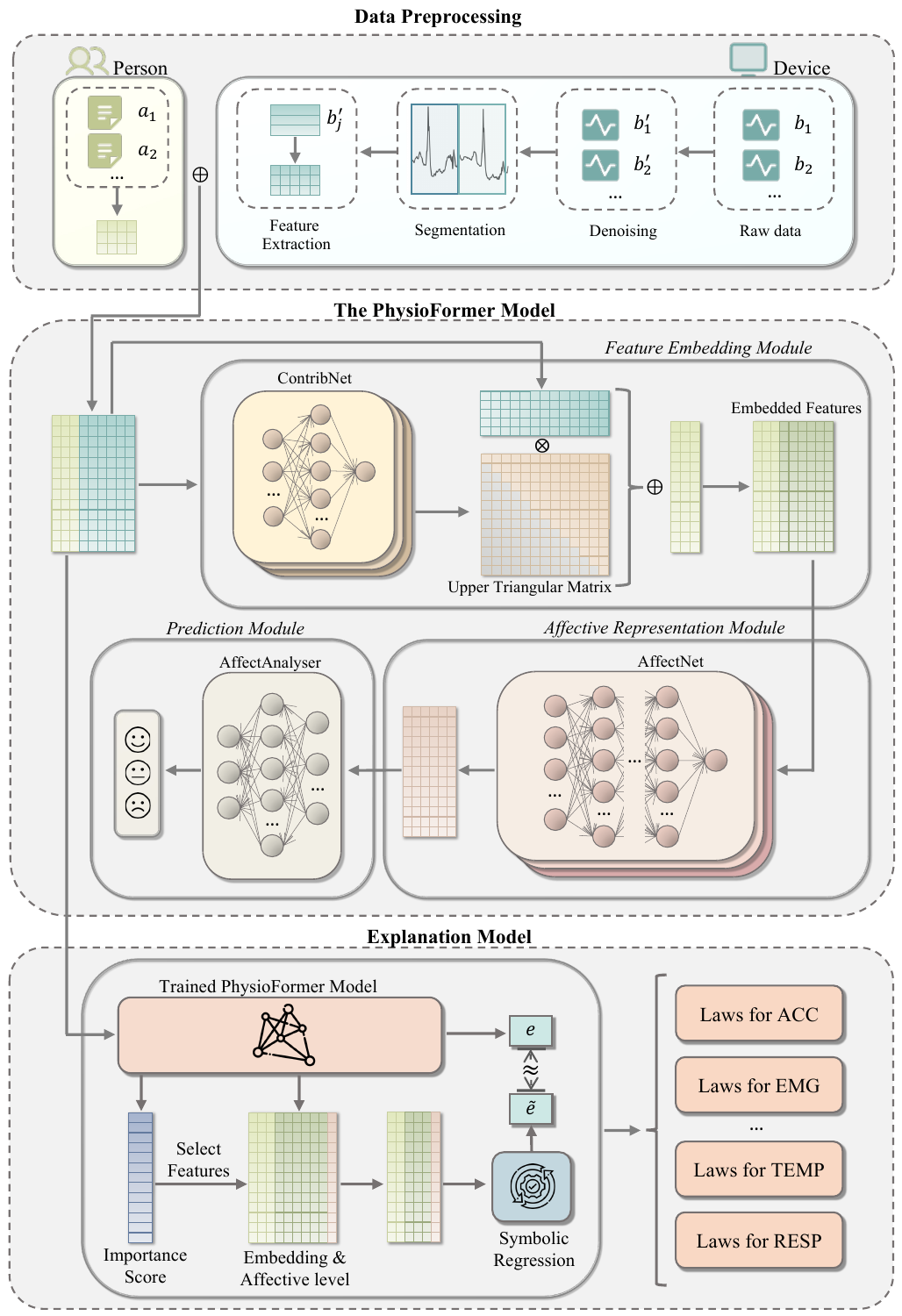}
	\caption{PhysioFormer model architecture consists of three submodules: the Feature Embedding Module, Affective Representation Module, and Prediction Module. The Feature Embedding Module encodes physiological data, the Affective Representation Module builds on these encoded features, and the Prediction Module forecasts the individual's current affective state. The Explanation model analyzes data within the trained model, generating feature importance scores and selecting key features, followed by symbolic regression to derive formulas that explain and quantify the influence of physiological indicators on affective states.}
	\label{1Physioformer}
\end{figure}

In this section, our task is to predict an individual's affective state using physiological signal data. The structure of the PhysioFormer model proposed in this paper is shown in the Figure \ref{1Physioformer}, and it can be divided into three submodules: (1) \textbf{Feature Embedding Module}, which is responsible for encoding the input features and generating feature representations containing physiological data; (2) \textbf{Affective Representation Module}, which models and represents the individual’s affective state based on the feature representations output by the Feature Embedding Module, capturing and describing the user’s affective state through neural networks; (3) \textbf{Prediction Module}, which uses the previously obtained features to predict the individual's affective state at the current moment.

Based on the aforementioned model architecture, we define the input and output of the entire model as follows:
\begin{itemize}
  \item \textbf{Input: }The input data consists of various physiological signals, including but not limited to HR, HRV, and EDA. These physiological signals are captured in real time through wearable devices or other physiological monitoring tools and are provided to the model in the form of time series. Each input feature vector not only contains these physiological signals but also includes individual attributes (such as age, gender, etc.), enabling the model to better capture inter-individual differences.
  \item \textbf{Output: }The output of the model is a prediction of the individual's current affective state, providing a classification of the affective state as one of three categories: tense, calm, or excited.
\end{itemize}

After data preprocessing, the raw physiological signal data is first processed by merging individual attributes features with physiological features to form a comprehensive feature vector. The \textit{ContribNet} is then constructed for each physiological indicator to calculate its contribution level to affective prediction, and an attention mechanism is introduced to re-encode the physiological indicators. The re-encoded physiological indicators are then concatenated with individual attributes again to form a new comprehensive feature vector. For each physiological indicator, we construct an \textit{AffectNet} to calculate the affective state level. Finally, the affective states of all physiological indicators are fed into \textit{AffectAnalyser}, which predicts the individual's affective state at the current moment. The detailed algorithm is shown in Algorithm \ref{algorithm1}.

\begin{algorithm}[htbp]
\caption{The proposed PhysioFormer model for affective computation}
\label{algorithm1}
\renewcommand{\algorithmicrequire}{\textbf{Input: }}
\renewcommand{\algorithmicensure}{\textbf{Output: }}
  \begin{algorithmic}[1]
  \REQUIRE $\mathcal{X}$ which is dataset after window segmentation and feature extraction
  \ENSURE Current moment affective categorization result $e$
  \STATE Overall features of subject $p_i$ under all windows: $PF_{p_i} \leftarrow Concat(A_{p_i},B_{p_i})$
  \REPEAT
  \STATE The contribution level $\alpha_{p_i}^{b_j}$ of a physiological indicator $b_j$: $\alpha_{p_i}^{b_j} \leftarrow ContribNet(BN(PF_{p_i}))$
    \STATE Physiological indicator $b_j$ after embedding: $\beta_{p_i}^{b_j} \leftarrow B_{p_i}^{b_j} \cdot Triu$
    \STATE Features after embedding under all windows: $\gamma_{p_i}^{b_j} \leftarrow Concat(A_{p_i}, \beta_{p_i}^{b_j})$
  \STATE Affective level reflected in physiological indicator $b_j$: $\theta_{p_i}^{b_j} \leftarrow AffectNet(BN(\gamma_{p_i}^{b_j}))$
  \UNTIL{Reaching the steps number equal to physiological indicators number, each get $\theta_{p_i}^{b_j}$}
  \STATE Splice all $\theta_{p_i}^{b_j}$ to obtain $\Theta_{p_i}$
  \STATE Sum with initial affective level $u_{p_i}$: $\Phi_{p_i} \leftarrow \Theta_{p_i} + u_{p_i}$
  \STATE Affective categorization result $e_{p_i} \leftarrow AffectAnalyser(\Phi_{p_i})$

  \RETURN $e_{p_i}$ 
  \end{algorithmic}
\end{algorithm}

\subsubsection{Data Preprocessing}

The raw dataset containing physiological signals typically consists of sampling times and their corresponding measurements. Before using this data as input for the neural network, proper preprocessing is required. The following section will describe in detail the data preprocessing methods employed in this study to ensure data quality.

\begin{enumerate}[A.]
  \item \textbf{Denoising. }During the process of data monitoring, various types of noise are often present, such as power line interference, motion artifacts, and other environmental noise \citep{signal_filter}. To address this, we employed a Butterworth low-pass filter to remove high-frequency noise. This is a smoothing filter with maximally flat response characteristics \citep{Bianchi2007}. The filter is used to eliminate high-frequency noise from the signal data and perform signal smoothing, effectively reducing noise and improving the signal-to-noise ratio. The smoothed signal more accurately reflects the participant's actual physiological state, preventing false fluctuations caused by noise.

  \item \textbf{Segmentation. }Since physiological signal datasets are time series data, they need to be segmented into fixed-size windows to generate the corresponding time series dataset. The Figure \ref{3window_damo} illustrates the process of data segmentation and window sliding used in this study. In the figure, Window1, Window2, and Window3 represent consecutive time windows. Each window contains time series data of the same length, and there is no partial overlap between the windows, allowing them to independently reflect physiological changes during each time period.
  \begin{figure}[htbp]
    \centering
    \includegraphics[width=.4\textwidth]{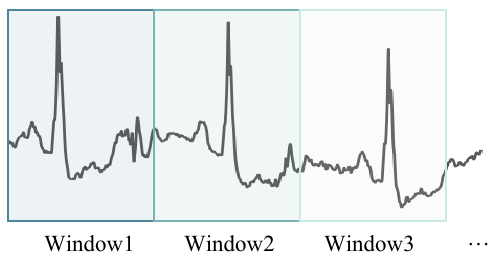}
    \caption{Example of time window segmentation in the dataset. This figure illustrates the time window segmentation process for the physiological signal dataset. The monitoring data is divided into multiple consecutive time windows, with each window being independent and having no overlapping parts.}
    \label{3window_damo}
  \end{figure}
  The total length of the monitoring data is $L$ seconds, and the length of each window is set to $T$ seconds. Therefore, the monitoring data of participant $p_i$ in the dataset can be divided into $\xi_{p_i} = \left\lfloor \frac{L}{T} \right\rfloor$ non-overlapping time windows. The set of all windows can be represented as:
  \begin{equation}
    \mathcal{X}_{p_i}=\{X_{1_{p_i}},X_{2_{p_i}},\ldots,X_{\xi_{p_i}}\}
  \end{equation}
  The data contained in the $n$-th window can be represented as:
  \begin{equation}
    X_{q_{p_i}}=\{x_{p_i}(t):t\in\left[\left(q-1\right)T,qT\right]\}
  \end{equation}
  Here, $x_{p_i}(t)$ represents the monitoring data at time $t$. Each window $X_{q_{p_i}}$ is independent, meaning there is no overlap between windows, which helps to avoid data overlap issues during feature extraction and subsequent analysis.

  \item \textbf{Feature Extraction. }Physiological signal data from the human body is complex and diverse, making it difficult to directly extract feature information. Therefore, after segmenting the dataset according to window size, feature extraction is necessary to transform the dataset into a format that can reveal affective states and be more easily processed by neural network models. To better understand and handle this data, we analyzed each type of physiological indicator and described the feature extraction methods for each indicator within each window after the fixed-size segmentation.

  For each indicator, we calculated its mean, standard deviation, maximum, and minimum values within each window, collectively referred to as the basic statistical features of the signal.
  \begin{itemize}
    \item \textbf{ACC: }The ACC data contains three dimensions (x, y, z), and the basic statistical features were calculated for each dimension. Additionally, the sum of the acceleration across the three dimensions was computed, followed by calculating the corresponding basic statistical features.
    \item \textbf{EDA: }A low-pass filter was first applied to the raw data, and the cvxEDA \citep{cvxEDA} algorithm was used to compute the relevant statistical features. From these, three features that reflect both short-term affective responses and long-term affective states, as well as patterns of autonomic nervous system activity, were selected, and their corresponding basic statistical features were calculated.
    \item \textbf{EMG: }A low-pass filter was first applied to the raw data, and then the basic statistical features were calculated within each window.
    \item \textbf{BVP: }For this indicator, the peak frequency within each window was computed. As this indicator reflects heart activity, HRV was also calculated within the window using the NeuroKit2 tool \citep{NeuroKit2}.
    \item \textbf{ECG: }The NeuroKit2 tool was used to compute HRV within the window, and HRV features such as standard deviation (SDNN) and root mean square of successive differences (RMSSD) were extracted.
    \item \textbf{TEMP: }TEMP is a numerical signal, and its basic statistical features were calculated within each window. Additionally, the slope of TEMP within the window was computed to reveal temperature trends over a specific time period.
    \item \textbf{RESP: }RESP is a numerical signal, and the basic statistical features of the respiratory rate within the window were calculated.
  \end{itemize}
For detailed information on the specific features, please refer to Appendix A, which provides the variable definitions and descriptions for each feature.
\end{enumerate}

\subsubsection{Feature Embedding Module}

The core task of feature embedding module is to extract features from the data obtained through sliding windows. This step not only effectively captures the temporal characteristics of physiological signals but also enhances the model's adaptability and reliability to individual differences by incorporating individual attributes. By merging individual attributes with physiological features to form a comprehensive feature vector, the model can gain a more holistic understanding of each individual's unique physiological responses in different affective states. The contribution level of physiological indicators is calculated using \textit{ContribNet}, and an attention mechanism is introduced to re-encode the physiological indicators. This allows the model to dynamically adjust its focus on various physiological signals, ensuring that the most informative features are fully utilized in affective prediction. This refined feature processing method not only improves the accuracy of affective computation but also enhances the model's ability to capture complex affective states, laying a solid foundation for subsequent affective representation and affective computation.

Specifically, the dataset is divided into multiple fixed-length windows, each containing a processed segment of continuous time-series data, which represents the overall features of each physiological indicator. This section uses two types of data to represent feature embedding. One part is the individual attributes features, containing basic information about the individual, such as age, gender, and height. Here, the individual attributes features of participant $p_i$ are denoted as $A_{p_i}$. The other part is the windowed monitoring data, consisting of the feature values calculated from the physiological indicator data within a given window. The monitoring data for participant $p_i$ is denoted as $B_{p_i}$. These feature data include both discrete and continuous data. For discrete data, one-hot encoding is used to represent categorical information in a format suitable for model processing. Continuous data are directly represented by their raw values to preserve their true quantitative information. By integrating individual attributes features $A_{p_i}$ with monitoring data $B_{p_i}$, the overall features of the participant across all windows can be represented as:
\begin{equation}
  PF_{p_i}=A_{p_i}\oplus B_{p_i}
\end{equation}
Here, $\oplus$ denotes the feature concatenation operation, and $PF_{p_i} \in \mathbb{R}^{k+m}$, where $m$ represents the dimension of the features calculated from the physiological indicator data.

Given that the overall static features $PF_{p_i}$ consist of multiple dimensions and exhibit complex linear relationships, this study constructs a \textit{ContribNet} network to process these high-dimensional features. The network possesses nonlinear modeling capabilities and can effectively capture and represent the complex relationships between input features through a multi-layer structure. To accommodate the characteristics and requirements of different indicators, we built an independent \textit{ContribNet} for each type of physiological indicator. In summary, the contribution level of physiological indicator $b_j$ can be represented as:
\begin{equation}
  \alpha_{p_i}^{b_j}=ContribNet\left(PF_{p_i}\right)
\end{equation}

In the feature encoding process, special attention must be paid to the temporal variation of physiological indicators. One of the key characteristics of time series data is its sequential nature, meaning that the physiological state at a given moment is often influenced by prior moments, with this influence being progressive. Therefore, accurately capturing these dynamic temporal changes is critical during feature extraction and encoding. In the feature encoding process, two types of data are used: individual attributes features and physiological indicators. Individual attributes features can be considered static features that remain constant over the entire time period, such as age and gender, and thus do not require temporal feature fusion. However, physiological indicators vary over time, and it is essential to capture their dynamic characteristics to more accurately reflect the changes in an individual's physiological state over time. To model temporal dependencies, we constructed an upper triangular matrix. This matrix ensures that the features at each time point depend only on the current and prior moments, without relying on future moments. This approach ensures that the model is sensitive to the temporal dynamics of physiological signals, allowing it to capture more subtle emotional changes. Based on the process described above, the encoded physiological indicator $b_j$ for participant $p_i$ can be represented as:
\begin{equation}
  \beta_{p_i}^{b_j}=\left(\left(B_{p_i}^{b_j}\right)^T\cdot Triu \left(\alpha_{p_i}^{b_j}\cdot1_{1\times\xi}\right)\right)^T
\end{equation}
Here, $Triu$ denotes the upper triangular matrix operation, and $\xi$ represents the number of windows.

Based on the aforementioned encoding process, the overall features of physiological indicator $b_j$ for individual $p_i$ across all windows can be represented as:
\begin{equation}
  \gamma_{p_i}^{b_j}=A_{p_i}\oplus\beta_{p_i}^{b_j}
\end{equation}

\subsubsection{Affective Representation Module}

The core task of affective representation module is to map the overall features of an individual across all time windows to affective state levels, which is crucial for accurately predicting affective states. By aggregating features from the time series, the affective representation module can identify and model the temporal patterns of affective states. The \textit{AffectNet} network is used here to map high-dimensional features into a lower-dimensional affective state space, simplifying data complexity while retaining key affective information and handling subtle affective changes. By establishing contextual relationships within the time series, the affective representation module captures the continuity and evolution of affective states, thus more accurately reflecting an individual's actual affective experience, laying the foundation for final affective classification and state prediction.

To achieve this task, we constructed an \textit{AffectNet} network, which has strong feature integration capabilities and can effectively fuse and process multiple types of feature information. This method allows for more precise capture and reflection of the dynamic changes in an individual's affective state, enhancing the accuracy and reliability of affective state predictions. When building the model, an individual \textit{AffectNet} network was constructed for each type of physiological indicator, allowing it to process its corresponding physiological data independently and dynamically adjust its parameters in response to variations in different physiological indicators. This enables more precise feature extraction and affective state prediction. In summary, the affective state level reflected by physiological indicator $b_j$ can be represented as:
\begin{equation}
  \theta_{p_i}^{b_j}=AffectNet(\gamma_{p_i}^{b_j})
\end{equation}

Here, $\Theta_{p_i} = \{\theta_{p_i}^{b_j} | \mathbf{i} \in \mathbf{N}, \ \mathbf{j} \in \mathbf{M}\}$ is used to represent the set of affective state levels mapped from all physiological indicators.

Additionally, to improve overall prediction accuracy and model reliability while reducing error accumulation, we introduced the initial affective state as a baseline, denoted as $u_{p_i}$, representing the initial affective state level before training. Thus, $\Theta_{p_i}$ is adjusted to:
\begin{equation}
  \Phi_{p_i}=\Theta_{p_i}+u_{p_i}
\end{equation}

Thus, the model is not only able to capture affective state changes caused by variations in physiological indicators but also takes into account the individual's initial state, providing a more comprehensive affective state assessment.

\subsubsection{Prediction Module and Model Training}

The core task of the Prediction module is to use the individual's affective state levels to predict their current affective state. This module analyzes and processes the affective state levels output by the affective representation module and applies the \textit{AffectAnalyser} network to map the affective state levels to specific affective categories.

We constructed an \textit{AffectAnalyser} network to accomplish this task. By mapping the input feature vector to a high-dimensional feature space, it captures the complex relationships between input features, enabling effective classification. Based on the adjusted affective state levels $\Phi_{p_i}$ obtained earlier, the final affective prediction is represented as:
\begin{equation}
  e_{p_i}=AffectAnalyser\left(\Phi_{p_i}\right)
\end{equation}

During the training process, it is essential to account for the complexity and diversity of physiological data. Therefore, we chose to use the cross-entropy loss function. The cross-entropy loss function is well-suited for classification problems, as it measures the difference between the predicted probability distribution and the true label distribution, providing a natural probabilistic explanations and exhibiting favorable gradient properties. Additionally, we introduced a regularization term into the loss function, which helps constrain the size of the model parameters, smoothens them, prevents overfitting, and improves the model’s generalization ability.

The complete loss function for the model is as follows:
\begin{equation}
  \mathcal{L}=-\frac{1}{N}\sum_{i=1}^{N}\log{\left(\frac{\exp{\left(e_{p_i}^{\left(\hat{e}_{p_i}\right)}\right)}}{\sum_{k=1}^{3}\exp{\left(e_{p_i}^{\left(k\right)}\right)}}\right)}+\lambda\cdot\frac{1}{N}\sum_{i=1}^{N}{\frac{1}{M}\sum_{j=1}^{M}\left(\alpha_{p_i}^{b_j}-1\right)^2}
\end{equation}

The above function consists of two parts: the first part is the Cross Entropy loss function, where $e_{p_i}$ represents the predicted affective state of the $i$-th individual, and $\hat{e}_{p_i}$ represents the true affective state of the $i$-th individual. This function is used to calculate the difference between the predicted and actual values in the classification task. The second part is the regularization term, where $\lambda$ is the regularization coefficient that controls the strength of the regularization term.

Based on the aforementioned loss function, we selected the \textbf{RMSprop} optimizer to optimize the model. RMSprop dynamically adjusts the learning rate for each parameter by taking the moving average of the squared gradients, thus providing an adaptive learning rate for each parameter. Given that the loss function includes both the cross-entropy loss and the regularization term, this may result in a non-stationary objective function, meaning that statistical data may exhibit time-varying and time-dependent properties during different periods \citep{Non_stationarity}. The RMSprop optimizer calculates the moving average of the squared gradients using exponential decay, effectively handling this non-stationary objective function and significantly improving the stability of the optimization process. Additionally, since RMSprop automatically adjusts the learning rate, it typically converges faster compared to fixed learning rate optimizers such as Stochastic Gradient Descent (SGD).

In this study, the model’s training mode adopts a cross-validation strategy. Specifically, in each epoch, the model sequentially uses data from each individual for training, with the goal of explicitly learning and adapting to the physiological signal characteristics of different individuals. This method allows the model to progressively capture the unique physiological features of each individual, enhancing its ability to generalize across individuals. Specifically, in each epoch, the model is trained through the following process.

For the $i$-th participant $p_i$, the prediction of their affective state can be expressed as:
\begin{equation}
  e_{p_i}=PhysioFormer\left(A_{p_i}||d_{p_i,b_j}\right)
\end{equation}
Here, $A_{p_i}$ represents the individual attributes features, and $d_{p_i,b_j}$ represents the physiological data features.

For each participant $p_i$, the loss function $\mathcal{L}$ is used to measure the difference between the predicted value $e_{p_i}$ and the true value $hat{e}_{p_i}$:
\begin{equation}
  \mathcal{L}(e_{p_i},\hat{e}_{p_i})=\mathcal{L}(PhysioFormer\left(A_{p_i}||d_{p_i,b_j}\right),\hat{e}_{p_i})
\end{equation}

In each epoch, the model sequentially uses data from each participant for training and updates the model parameters using the gradient descent method. For the $i$-th participant in the $t$-th epoch, the parameter update formula is:
\begin{equation}
  \varphi^{\left(t\right)}\gets\varphi^{\left(t-1\right)}-\eta\cdot\nabla_\varphi\mathcal{L}(e_{p_i},\hat{e}_{p_i})
\end{equation}
Here, $\varphi^{\left(t\right)}$ represents the model parameters at the end of the current epoch, and $\varphi^{\left(t-1\right)}$ represents the model parameters at the end of the previous epoch.

In each epoch, the model is trained on data from all $N$ participants. The entire training process can be represented as:
\begin{equation}
  \mathrm{For\ each\ epoch\ }t: \mathrm{For\ each\ }i\in\{1,2,\ldots,N\}:\varphi^{\left(t\right)}\gets\varphi^{\left(t-1\right)}-\eta\cdot\nabla_\varphi\mathcal{L}\left(e_{p_i},\hat{e}_{p_i}\right)
\end{equation}
Here, $\mathcal{L}(e_{p_i}, \hat{e}_{p_i})$ represents the loss function for each participant, and $\eta$ represents the learning rate.

\subsection{Symbolic Laws Extraction}

\subsubsection{Extraction Process Overview}

The physiological signal data involved in affective computing tasks are complex and multidimensional, often containing highly nonlinear relationships. Although the PhysioFormer model can capture these intricate dynamics, its "black box" nature tends to limit explainability, making it difficult to clearly understand the relationship between inputs and outputs. In this task, we aim to uncover the specific roles of various physiological indicators in affective computation, as well as the complex interactions between physiological signals. This will not only help us better understand how multiple physiological indicators jointly influence affective states, but also enhance the trust of users and researchers in the model's predictions.

\begin{figure}[htbp]
  \centering
  \includegraphics[width=0.75\textwidth]{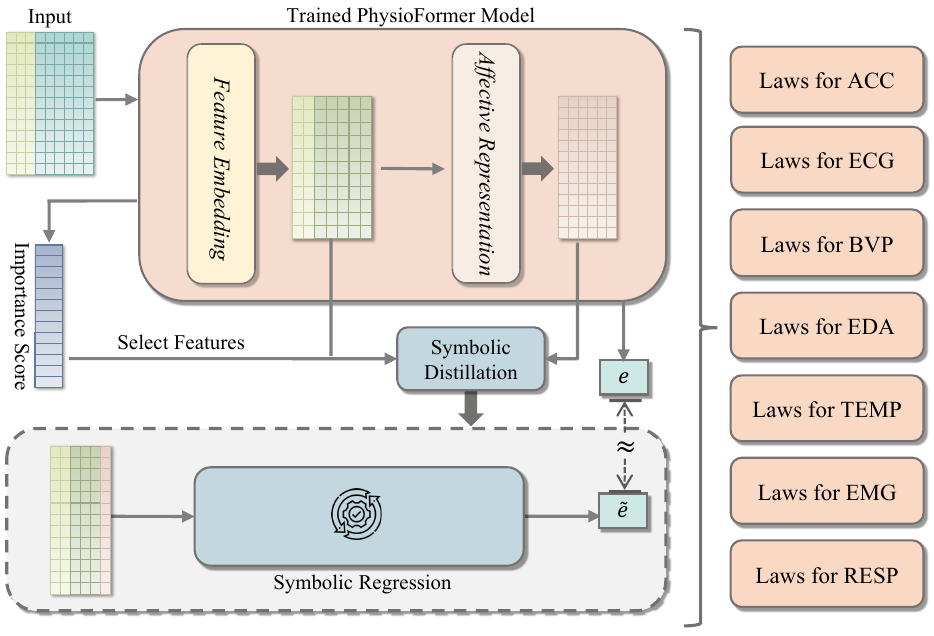}
  \caption{The figure shows the process of Explanation model. First, the input data undergoes feature extraction and affective state representation through the Feature Embedding and Affective Representation modules. The processed features and state information are used to symbolic distillation, where feature importance scores generated by the PhysioFormer model are used to select key features. Next, symbolic regression is employed to generate the predicted value $\tilde{e}$, which is compared with the model's predicted value $e$ thereby extracting and generating symbolic laws for the physiological indicators.}
  \label{2symbolic}
\end{figure}

To achieve this goal, we designed an Explanation Model that combines symbolic distillation with deep learning to improve both the explainability and performance of the model. The detailed process structure is shown in the Figure \ref{2symbolic}. First, the input data is processed through the Feature Embedding and Affective Representation modules, generating features and state information. The trained PhysioFormer model generates feature importance scores, based on which key features are selected. Using symbolic regression, predicted values $\tilde{e}$ are generated and compared to the model's predicted values $e$. This process extracts symbolic laws for each physiological indicator, making the internal mechanisms of the model more transparent and providing explanations and quantifications of the influence of different physiological indicators on affective states.

\subsubsection{Symbolic Distillation for Physiological Indicators}
To enhance the explainability of the internal mechanisms of the PhysioFormer model, we introduced a Explanation Model, transforming the influence of various physiological indicators on affective state prediction into symbolic expressions. In this study, we employed symbolic regression techniques with the aim of finding the optimal mathematical formula that fits the given data, thereby revealing the model's intrinsic decision logic and the interactions between physiological indicators. Let $P$ represent the network model to be analyzed, $S$ represent the symbolic regression model, and $X \in \mathbb{R}^{\xi \times (k+m)}$ represent the input data to the model. According to the objective of symbolic regression, we can define the optimization goal of symbolic regression as:
\begin{equation}
  \varepsilon={\underset{\varepsilon}{\arg\min}\left|P(X)-S(\varepsilon,X)\right|}
\end{equation}
Here, $\varepsilon$ represents the mathematical formula calculated by symbolic regression, $X$ denotes the range of input data, $\xi$ represents the number of windows into which the individual's physiological data is segmented, $k$ denotes the feature dimension of the individual attributes features, and $m$ represents the dimension of the features computed from the physiological data.

Since symbolic regression is a typical combinatorial optimization problem, the solution space grows exponentially with the number of variables. Using all the features for computation would incur enormous computational costs and reduce the model's efficiency and stability. To address this issue, we applied a gradient-based feature importance estimation method to select a subset of features, which were then used for the symbolic regression task. Specifically, this method calculates the gradient of the intermediate outputs of the model with respect to the input features. These gradients reflect how small changes in the input features influence the output \citep{grad_importance}. The resulting gradients may be either positive or negative, but we are concerned with the magnitude of change rather than the direction. Therefore, the absolute value of the gradients is taken to represent feature importance, and these absolute values are normalized to obtain relative importance scores. This process can be expressed by the following formula:
\begin{equation}
  S=\sum_{i=1}^{\xi}y_i
\end{equation}
Here, $y_i$ represents the intermediate output of the model.

The gradient of the output $S$ with respect to the input features $X$ is calculated and then the absolute value is taken:
\begin{equation}
  \left|\frac{\partial S}{\partial X}\right|=\left|\left[\frac{\partial S}{\partial X_{11}},\frac{\partial S}{\partial X_{12}},\ldots,\frac{\partial S}{\partial X_{\xi\left(k+m\right)}}\right]\right|
\end{equation}

The importance score for each feature is calculated:
\begin{equation}
  I_j=\frac{\sum_{i=1}^{\xi}G_{ij}}{\sum_{j=1}^{\left(k+m\right)}\sum_{i=1}^{\xi}G_{ij}}
\end{equation}
Here, $I_j$ represents the importance score of the $j$-th feature.

Based on the calculated feature importance scores, a subset of features is selected for symbolic regression, denoted as $\rho(X)$. Based on the aforementioned process, the optimization objective of symbolic regression can be defined as:
\begin{equation}
  \varepsilon={\underset{\varepsilon}{\arg\min}\left|P(X)-S(\varepsilon,\rho(X))\right|}
\end{equation}

During the model optimization process, the absolute error loss function was used to optimize the model. Since each indicator's \textit{ContribNet} and \textit{AffectNet} operate independently, these components can perform symbolic regression in parallel. However, considering the coupling relationships and sequential dependencies between the components, the analysis process must be carried out step-by-step according to a predefined order to ensure overall model consistency and optimal performance.

\section{Experiments}

This section outlines the experimental design for comparing the performance of the PhysioFormer model with nine existing affective computation networks on two sub-datasets, Wrist and Chest, from the publicly available WESAD dataset. The goal is to evaluate the effectiveness of the proposed framework in affective computation tasks. Through this experiment, we aim to address five core research questions (RQ), providing empirical evidence to thoroughly understand the advantages and contributions of PhysioFormer and further validate its potential application in the field of affective computation.
\begin{itemize}
  \item RQ1: How does the performance of the PhysioFormer model compare to existing SOTA models in affective computation tasks? This research question seeks to evaluate the advantages of the PhysioFormer model in processing multimodal physiological signals.
  \item RQ2: Does the choice of different window sizes during data preprocessing significantly affect the model's prediction performance? This question explores how the window size, a critical preprocessing parameter, impacts the model’s effectiveness and accuracy.
  \item RQ3: How does the number of hidden layer neurons at different scales affect the performance of the PhysioFormer model during training? This research question aims to assess the balance between the number of neurons in the model's structure, model complexity, and performance.
  \item RQ4: Does the feature embedding module have a significant impact on the model's prediction performance? This question seeks to uncover the role of feature embedding in extracting useful information and enhancing model performance, providing insights into its importance.
  \item RQ5: Does the inclusion of individual attributes features (e.g., age, gender) significantly improve the model's prediction performance? This question investigates the role of individual characteristics in affective computation tasks and further explains their impact on the model's generalization ability.
\end{itemize}

\subsection{Dataset}
The dataset used in this study is the publicly available WESAD dataset, designed for affective computation. It contains physiological and motion data recorded by two devices: one from a chest-worn device (RespiBAN), which includes six monitoring indicators collected at a frequency of 700Hz; and another from a wrist-worn device (Empatica E4), which includes four monitoring indicators with different sampling frequencies \citep{WESAD_intro}. The physiological indicators included in this dataset are introduced below:
\begin{itemize}
  \item Accelerometer, ACC: ACC data records changes in acceleration in three-dimensional space (x, y, z axes), reflecting the intensity, direction, and frequency of body movement.
  \item Electrodermal Activity, EDA: EDA data records changes in skin conductance, which are often closely related to psychological states.
  \item Electromyography, EMG: EMG data captures the electrical activity of muscle fibers during contraction and relaxation, providing information on muscle function and nervous system control.
  \item Blood Volume Pulse, BVP: BVP data records changes in blood volume through peripheral vessels with each heartbeat, which can be used to monitor cardiovascular health and analyze heart rate variability (HRV).
  \item Electrocardiogram, ECG: ECG data records the electrical activity generated by the heart during each heartbeat, providing insights into heart health.
  \item Temperature, TEMP: TEMP data records measurements of the body’s core temperature, reflecting the individual's temperature status at specific time points.
  \item Respiration, RESP: RESP data captures the respiratory cycle and airflow changes during breathing, reflecting an individual's respiratory rate and depth.
\end{itemize}

\subsection{Compared Method}

In this subsection, we provide an overview of the nine models used in the comparison experiments, which include both traditional machine learning methods and deep learning approaches. Specifically, the machine learning methods include Random Forest, SVM, KNN, AdaBoost, Decision Tree, and Linear Discriminant Analysis (LDA); the deep learning methods include Convolutional Neural Network (CNN), Recurrent Neural Network (RNN), and Long Short-Term Memory (LSTM). These models have been widely applied in previous affective computation studies and have achieved significant results. To ensure fairness and reliability in the comparison experiments, we reconstructed these models in this study and conducted experiments on the two sub-datasets (Wrist and Chest) of the WESAD dataset. Through this process, we are able to evaluate the performance differences between the PhysioFormer model and existing models under the same data conditions, thereby validating the effectiveness of the proposed approach and its advantages in affective computation tasks.

\begin{itemize}
  \item Bobade et al. \cite{Bobade2020} proposed a stress detection system based on multimodal physiological data, utilizing machine learning to identify individuals' stress states. The models used for classification tasks include KNN, LDA, Random Forest, Decision Tree, AdaBoost, and SVM, along with a simple feedforward neural network (ANN). The classification tasks were divided into two categories: a three-class task (pleasant, neutral, and stress) and a binary task (stress vs. non-stress). The experimental results showed that in the three-class task, the best accuracy achieved by machine learning methods was 81.65\%, while in the binary task, the accuracy reached 93.20\%. With deep learning methods, the accuracy for the three-class and binary tasks increased to 84.32\% and 95.21\%, respectively. Additionally, the study explored different feature extraction and preprocessing steps, including Principal Component Analysis (PCA) and normalization techniques, to optimize classification performance. Ultimately, the deep learning-based ANN model outperformed traditional machine learning methods in both classification tasks, becoming the best model. In my research, I have adopted the Decision Tree, SVM, and AdaBoost algorithms from this study as baseline models for comparison experiments, in order to further evaluate the performance of the PhysioFormer model in multimodal affective computation tasks.
  \item Siirtola et al. \citep{Siirtola2019} conducted continuous stress detection using sensors from a commercial smartwatch, focusing on how stress recognition can be achieved through physiological signals such as skin temperature (ST), BVP, and HR without relying on EDA signals or user dependence. The study utilized the WESAD dataset and conducted experimental comparisons using three classifiers: LDA, Quadratic Discriminant Analysis (QDA), and Random Forest. The experimental results showed that the LDA classifier combined with ST, BVP, and HR signals performed best in the stress recognition task, achieving a balanced accuracy of 87.4\%. In this study, I drew on the work of Siirtola et al. by using Random Forest and LDA classifiers as the basis for comparison experiments. I further expanded the experiments by testing the impact of different combinations of physiological signals on the model's performance.
  \item Ferdinando et al. \citep{Ferdinando2018} aimed to process EDA signals using the cvxEDA method to extract key features for affective recognition and used the KNN classifier to address a three-class affective recognition problem on the MAHNOB-HCI dataset. The experimental results showed that under subject-dependent conditions, the recognition accuracies for valence and arousal reached 74.6\% and 77.3\%, respectively. In my research, I adopted the KNN model from this study as part of the comparison experiments and combined different physiological signal sets to validate the performance of my model.
  \item Yu et al. \citep{Yu2020} conducted an in-depth exploration of affective recognition based on EDA signals, focusing on the structure and performance of three deep neural network (DNN) models: ResNet, LSTM, and a hybrid model combining ResNet and LSTM. The study was based on the MAHNOB-HCI dataset and experimented on a three-class affective recognition task for valence and arousal dimensions. The experimental results showed that the ResNet model outperformed both LSTM and the hybrid model in affective recognition tasks, achieving an accuracy of 86.73\% and an F1 score of 85.71\% for valence recognition, and 86.92\% accuracy and 85.96\% F1 score for arousal recognition. In my research, I adopted the CNN (ResNet) model, which performed exceptionally well in capturing both static and dynamic features. Inspired by LSTM's strength in handling time-series data, I further introduced a RNN as a comparison model. By combining different physiological signal sets and evaluating the performance of ResNet, LSTM, and RNN models on the Wrist and Chest sub-datasets of the WESAD dataset, I conducted an in-depth analysis of the performance differences between these models in multimodal affective computation tasks.
\end{itemize}

\subsection{Evaluation Metrics and Basic Parameterization}

In this experiment, we used three metrics to evaluate the model's performance: accuracy (ACC), F1-Score, and Mean Squared Error (MSE). To calculate ACC and F1-Score, we first need to compute the confusion matrix. The formula is as follows:
\begin{equation}
  Confusion\ Matrix\ =\ \left(\begin{matrix}TP_A&FP_B&FP_C\\FN_A&TP_B&FP_C\\FN_A&FN_B&TP_C\\\end{matrix}\right)
\end{equation}
Here, TP represents the number of samples correctly predicted as the correct class, FP represents the number of samples incorrectly predicted as the correct class, and FN represents the number of samples incorrectly predicted as the wrong class.

ACC calculates the proportion of correctly predicted samples among all test samples. Based on the confusion matrix, the formula for calculating ACC is as follows:
\begin{equation}
  ACC=\frac{TP+TN}{TP+TN+FP+FN}
\end{equation}

F1-Score is the harmonic mean of Precision and Recall, and it can effectively handle situations with imbalanced data distributions. For each class $i \in \{A, B, C\}$, the formula for calculating F1-Score is as follows:
\begin{align}
  {Recall}_i &= \frac{{TP}_i}{{TP}_i+{FN}_i}\\
  {Precision}_i &= \frac{{TP}_i}{{TP}_i+{FP}_i}\\
  {\mathrm{F1-Score}}_i &= 2\cdot\frac{{Precision}_i\cdot{Recall}_i}{{Precision}_i+{Recall}_i}
\end{align}

MSE calculates the average of the squared differences between the predicted values and the actual values, providing a measure of the degree of difference between the model's predictions and the true values. The formula for MSE is:
\begin{equation}
  \mathrm{MSE}=\frac{1}{n}\sum_{i=1}^{n}\left(y_i-\widehat{y_i}\right)^2
\end{equation}
Here, $y_i$ and $\hat{y}_i$ represent the predicted value and the true value, respectively.

The code framework for the experiments in this paper was implemented using Python 3.10 and PyTorch 2.2.2. The experiments were conducted on a Mac with an Apple M1 Pro CPU (8 cores) with 16GB of memory and a Windows machine equipped with a GeForce GTX 1650 GPU with 4GB of memory. 

The basic configuration parameters are as follows:
\begin{itemize}
  \item The maximum number of epochs was set to 150, with an early stopping mechanism to prevent overfitting.
  \item The initial learning rate was set to 1e-4, and a StepLR scheduler was used to dynamically adjust the learning rate.
  \item The batch size was adjusted depending on the dataset and window size.
\end{itemize}

\subsection{Model Experiment Result and Analysis}

\subsubsection{Performance Evaluation Results (RQ1)}

In this experiment, we used a 30-second window to segment the dataset and conducted comparative experiments between the PhysioFormer model and various traditional machine learning and deep learning models. The traditional machine learning models include Random Forest, SVM, KNN, AdaBoost, Decision Tree, and LDA, while the deep learning models include CNN, RNN, and LSTM. The specific experimental results are shown in the Table \ref{5mainresult}:

\begin{table}[htbp]
  \caption{The performance comparation beween the proposed PhysioFormer Model and the state-of-the-art (SOTA) methods in terms of AUC (\%), F1-Socre and MSE on WESAD dataset.}
  \label{5mainresult}
  \resizebox{0.6\linewidth}{!}{
  \begin{tabular}{c|cc|cc|cc}
  \toprule
  \multirow{2}{*}{} & \multicolumn{2}{c|}{\textbf{ACC}}           & \multicolumn{2}{c|}{\textbf{F1-Score}}      & \multicolumn{2}{c}{\textbf{MSE}}            \\ \cline{2-7} 
                    & \multicolumn{1}{c|}{\textbf{Wrist}} & \textbf{Chest} & \multicolumn{1}{c|}{\textbf{Wrist}} & \textbf{Chest} & \multicolumn{1}{c|}{\textbf{Wrist}} & \textbf{Chest} \\ \hline
  \textbf{Random Forest}     & \multicolumn{1}{c|}{97.46} & 97.71 & \multicolumn{1}{c|}{97.40} & 97.71 & \multicolumn{1}{c|}{0.151} & 0.122 \\ \hline
  \textbf{SVM}               & \multicolumn{1}{c|}{97.88} & 98.62 & \multicolumn{1}{c|}{97.88} & 98.58 & \multicolumn{1}{c|}{0.134} & 0.128 \\ \hline
  \textbf{KNN}               & \multicolumn{1}{c|}{97.45} & 96.79 & \multicolumn{1}{c|}{97.62} & 96.20 & \multicolumn{1}{c|}{0.138} & 0.146 \\ \hline
  \textbf{AdaBoost}          & \multicolumn{1}{c|}{90.50} & 91.74 & \multicolumn{1}{c|}{90.53} & 91.76 & \multicolumn{1}{c|}{0.175} & 0.179 \\ \hline
  \textbf{Decision Tree}     & \multicolumn{1}{c|}{88.27} & 95.41 & \multicolumn{1}{c|}{88.29} & 95.39 & \multicolumn{1}{c|}{0.205} & 0.146 \\ \hline
  \textbf{LDA}               & \multicolumn{1}{c|}{85.20} & 72.93 & \multicolumn{1}{c|}{84.51} & 68.69 & \multicolumn{1}{c|}{0.303} & 0.298 \\ \hline
  \textbf{CNN}               & \multicolumn{1}{c|}{98.33} & 81.46 & \multicolumn{1}{c|}{97.95} & 75.07 & \multicolumn{1}{c|}{0.065} & 0.185 \\ \hline
  \textbf{RNN}               & \multicolumn{1}{c|}{84.93} & 79.26 & \multicolumn{1}{c|}{84.98} & 80.03 & \multicolumn{1}{c|}{0.391} & 0.437 \\ \hline
  \textbf{LSTM}              & \multicolumn{1}{c|}{96.87} & 87.00 & \multicolumn{1}{c|}{97.02} & 87.39 & \multicolumn{1}{c|}{0.184} & 0.400 \\ \hline
  \textbf{PhysioFormer}      & \multicolumn{1}{c|}{\textbf{99.54}} & \textbf{99.21} & \multicolumn{1}{c|}{\textbf{99.54}} & \textbf{99.17} & \multicolumn{1}{c|}{\textbf{0.025}} & \textbf{0.057} \\ \bottomrule
  \end{tabular}
  }
\end{table}

The experimental results show that the PhysioFormer model outperformed all other models across all metrics. Its ACC on the Wrist and Chest datasets reached 99.54\% and 99.21\%, respectively, with F1-scores of 99.54 and 99.17, and the lowest MSE of 0.025 and 0.057, respectively. This demonstrates that PhysioFormer exhibits exceptionally high predictive accuracy and stability when processing these datasets.

In contrast, SVM performed exceptionally well among traditional machine learning algorithms, particularly on the Chest dataset, where both ACC and F1-score exceeded 98\%, and MSE remained low at 0.134 and 0.128, respectively. This indicates that SVM handles high-dimensional data very effectively. The Random Forest model also performed well on both the Wrist and Chest datasets, with ACC and F1-scores approaching 97\%, and MSEs of 0.151 and 0.122, respectively, demonstrating its high accuracy and low error when processing physiological data. KNN's performance on the Wrist dataset was close to that of Random Forest, but it underperformed slightly on the Chest dataset, particularly in terms of F1-score and MSE. This could be related to KNN’s sensitivity to data distribution. AdaBoost and Decision Tree models performed slightly worse than the previously mentioned models, particularly in terms of ACC and F1-score. For example, AdaBoost achieved ACC values of 90.50\% and 91.74\% on the Wrist and Chest datasets, respectively, while Decision Tree performed noticeably better on the Chest dataset (95.41\%) than on the Wrist dataset (88.27\%). LDA performed the worst, especially on the Chest dataset, where ACC and F1-score were 72.93\% and 68.69\%, respectively, and the error was high, indicating that LDA struggles with this type of complex data. 

Among deep learning models, CNN performed relatively well, particularly on the Wrist dataset, where ACC reached 98.33\%, F1-score was 97.95\%, and MSE was low at 0.065, showing CNN's strength in handling physiological data with spatial features. However, CNN's performance significantly dropped on the Chest dataset, with ACC at only 81.46\%, F1-score falling to 75.07\%, and MSE increasing to 0.185. This may suggest that CNN has limitations in adapting to different data channels. LSTM showed stability in handling time-series data, achieving 96.87\% ACC, 97.02\% F1-score, and 0.184 MSE on the Wrist dataset. Although its performance declined on the Chest dataset (ACC: 87.00\%, F1-score: 87.39\%), it was still more stable than RNN. RNN showed relatively poor performance on both datasets, especially on the Chest dataset, where ACC and F1-score were 79.26\% and 80.03\%, respectively, with a high MSE of 0.437, indicating that RNN may have limitations in capturing long-term dependencies and handling more complex physiological signals.

Based on the experimental results, it is evident that the PhysioFormer model addresses many of the shortcomings present in traditional machine learning and deep learning models, demonstrating its unique advantages and outstanding performance. Firstly, PhysioFormer model, through its advanced architecture design, effectively overcomes the limitations of traditional machine learning models in handling multimodal physiological signals. By integrating multi-level feature embedding and affective representation modules, it extracts more representative and distinctive features from multimodal signals, excelling across diverse data types. Secondly, PhysioFormer model excels in addressing the stability issues found in deep learning models. Its adaptive feature processing mechanism not only captures dynamic changes in time-series data but also effectively fuses information across different modalities, ensuring consistent performance across various datasets. This stability gives PhysioFormer model a significant advantage in handling complex real-world scenarios. Finally, PhysioFormer model's low MSE indicates minimal prediction error, which is attributed to its innovative design in multimodal fusion and time-series modeling. By introducing a feature embedding module, PhysioFormer model is better equipped to handle long-term dependencies and sequential information, making it far superior to other models in affective computation tasks.

\subsubsection{Convergence Rate}

\begin{figure}[htbp]
  \centering
  \includegraphics[width=0.9\textwidth]{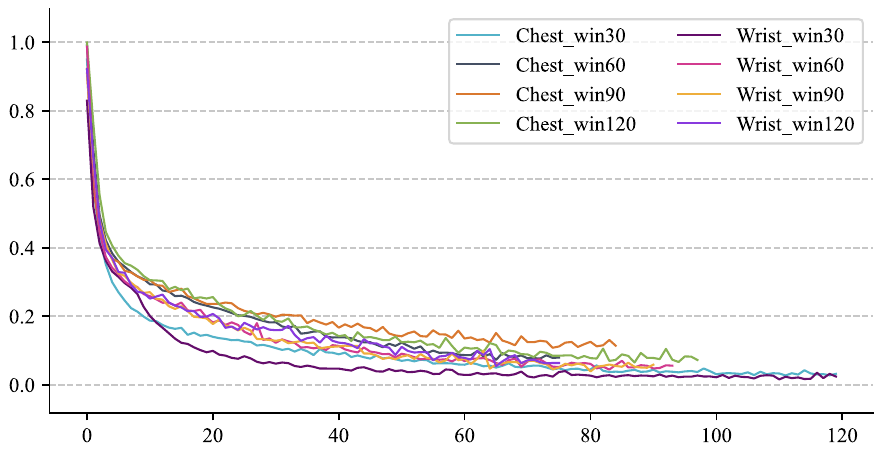}
  \caption{Convergence trends of the PhysioFormer model across datasets by splitting the WESAD dataset through windows of different sizes. Although convergence trends are shown on all datasets, there are differences in the speed of convergence and the magnitude of losses.}
  \label{4combined_loss_plot}
\end{figure}

For each dataset, we trained the PhysioFormer model and recorded its loss values during the training process. The variation in the loss values over time during training is shown in the Figure \ref{4combined_loss_plot}.

The results show that all models exhibit a rapid decline in loss values during the initial stages of training, indicating that the models quickly learn effective features, significantly reducing prediction error. More importantly, all models demonstrate good convergence in the later stages of training, with the final loss values stabilizing, suggesting that the PhysioFormer model is capable of achieving convergence when processing both the Wrist and Chest datasets with different time windows.

Notably, the datasets with different time windows show some variation in convergence speed and final loss values. Although models with all time windows eventually converge, the dataset with a 30-second window exhibits faster convergence and lower final loss values. This indicates that the model can reach an optimal state more quickly with shorter time windows, while also reducing the risk of overfitting during training. In contrast, as the time window length increases, the model's convergence slows down slightly, and the final loss value is slightly higher. This may be attributed to the longer time windows introducing more temporal dependencies; while such information helps the model capture long-term trends, it also increases optimization difficulty, requiring more time for the model to converge.

Overall, the experimental results indicate that the PhysioFormer model consistently converges across datasets with different time windows, and it shows faster convergence and lower final loss values with shorter time windows. This further validates the efficiency and reliability of the PhysioFormer model in processing multimodal physiological signals.

\subsubsection{Sensitivity Tests}

When designing and optimizing complex models, sensitivity tests is a critical step. Through sensitivity experiments, we can systematically evaluate the model's response to different parameters, revealing how these parameters impact the model's performance. This not only helps to understand the internal mechanisms of the model but also provides a scientific basis for optimizing model parameters. Specifically, in this experiment, we explored the effects of adjusting the window size and the number of hidden layer neurons in the PhysioFormer model on its performance across different datasets. This allowed us to identify the optimal model configuration.

\textbf{The effect of window size. (RQ2)}This experiment aimed to evaluate the impact of different window sizes on the model's performance. We selected four different window lengths: 30 seconds, 60 seconds, 90 seconds, and 120 seconds, and tested them on the sensor data from both the Wrist and Chest positions. For each window size, we calculated the model's ACC, F1-Score, and MSE to comprehensively assess its performance. The experimental results are shown in the Table \ref{5windows}.

\begin{table}[htbp]
  \caption{The performance comparation of the proposed PhysioFormer model on different datasets divided by different window sizes in terms of AUC (\%), F1-Socre and MSE.}
  \label{5windows}
  \resizebox{0.5\linewidth}{!}{
  \begin{tabular}{c|cc|cc|cc}
  \toprule
  \multirow{2}{*}{} & \multicolumn{2}{c|}{\textbf{ACC}}           & \multicolumn{2}{c|}{\textbf{F1-Score}}      & \multicolumn{2}{c}{\textbf{MSE}}            \\ \cline{2-7} 
                    & \multicolumn{1}{c|}{\textbf{Wrist}} & \textbf{Chest} & \multicolumn{1}{c|}{\textbf{Wrist}} & \textbf{Chest} & \multicolumn{1}{c|}{\textbf{Wrist}} & \textbf{Chest} \\ \hline
  \textbf{30s}               & \multicolumn{1}{c|}{\textbf{99.54}} & \textbf{99.21} & \multicolumn{1}{c|}{\textbf{99.54}} & \textbf{99.17} & \multicolumn{1}{c|}{\textbf{0.025}} & \textbf{0.057} \\ \hline
  \textbf{60s}               & \multicolumn{1}{c|}{97.61} & 97.52 & \multicolumn{1}{c|}{97.59} & 97.53 & \multicolumn{1}{c|}{0.108} & 0.117 \\ \hline
  \textbf{90s}               & \multicolumn{1}{c|}{97.87} & 95.45 & \multicolumn{1}{c|}{97.86} & 95.38 & \multicolumn{1}{c|}{0.081} & 0.203 \\ \hline
  \textbf{120s} & \multicolumn{1}{c|}{97.70} & 96.85 & \multicolumn{1}{c|}{97.68} & 96.50 & \multicolumn{1}{c|}{0.089} & 0.141 \\ \bottomrule
  \end{tabular}
  }
\end{table}

In this experiment, we evaluated the impact of different window sizes (30 seconds, 60 seconds, 90 seconds, and 120 seconds) on the performance of the PhysioFormer model. The results show that window size significantly affects the model's predictive performance. With a 30-second window, the model performed best, achieving ACC of 99.54\% and 99.21\%, and F1-Score of 99.54 and 99.17 on the Wrist and Chest datasets, respectively, with the lowest MSE of 0.025 and 0.057. This suggests that shorter time windows can more effectively capture subtle variations in physiological signals, resulting in higher predictive accuracy and stability.

However, as the window size increased, the model's performance declined. With a 60-second window, although the model maintained relatively high accuracy and F1-Score, MSE increased, indicating that handling longer time sequences introduced more temporal dependencies, adding complexity and increasing prediction error. For the 90-second and 120-second windows, particularly for the 90-second Chest dataset, accuracy and F1-Score dropped significantly, and MSE increased further, suggesting that excessively long windows may make it difficult for the model to handle complex temporal dependencies, affecting prediction performance.

In summary, the 30-second window provides the optimal balance, capturing sufficient information while avoiding redundancy and complexity, ensuring higher predictive accuracy and lower error.

\textbf{The effect of the number of neurons in the hidden layer. (RQ3)}This experiment aimed to evaluate the impact of the number of hidden layer neurons in the \textit{ContribNet} and \textit{AffectNet} models within the PhysioFormer framework on overall model performance. To achieve this, we selected four different scales of neuron counts (50, 100, 200, 500) and conducted experiments on both the Wrist and Chest datasets to assess the performance of the \textit{ContribNet} and \textit{AffectNet} models within PhysioFormer. The experimental results are shown in the Table \ref{5wrist_neurons} and Table \ref{5chest_neurons}.

\begin{table}[htbp]
  \caption{The performance comparison of the proposed PhysioFormer model on the Wrist dataset with different numbers of hidden layer neurons in the \textit{ContribNet} and \textit{AffectNet} models in terms of AUC (\%).}
  \label{5wrist_neurons}
  \resizebox{0.45\linewidth}{!}{
  \begin{tabular}{cccccc}
  \toprule
  \multicolumn{6}{c}{\textbf{Wrist}}  \\ \toprule
  \multicolumn{1}{c|}{\textbf{}} & \multicolumn{5}{c}{\textbf{ContribNet}} \\ \hline
  \multicolumn{1}{c|}{\multirow{5}{*}{\textbf{AffectNet}}} &
    \multicolumn{1}{c|}{} &
    \multicolumn{1}{c|}{50} &
    \multicolumn{1}{c|}{100} &
    \multicolumn{1}{c|}{200} &
    500 \\ \cline{2-6} 
  \multicolumn{1}{c|}{}          & \multicolumn{1}{c|}{50}  & \multicolumn{1}{c|}{96.97} & \multicolumn{1}{c|}{97.39} & \multicolumn{1}{c|}{97.87} & 97.45 \\ \cline{2-6} 
  \multicolumn{1}{c|}{}          & \multicolumn{1}{c|}{100} & \multicolumn{1}{c|}{97.51} & \multicolumn{1}{c|}{\textbf{99.54}} & \multicolumn{1}{c|}{96.97} & 98.00 \\ \cline{2-6} 
  \multicolumn{1}{c|}{}          & \multicolumn{1}{c|}{200} & \multicolumn{1}{c|}{98.06} & \multicolumn{1}{c|}{98.30} & \multicolumn{1}{c|}{97.69} & 98.87 \\ \cline{2-6} 
  \multicolumn{1}{c|}{}          & \multicolumn{1}{c|}{500} & \multicolumn{1}{c|}{99.03} & \multicolumn{1}{c|}{98.56} & \multicolumn{1}{c|}{98.84} & 98.78 \\ \bottomrule
  \end{tabular}
  }
\end{table}

\begin{table}[htbp]
  \caption{The performance comparison of the proposed PhysioFormer model on the Wrist dataset with different numbers of hidden layer neurons in the \textit{ContribNet} and \textit{AffectNet} models in terms of AUC (\%).}
  \label{5chest_neurons}
  \resizebox{0.45\linewidth}{!}{
  \begin{tabular}{cccccc}
  \toprule
  \multicolumn{6}{c}{\textbf{Chest}}  \\ \toprule
  \multicolumn{1}{c|}{\textbf{}} & \multicolumn{5}{c}{\textbf{ContribNet}} \\ \hline
  \multicolumn{1}{c|}{\multirow{5}{*}{\textbf{AffectNet}}} &
    \multicolumn{1}{c|}{} &
    \multicolumn{1}{c|}{50} &
    \multicolumn{1}{c|}{100} &
    \multicolumn{1}{c|}{200} &
    500 \\ \cline{2-6} 
  \multicolumn{1}{c|}{}          & \multicolumn{1}{c|}{50}  & \multicolumn{1}{c|}{99.08} & \multicolumn{1}{c|}{96.78} & \multicolumn{1}{c|}{96.96} & 97.24 \\ \cline{2-6} 
  \multicolumn{1}{c|}{}          & \multicolumn{1}{c|}{100} & \multicolumn{1}{c|}{98.80} & \multicolumn{1}{c|}{\textbf{99.21}} & \multicolumn{1}{c|}{97.33} & 97.61 \\ \cline{2-6} 
  \multicolumn{1}{c|}{}          & \multicolumn{1}{c|}{200} & \multicolumn{1}{c|}{98.52} & \multicolumn{1}{c|}{97.98} & \multicolumn{1}{c|}{99.08} & 98.80 \\ \cline{2-6} 
  \multicolumn{1}{c|}{}          & \multicolumn{1}{c|}{500} & \multicolumn{1}{c|}{99.08} & \multicolumn{1}{c|}{98.99} & \multicolumn{1}{c|}{98.90} & 98.90 \\ \hline
  \end{tabular}
  }
\end{table}

On the Wrist dataset, the performance of the model fluctuated with an increase in the number of neurons in the \textit{ContribNet} and \textit{AffectNet} models. For instance, when the number of neurons in \textit{ContribNet} was fixed at 50, the combination with 500 neurons in \textit{AffectNet} performed best, achieving an accuracy of 99.03\%. However, when the number of neurons in \textit{ContribNet} was increased to 100, the model reached its peak performance, with an accuracy of 99.54\% when \textit{AffectNet} also had 100 neurons. Overall, the results from the Wrist dataset indicate that increasing the number of neurons beyond a certain point did not continue to improve the model's accuracy. In some cases, performance even declined, which could be related to overfitting due to the increased complexity of the model.

On the Chest dataset, the results similarly showed the impact of neuron count on model performance. When the number of neurons in both \textit{ContribNet} and \textit{AffectNet} was set to 50, the model performed relatively well, achieving an accuracy of 99.08\%. The best performance, with an accuracy of 99.21\%, was observed when both \textit{AffectNet} and \textit{ContribNet} had 100 neurons. As the neuron count increased, particularly with configurations of 500 neurons, the model's performance stabilized but did not significantly surpass the performance with fewer neurons.

Overall, the results suggest that while increasing the number of hidden layer neurons can improve the model's performance, this improvement is limited. Too many neurons can lead to increased model complexity, which raises the risk of overfitting, especially with relatively limited data. The best model performance often occurred with moderate neuron configurations, such as 100 or 200 neurons. In conclusion, setting the number of neurons in both \textit{ContribNet} and \textit{AffectNet} to 100 offers optimal performance for the PhysioFormer model.

\subsubsection{Ablation Studies}

Ablation studies are used to assess the contribution of various components or features within a model. They help evaluate the actual impact of each model component. In this experiment, we gradually removed specific modules from the model to evaluate their influence on overall performance, thereby validating their effectiveness. Specifically, we designed two ablation experiments: one to verify the importance of the Feature Embedding module and another to assess the role of individual attributes in the model. The results of these experiments will provide scientific evidence for model optimization and improvement.

\begin{figure}[htbp]
  \centering
  \includegraphics[width=0.85\textwidth]{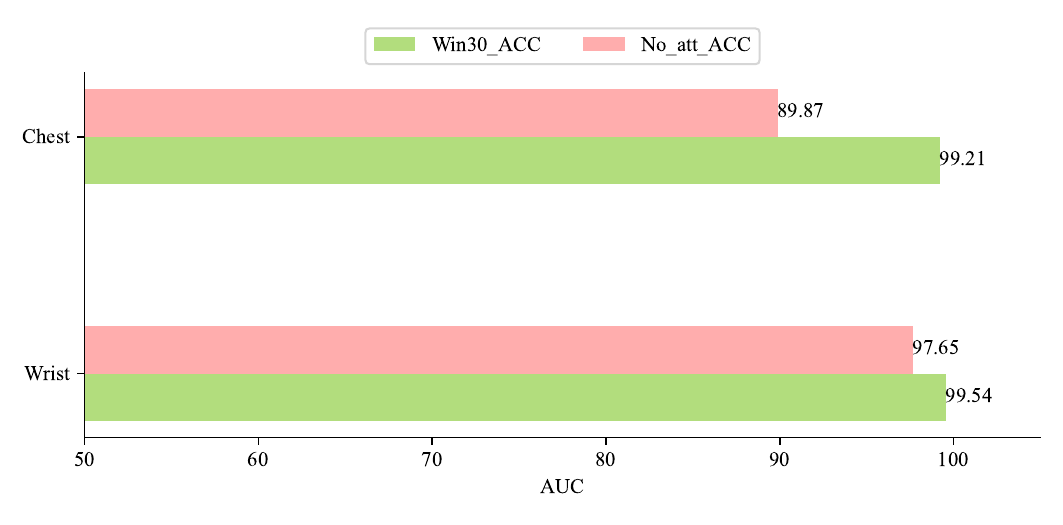}
  \caption{The role of feature embedding in affective computation tasks on both datasets, with the results presented in terms of ACC. $No\_att\_ACC$ represents the performance of the PhysioFormer model without using the feature embedding module, while $Win30\_ACC$ indicates the performance of the PhysioFormer model after applying the feature embedding module.}
  \label{5no_att}
\end{figure}

\textbf{The Validity of Feature Embedding. (RQ4)}In this experiment, we evaluated the performance differences between the model with the feature embedding module and the model without it on the Wrist and Chest datasets. To assess the effectiveness of the feature embedding module, we conducted two sets of comparison experiments: one using the complete model with the feature embedding module, and the other by removing it. ACC was used as the primary evaluation metric. The experimental results are shown in the Figure \ref{5no_att}.

According to the experimental results, the model with the feature embedding module achieved an accuracy of 99.54\% on the Wrist dataset, compared to 97.65\% for the model without it, showing an improvement of approximately 1.89\%. On the Chest dataset, the model with the feature embedding module achieved an accuracy of 99.21\%, while the model without the feature embedding module had an accuracy of 89.87\%, reflecting an improvement of about 9.34\%.

These results demonstrate that the model with the physiological data feature embedding module outperforms the one without it on both the Wrist and Chest datasets, with a particularly notable improvement on the Chest dataset. This indicates that the feature embedding module effectively quantifies and highlights features that significantly contribute to physiological signal prediction, better capturing key characteristics in the data and enhancing the model’s predictive capability.

\begin{figure}[htbp]
  \centering
  \includegraphics[width=0.85\textwidth]{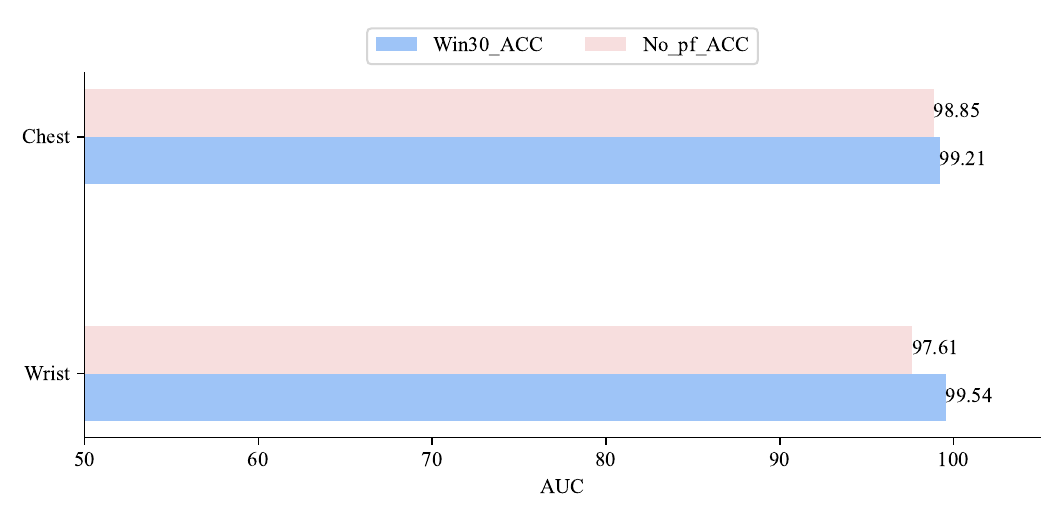}
  \caption{The role of individual attributes features in affective computation tasks on both datasets, with the results presented in terms of ACC. $No\_pf\_ACC$ represents the performance of the PhysioFormer model without combining individual attributes features, while $Win30\_ACC$ indicates the performance of the PhysioFormer model after combining individual attributes features.}
  \label{6no_pf}
\end{figure}

\textbf{The Validity of Individual Attributes Features. (RQ5)}In this experiment, we evaluated the performance differences between models that include individual attributes features (such as age, gender, etc.) alongside physiological monitoring data and models that only use physiological monitoring data on the Wrist and Chest datasets. The experiment was designed in two parts: one model used the complete input data, including individual attributes features and physiological monitoring data, while the other model excluded the individual attributes features and used only the physiological monitoring data for training and testing. This comparative experiment allowed us to assess the impact of individual attributes features on the model's performance. ACC was used as the primary evaluation metric. The experimental results are shown in the Figure \ref{6no_pf}.

The model including individual attributes features data achieved an accuracy of 99.54\% on the Wrist dataset, compared to 97.61\% for the model without individual attributes features, reflecting an improvement of approximately 1.93\%. On the Chest dataset, the model with individual attributes features reached an accuracy of 99.21\%, while the model without these features achieved 98.85\%, showing an improvement of about 0.36\%.

The results indicate that the model incorporating individual attributes features data outperforms the one without it on both the Wrist and Chest datasets. Although the improvement on the Chest dataset is relatively modest, it still demonstrates the positive impact of individual attributes features data on model performance. This suggests that individual attributes features data contributes to enhancing the model's accuracy by providing additional context, helping the model capture and predict physiological signal changes more precisely. By combining individual attributes features with physiological monitoring data, the model gains a more comprehensive understanding of physiological states, resulting in improved overall prediction performance.

\subsection{Discovered Laws and Analysis}

\subsubsection{Features Distribution and Importance}

\begin{figure}[htbp]
  \centering
  \includegraphics[width=0.75\textwidth]{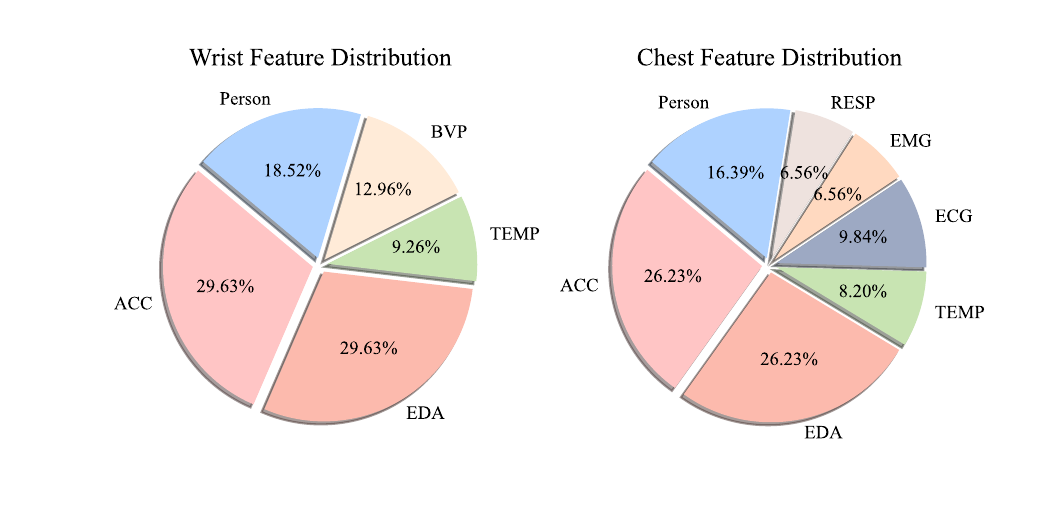}
  \caption{Distribution map of the number of features. The figure shows the distribution of the number of features calculated from various physiological indicators in the Wrist dataset and the Chest dataset, respectively.}
  \label{7feature_distribution}
\end{figure}

The Wrist dataset contains four types of physiological indicators, while the Chest dataset contains six types of physiological indicators. The distribution of the number of features computed from the physiological indicators in both datasets is shown in the Figure \ref{7feature_distribution}.

From the distribution chart, it is clear that ACC and EDA are the primary features in the Wrist dataset, each accounting for nearly one-third of the total features. Individual attributes also make up a significant portion, while BVP and TEMP have smaller proportions. In the Chest dataset, ACC and EDA similarly occupy major positions, each representing more than a quarter of the total features. Individual attributes also hold a large share, while ECG and TEMP occupy a notable portion, with EMG and RESP having smaller proportions. In summary, ACC and EDA are the dominant features in both datasets, but the Chest dataset includes a greater variety of physiological indicators.

\begin{figure}[htbp]
  \centering
  \includegraphics[width=\textwidth]{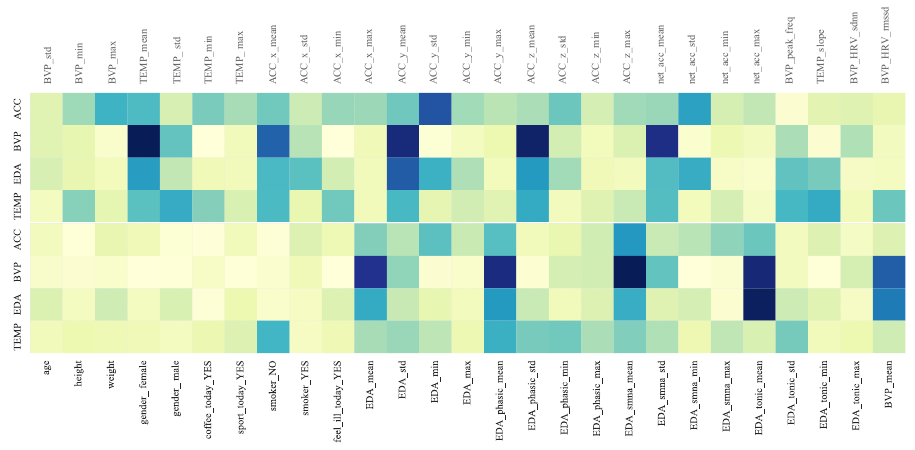}
  \caption{The visualization of the importance scores for all features in the Wrist dataset, with darker colors indicating higher importance scores. In the subsequent symbolic distillation task, the top ten features with the highest importance scores will be selected for further analysis and modeling.}
  \label{8wrist_importance}
\end{figure}

In the PhysioFormer model, the importance scores of feature data for the four physiological indicators in the Wrist dataset (ACC, EDA, BVP, and TEMP) are shown in the Figure \ref{8wrist_importance}. By analyzing the figure, it is evident that ACC-related features significantly influence all physiological indicators. Specifically, features like $ACC\_y\_mean$ and $Net\_ACC\_mean$ not only have high importance for ACC itself but also show strong correlations with EDA, BVP, and TEMP. This suggests that physical activity intensity plays a key role in monitoring and predicting physiological states, likely due to the effects of movement on cardiovascular activity, electrodermal activity, and body temperature, making it important across multiple physiological indicators.

Additionally, individual attributes features (such as smoking status and weight) play a crucial role in TEMP and EDA. Smoking status shows high importance for TEMP, indicating a significant impact of smoking on temperature regulation. Similarly, weight has a notable influence on EDA, possibly due to its effect on skin conductance. Furthermore, features like temperature change rate and standard deviation are highly important for the TEMP indicator, reflecting the dynamic changes involved in the process of temperature regulation.

\begin{figure}[htbp]
  \centering
  \includegraphics[width=\textwidth]{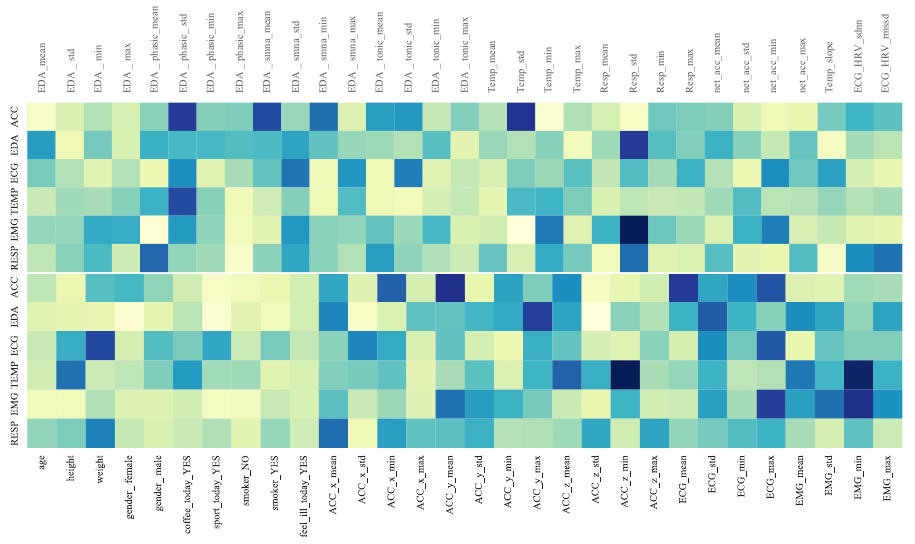}
  \caption{The visualization of the importance scores for all features in the Chest dataset, with darker colors indicating higher importance scores. In the subsequent symbolic distillation task, the top ten features with the highest importance scores will be selected for further analysis and modeling.}
  \label{9chest_importance}
\end{figure}

The importance scores of feature data for the six physiological indicators in the Chest dataset (ACC, EDA, ECG, TEMP, EMG, and RESP) are shown in the Figure \ref{9chest_importance}. In the subsequent symbolic regression task, the top 10 features with the highest importance scores were extracted for further analysis, as detailed in Appendix B. ACC-related features still show a dominant influence across all six physiological indicators, further underscoring the critical role of physical activity in the monitoring and prediction of various physiological states.

Additionally, in the Chest dataset, the impact of individual attributes on physiological features becomes more pronounced. Specifically, weight significantly affects ECG and RESP indicators, while height has a notable impact on TEMP and ECG. Weight may influence the function of the cardiovascular and respiratory systems; changes in weight can lead to variations in blood pressure and heart rate, which are reflected in ECG features. Furthermore, weight gain can lead to breathing difficulties, which can be captured by the RESP indicator. Height may affect the ratio of body surface area to volume, influencing heat dissipation and thermoregulation mechanisms, which can be observed in TEMP features.

\subsubsection{Discovered Laws Expression}

In this section, symbolic regression was used to model each physiological indicator in both the Wrist and Chest datasets. The symbolic regression algorithm employed was implemented using the open-source symbolic regression library PySR, with the predefined operators including ``$+$", ``$-$", ``$\times$", ``$sin$", ``$cos$", ``$log$", ``$exp$", and ``$pow$". 

Through symbolic regression, we obtained several formulas (with complexity less than 15). The next step is to select an optimal formula to explain the physiological indicator. In the selection process, we first focus on formulas that have converged, as these indicate that the model reached a stable state during training, with the loss value no longer significantly decreasing, suggesting that the model has well-fitted the data and found a relatively optimal parameter combination. 

Among the converged formulas, we selected the one with the fewest variables. This is because formulas with fewer variables simplify the structure to some extent, improving the explainability of the model. Although these formulas may still be complex, they are able to capture subtle patterns and complex relationships in the data, which is critical for accurate modeling and prediction. Additionally, choosing formulas with fewer variables helps reduce the risk of overfitting, enhancing the model's generalization ability, and making it more stable and reliable when applied to new data.

\begin{figure}[htbp]
  \centering
  \begin{subfigure}{.47\textwidth}
    \centering
    \includegraphics[width=\linewidth]{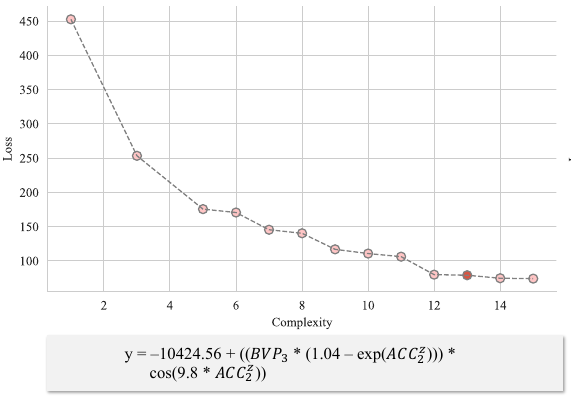}
    \caption{Complexity-Loss curve of ACC in Wrist}
  \end{subfigure}
  \begin{subfigure}{.47\textwidth}
    \centering
    \includegraphics[width=\linewidth]{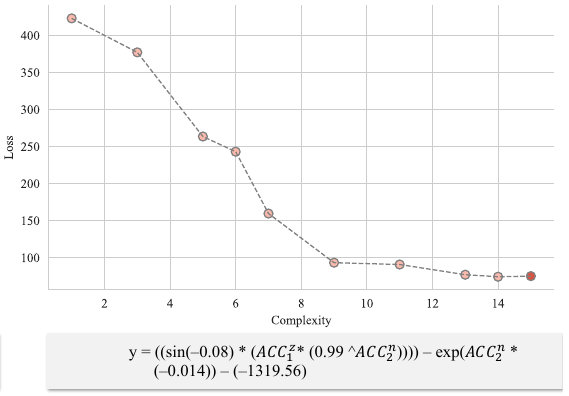}
    \caption{Complexity-Loss curve of EDA in Wrist}
  \end{subfigure}
  \begin{subfigure}{.47\textwidth}
    \centering
    \includegraphics[width=\linewidth]{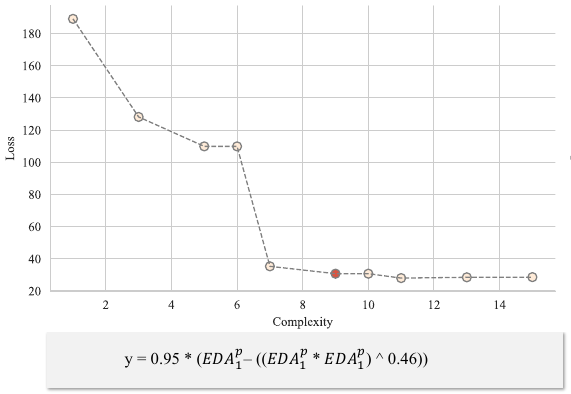}
    \caption{Complexity-Loss curve of BVP in Wrist}
  \end{subfigure}
  \begin{subfigure}{.47\textwidth}
    \centering
    \includegraphics[width=\linewidth]{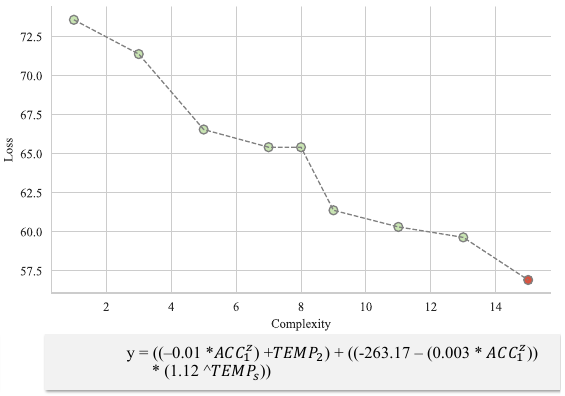}
    \caption{Complexity-Loss curve of TEMP in Wrist}
  \end{subfigure}
  \caption{Complexity-Loss curve in Wrist dataset. The red dot indicate the selected formula, and the corresponding formula are shown below the charts.}
  \label{10wrist_cx_loss}
\end{figure}

\begin{figure}[htbp]
  \centering
  \begin{subfigure}{.47\textwidth}
    \centering
    \includegraphics[width=\linewidth]{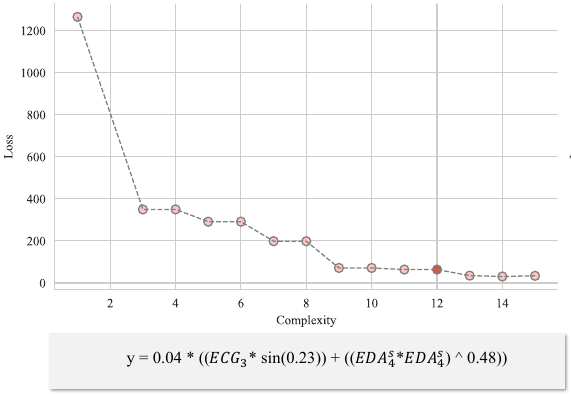}
    \caption{Complexity-Loss curve of ACC in Chest}
  \end{subfigure}
  \begin{subfigure}{.47\textwidth}
    \centering
    \includegraphics[width=\linewidth]{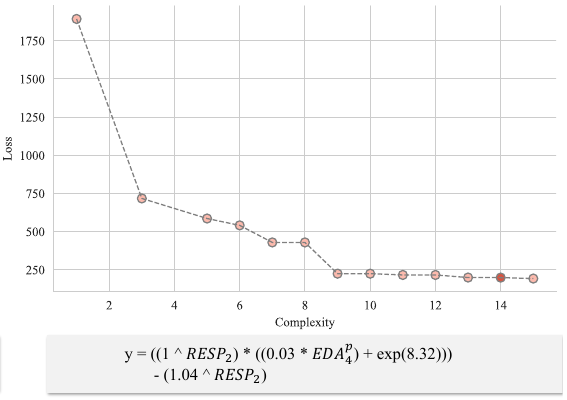}
    \caption{Complexity-Loss curve of EDA in Chest}
  \end{subfigure}
  \begin{subfigure}{.47\textwidth}
    \centering
    \includegraphics[width=\linewidth]{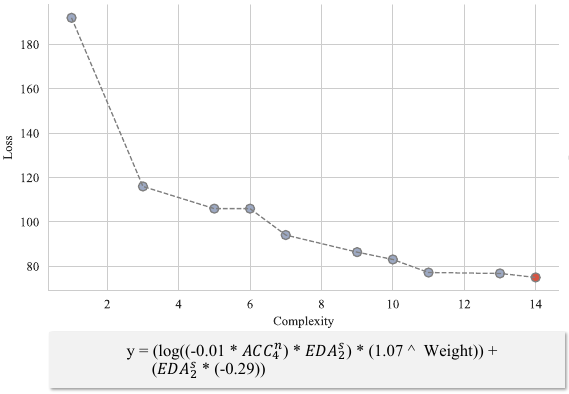}
    \caption{Complexity-Loss curve of ECG in Chest}
  \end{subfigure}
  \begin{subfigure}{.47\textwidth}
    \centering
    \includegraphics[width=\linewidth]{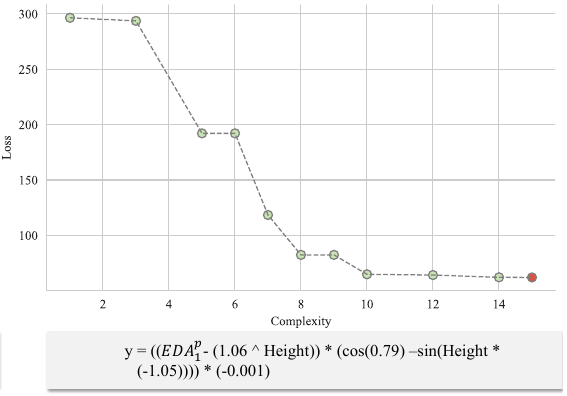}
    \caption{Complexity-Loss curve of TEMP in Chest}
  \end{subfigure}
  \begin{subfigure}{.47\textwidth}
    \centering
    \includegraphics[width=\linewidth]{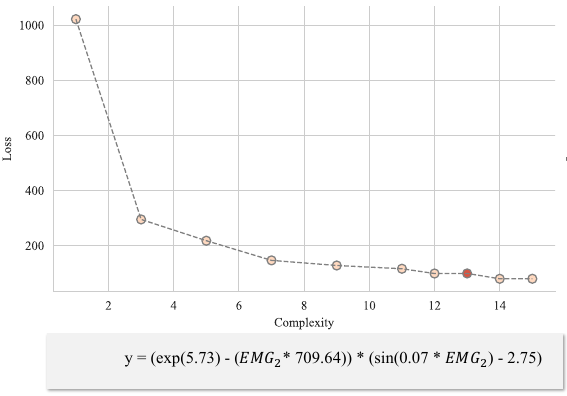}
    \caption{Complexity-Loss curve of EMG in Chest}
  \end{subfigure}
  \begin{subfigure}{.47\textwidth}
    \centering
    \includegraphics[width=\linewidth]{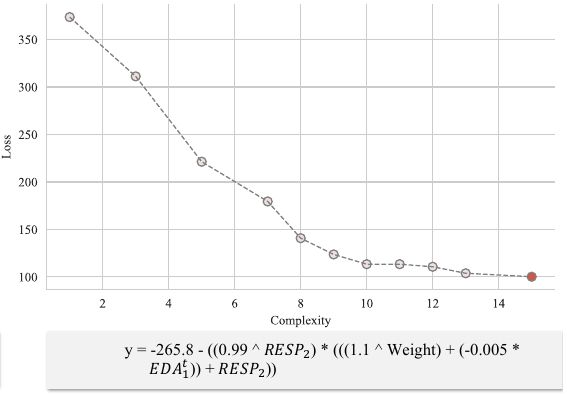}
    \caption{Complexity-Loss curve of RESP in Chest}
  \end{subfigure}
  \caption{Complexity-Loss curve in Chest dataset. The red dot indicate the selected formula, and the corresponding formula are shown below the charts.}
  \label{12chest_cx_loss}
\end{figure}

Based on the training process and the formula selection method described above, the Complexity-Loss curve and the selected optimal formula are shown in the Figure \ref{10wrist_cx_loss} and Figure \ref{12chest_cx_loss}, with the red dot indicating the chosen formula. The specific formulas generated through symbolic regression can be found in Appendix C.

From the experimental results, these curves exhibit a typical pattern: as the complexity of the formulas increases, the loss value rapidly decreases, but after reaching a certain level of complexity, the downward trend slows, showing convergence or slight fluctuations. Specifically, across all physiological indicator curves, as the formula complexity increases from low to moderate (usually in the complexity range of 2 to 6), the loss value drops rapidly. This indicates that increasing complexity in the initial stages significantly improves the model’s ability to fit the data, allowing it to better capture the relationship between physiological signals and affective states. When the formula complexity reaches a moderate level (typically between 8 and 10), the loss value continues to decrease, but the rate of decline slows. For example, in the ACC and EDA curves, the loss value stabilizes as complexity reaches 8 to 10. This phase suggests that increasing formula complexity has diminishing returns in reducing the loss, and the model begins to converge. As the complexity increases further (usually beyond 12), the curve flattens, with minimal changes in the loss value, and in some cases, slight fluctuations are observed. For instance, in the BVP and TEMP curves, the loss value shows little to no significant reduction beyond complexity 12, indicating that the model's complexity has reached a sufficient level. Further increases in complexity may introduce the risk of overfitting, without significant improvement in model performance.

Additionally, the complexity-loss curves reveal that formulas for certain physiological indicators converge more slowly as complexity increases. This phenomenon may be attributed to several factors: First, some physiological indicator data may contain higher levels of noise and variability, making it more challenging for the model to accurately capture patterns in the data. Second, interactions between different physiological indicators may involve multiple physiological mechanisms, requiring more complex mathematical expressions to explain these relationships. Finally, limitations in the search space and computational resources may also affect the ability to efficiently find optimal formulas.

Overall, these curves illustrate the typical trade-off between model complexity and performance. Increasing complexity initially leads to significant improvements in model accuracy, but after reaching a certain level, the gains from further complexity diminish and may even result in overfitting. Therefore, selecting an appropriate complexity level for the final model formula ensures a low loss value while avoiding unnecessary complexity.

\begin{figure}[!h]
  \centering
  \includegraphics[width=0.75\textwidth]{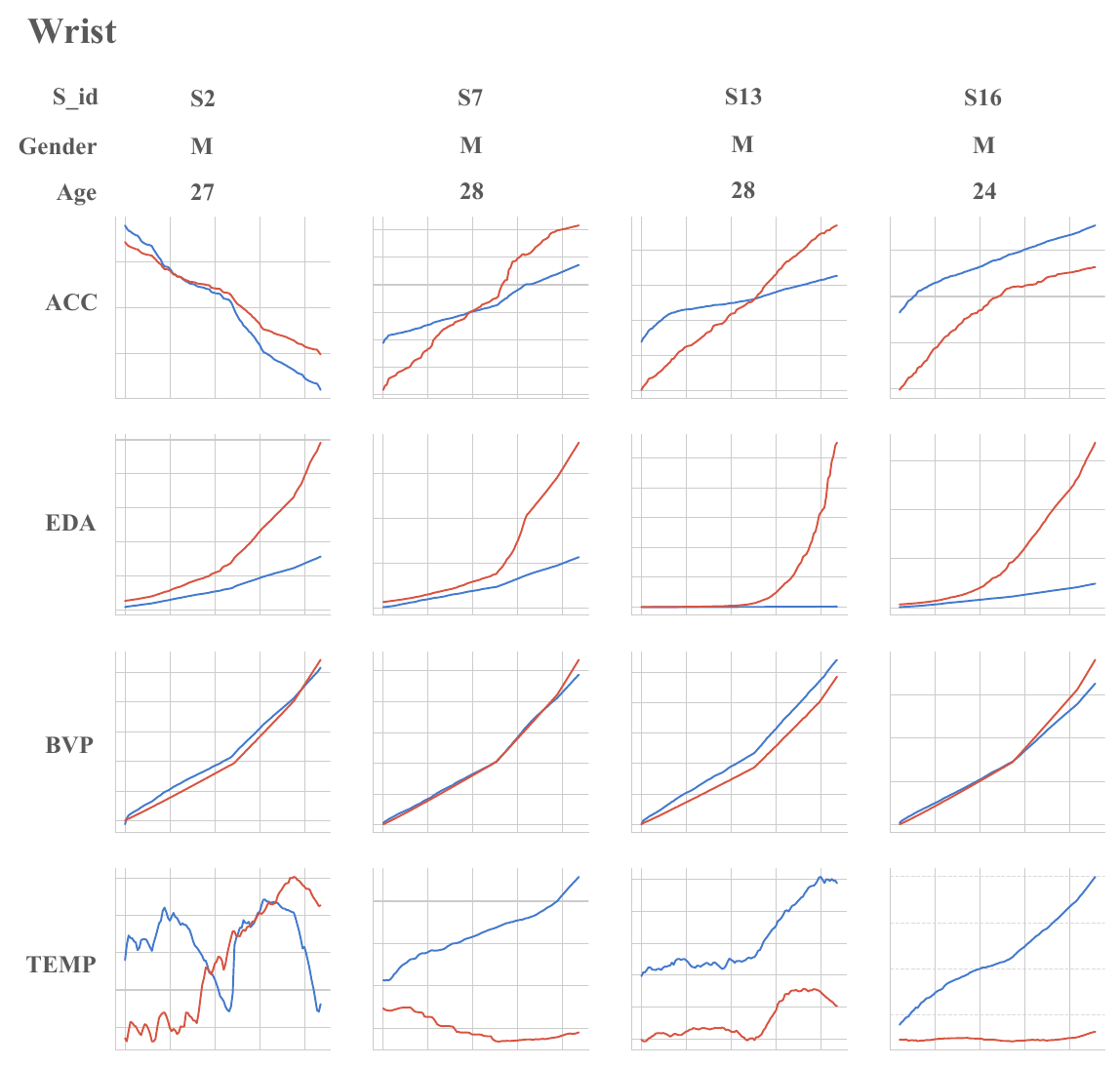}
  \caption{The comparison of the fitted curves for affective indicators based on the selected formulas for four randomly chosen individuals. Using the selected formulas and the output from the PhysioFormer model, the fitting curves for each affective indicator in the Wrist dataset were plotted. In the figure, the blue curve represents the model's output, while the red curve represents the results calculated from the formulas.}
  \label{11wrist_fitting}
\end{figure}

\begin{figure}[htbp]
  \centering
  \includegraphics[width=0.75\textwidth]{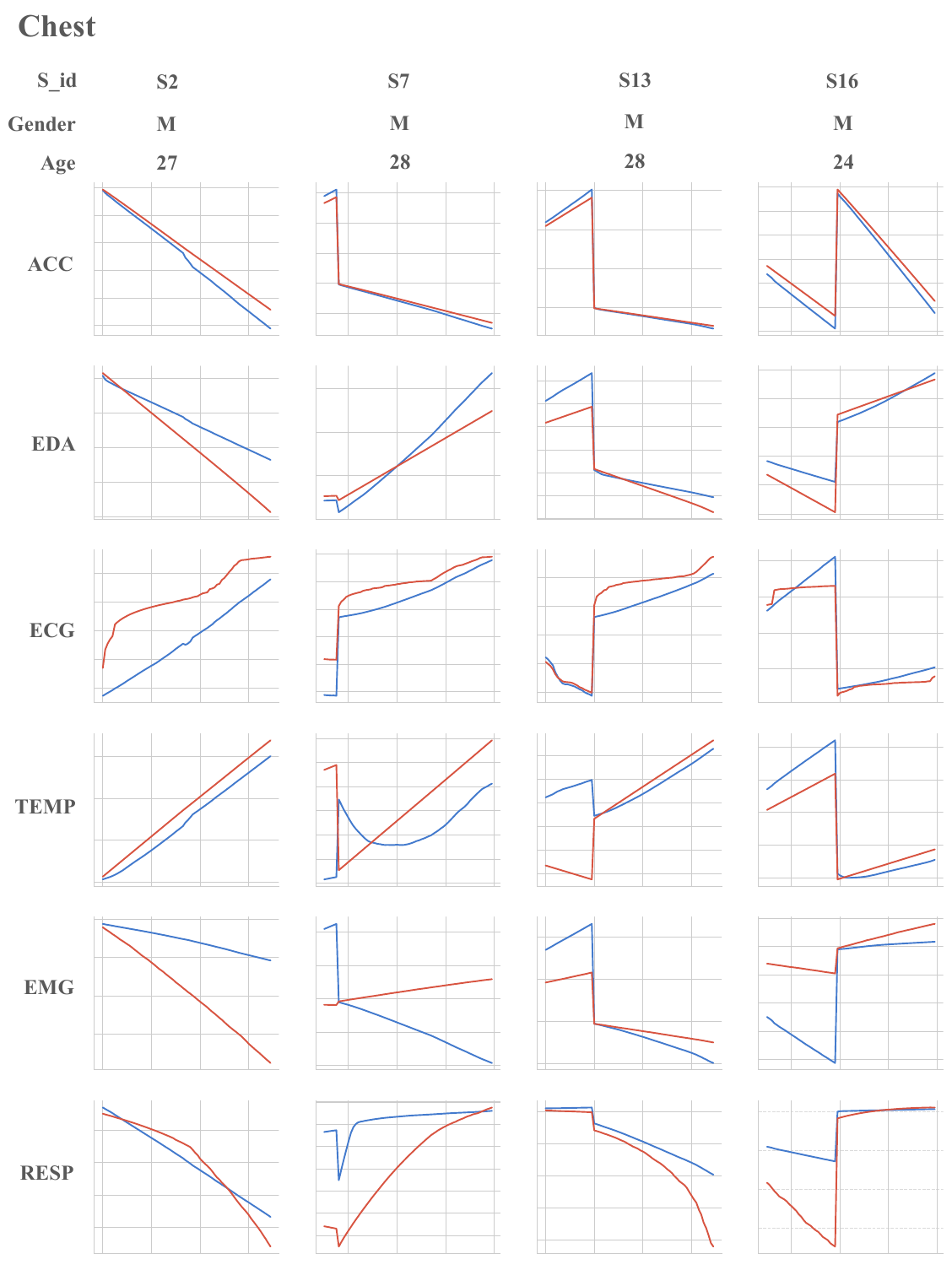}
  \caption{The comparison of the fitted curves for affective indicators based on the selected formulas for four randomly chosen individuals. Using the selected formulas and the output from the PhysioFormer model, the fitting curves for each affective indicator in the Chest dataset were plotted. In the figure, the blue curve represents the model's output, while the red curve represents the results calculated from the formulas.}
  \label{13chest_fitting}
\end{figure}

Based on the selected formulas, we randomly chose four individuals' affective indicators. Using the selected formulas and the output from the PhysioFormer model, we plotted the fitting curves for each physiological indicator across both datasets, as shown in the Figure \ref{11wrist_fitting} and Figure \ref{13chest_fitting}.

To evaluate the fit of each formula, we used the $R^2$ metric, which measures the goodness of fit between the predicted values and the actual values. The closer the $R^2$ value is to 1, the stronger the model's ability to explain the data. We calculated the $R^2$ values for all selected formulas, and the results are shown in the Table \ref{R_result}.

\begin{table}[htbp]
  \caption{The table presents the evaluation results of the selected formulas using the $R^2$ metric, which measures the goodness of fit between the predicted values and the actual values. The $R^2$ values for all the selected formulas are listed, providing a clear view of the fitting performance of each formula.}
  \label{R_result}
  \resizebox{0.55\linewidth}{!}{
  \begin{tabular}{c|c|c|c|c|c|c|c}
  \toprule
  \textbf{}      & \textbf{ACC} & \textbf{EDA} & \textbf{TEMP} & \textbf{BVP} & \textbf{ECG} & \textbf{EMG} & \textbf{RESP} \\ \hline
  \textbf{Wrist} & 0.98 & 0.15 & 0.39 & 0.99 & -- & -- & -- \\ \hline
  \textbf{Chest} & 0.98 & 0.63 & 0.76 & -- & 0.73 & 0.15 & 0.67 \\ \bottomrule
  \end{tabular}
  }
\end{table}

According to the results shown in the table, the high-fitting formulas include ACC and BVP from the Wrist dataset, as well as ACC, TEMP, and ECG from the Chest dataset, with $R^2$ values close to or reaching 0.98 to 0.99. This indicates that these formulas predict the corresponding data very well, and the model is able to explain the variability in the data accurately. These physiological indicators are well predicted and explained under the current model and formula combinations.

Moderate-fitting formulas include TEMP from the Wrist dataset and EDA and RESP from the Chest dataset, with $R^2$ values ranging from 0.39 to 0.76. These formulas perform reasonably well but still have room for improvement. While they capture some important patterns in the data, they do not fully explain the data’s variability.

Low-fitting formulas include EDA from the Wrist dataset and EMG from the Chest dataset, both with low $R^2$ values of 0.15, indicating poor predictive performance for these formulas. The model struggles to capture the patterns in the data for these indicators, which may be due to the model’s insufficient complexity or interference from noise and outliers in the data.

The results suggest that the model performs satisfactorily for most physiological indicators, but its predictive ability needs improvement for certain indicators. Future work can focus on the following areas: first, introducing more relevant features to bring additional information and enhance the model's predictive ability; second, conducting data cleaning and preprocessing for indicators with low $R^2$ values to remove noise and outliers, thereby improving data quality.

\section{Conclusion and Future Work}

In this study, we proposed and implemented the PhysioFormer model for affective computation based on multimodal physiological signals. PhysioFormer model effectively addresses the variability in physiological signals across individuals by integrating individual attributes and multimodal signals. This integration allows the model to demonstrate high reliability and generalization in cross-individual affective computation tasks, ensuring its stability across different affective computation tasks, making it highly applicable and reliable in real-world scenarios. By incorporating feature embedding and affective representation modules, PhysioFormer is able to capture the temporal dependencies and multimodal features of physiological signals, significantly enhancing its accuracy. Extensive experiments on the Wrist and Chest subsets of the WESAD dataset demonstrated PhysioFormer's superior performance in affective computation tasks. The experimental results showed that PhysioFormer achieved over 99\% accuracy on both subsets, outperforming the current SOTA models. Additionally, we introduced an Explanation model to the PhysioFormer model, utilizing symbolic regression techniques to extract symbolic laws that reveal the relationships between physiological signals and affective states. This enhancement improves the model's explainability, offering new insights into the intrinsic connections between physiological signals and affective states.

However, despite the promising results, this study reveals several key areas for future research. First, the scalability of PhysioFormer to larger datasets and real-world applications remains to be validated, particularly how to balance the complexity and computational efficiency in real-time affective recognition systems. Second, addressing cross-individual variability, future work should further optimize the feature embedding module to improve the model’s generalization across different individuals, ensuring reliability in broader populations. Additionally, integrating more physiological signals (e.g., blood pressure, EEG) and environmental factors (e.g., noise, light intensity) could help build a more comprehensive affective recognition system, further enhancing the model’s performance and adaptability.

Future research can also expand and refine the symbolic distillation approach to extract more explainable mathematical formulas, thereby improving model explainability and transparency. This is particularly important for applying the PhysioFormer model in emotion-driven real-world applications such as health monitoring and psychological interventions. Moreover, further exploration of how symbolic regression can be used to explain the relationship between multimodal physiological signals and complex affective states will provide new directions and inspiration for future research in affective computation.

\bibliographystyle{elsarticle-num}
\bibliography{cas-refs,All}

\begin{thebibliography}{10}
\expandafter\ifx\csname url\endcsname\relax
  \def\url#1{\texttt{#1}}\fi
\expandafter\ifx\csname urlprefix\endcsname\relax\def\urlprefix{URL }\fi
\expandafter\ifx\csname href\endcsname\relax
  \def\href#1#2{#2} \def\path#1{#1}\fi

\bibitem{Chen2023b}
Z.~Chen, M.~Lin, Z.~Wang, Q.~Zheng, C.~Liu, Spatio-temporal representation
  learning enhanced speech emotion recognition with multi-head attention
  mechanisms, Knowledge-Based Systems 281 (2023) 111077.
\newblock \href {https://doi.org/10.1016/j.knosys.2023.111077}
  {\path{doi:10.1016/j.knosys.2023.111077}}.

\bibitem{Wang2024b}
Z.~Wang, W.~Wu, C.~Zeng, H.~Luo, J.~Sun, Psychological factors enhanced
  heterogeneous learning interactive graph knowledge tracing for understanding
  the learning process, Frontiers in Psychology 15 (May 2024).
\newblock \href {https://doi.org/10.3389/fpsyg.2024.1359199}
  {\path{doi:10.3389/fpsyg.2024.1359199}}.

\bibitem{Saloni2023}
S.~Dattani, L.~Rodés-Guirao, H.~Ritchie, M.~Roser, Mental health, Our World in
  DataHttps://ourworldindata.org/mental-health (2023).

\bibitem{McEwen2012}
B.~S. McEwen, Brain on stress: {{How}} the social environment gets under the
  skin, Proceedings of the National Academy of Sciences 109~(supplement\_2)
  (2012) 17180--17185.
\newblock \href {https://doi.org/10.1073/pnas.1121254109}
  {\path{doi:10.1073/pnas.1121254109}}.

\bibitem{Ertin2011}
E.~Ertin, N.~Stohs, S.~Kumar, A.~Raij, M.~al'Absi, S.~Shah,
  \href{https://doi.org/10.1145/2070942.2070970}{Autosense: unobtrusively
  wearable sensor suite for inferring the onset, causality, and consequences of
  stress in the field}, in: Proceedings of the 9th ACM Conference on Embedded
  Networked Sensor Systems, SenSys '11, Association for Computing Machinery,
  New York, NY, USA, 2011, p. 274–287.
\newblock \href {https://doi.org/10.1145/2070942.2070970}
  {\path{doi:10.1145/2070942.2070970}}.
\newline\urlprefix\url{https://doi.org/10.1145/2070942.2070970}

\bibitem{Nadai2016}
S.~De~Nadai, M.~D'Incà, F.~Parodi, M.~Benza, A.~Trotta, E.~Zero, L.~Zero,
  R.~Sacile, Enhancing safety of transport by road by on-line monitoring of
  driver emotions, in: 2016 11th System of Systems Engineering Conference
  (SoSE), 2016, pp. 1--4.
\newblock \href {https://doi.org/10.1109/SYSOSE.2016.7542941}
  {\path{doi:10.1109/SYSOSE.2016.7542941}}.

\bibitem{Hosseini2023}
E.~Hosseini, R.~Fang, R.~Zhang, S.~Rafatirad, H.~Homayoun, Emotion and {{Stress
  Recognition Utilizing Galvanic Skin Response}} and {{Wearable Technology}}:
  {{A Real-time Approach}} for {{Mental Health Care}}, in: 2023 {{IEEE
  International Conference}} on {{Bioinformatics}} and {{Biomedicine}}
  ({{BIBM}}), IEEE, Istanbul, Turkiye, 2023, pp. 1125--1131.
\newblock \href {https://doi.org/10.1109/BIBM58861.2023.10386049}
  {\path{doi:10.1109/BIBM58861.2023.10386049}}.

\bibitem{Andreou2011}
E.~Andreou, E.~C. Alexopoulos, C.~Lionis, L.~Varvogli, C.~Gnardellis, G.~P.
  Chrousos, C.~Darviri,
  \href{https://www.mdpi.com/1660-4601/8/8/3287}{Perceived stress scale:
  Reliability and validity study in greece}, International Journal of
  Environmental Research and Public Health 8~(8) (2011) 3287--3298.
\newblock \href {https://doi.org/10.3390/ijerph8083287}
  {\path{doi:10.3390/ijerph8083287}}.
\newline\urlprefix\url{https://www.mdpi.com/1660-4601/8/8/3287}

\bibitem{Han2017}
H.~Mei, X.~Xu, Eeg-based emotion classification using convolutional neural
  network, in: 2017 International Conference on Security, Pattern Analysis, and
  Cybernetics (SPAC), 2017, pp. 130--135.
\newblock \href {https://doi.org/10.1109/SPAC.2017.8304263}
  {\path{doi:10.1109/SPAC.2017.8304263}}.

\bibitem{Song2020}
T.~Song, W.~Zheng, P.~Song, Z.~Cui, Eeg emotion recognition using dynamical
  graph convolutional neural networks, IEEE Transactions on Affective Computing
  11~(3) (2020) 532--541.
\newblock \href {https://doi.org/10.1109/TAFFC.2018.2817622}
  {\path{doi:10.1109/TAFFC.2018.2817622}}.

\bibitem{Ashwin2022}
V.~Ashwin, R.~Jegan, P.~Rajalakshmy, Stress detection using wearable
  physiological sensors and machine learning algorithm, in: 2022 6th
  International Conference on Electronics, Communication and Aerospace
  Technology, 2022, pp. 972--977.
\newblock \href {https://doi.org/10.1109/ICECA55336.2022.10009326}
  {\path{doi:10.1109/ICECA55336.2022.10009326}}.

\bibitem{Zeng2023b}
C.~Zeng, K.~Yan, Z.~Wang, Y.~Yu, S.~Xia, N.~Zhao, Abs-cam: A gradient
  optimization interpretable approach for explanation of convolutional neural
  networks, Signal, Image and Video Processing 17~(4) (2023) 1069--1076.
\newblock \href {https://doi.org/10.1007/s11760-022-02313-0}
  {\path{doi:10.1007/s11760-022-02313-0}}.

\bibitem{Udrescu2020}
S.-M. Udrescu, M.~Tegmark, {{AI Feynman}}: {{A}} physics-inspired method for
  symbolic regression, Science Advances 6~(16) (2020) eaay2631.
\newblock \href {https://doi.org/10.1126/sciadv.aay2631}
  {\path{doi:10.1126/sciadv.aay2631}}.

\bibitem{Garc2024}
R.~A. García-Hernández, H.~Luna-García, J.~M. Celaya-Padilla,
  A.~García-Hernández, L.~C. Reveles-Gómez, L.~A. Flores-Chaires, J.~R.
  Delgado-Contreras, D.~Rondon, K.~O. Villalba-Condori,
  \href{https://www.mdpi.com/2076-3417/14/16/7165}{A systematic literature
  review of modalities, trends, and limitations in emotion recognition,
  affective computing, and sentiment analysis}, Applied Sciences 14~(16)
  (2024).
\newblock \href {https://doi.org/10.3390/app14167165}
  {\path{doi:10.3390/app14167165}}.
\newline\urlprefix\url{https://www.mdpi.com/2076-3417/14/16/7165}

\bibitem{Bernhard2018}
B.~Kratzwald, S.~Ilic, M.~Kraus, S.~Feuerriegel, H.~Prendinger, Decision
  support with text-based emotion recognition: Deep learning for affective
  computing, CoRR abs/1803.06397 (2018).
\newblock \href {http://arxiv.org/abs/1803.06397} {\path{arXiv:1803.06397}}.

\bibitem{Quaedflieg2013}
C.~W. E.~M. Quaedflieg, T.~Meyer, T.~Smeets, The imaging maastricht acute
  stress test (imast): a neuroimaging compatible psychophysiological stressor,
  Psychophysiology 50~(8) (2013) 758--766.
\newblock \href {https://doi.org/10.1111/psyp.12058}
  {\path{doi:10.1111/psyp.12058}}.

\bibitem{Dedovic2005}
K.~Dedovic, R.~Renwick, N.~K. Mahani, V.~Engert, S.~J. Lupien, J.~C. Pruessner,
  The montreal imaging stress task: using functional imaging to investigate the
  effects of perceiving and processing psychosocial stress in the human brain,
  Journal of Psychiatry and Neuroscience 30~(5) (2005) 319--325.
\newblock \href {https://doi.org/10.1016/S0895-4356(01)00408-1}
  {\path{doi:10.1016/S0895-4356(01)00408-1}}.

\bibitem{Sarkar2022}
P.~Sarkar, A.~Etemad, Self-supervised {{ECG Representation Learning}} for
  {{Emotion Recognition}}, IEEE Transactions on Affective Computing 13~(3)
  (2022) 1541--1554.
\newblock \href {http://arxiv.org/abs/2002.03898} {\path{arXiv:2002.03898}},
  \href {https://doi.org/10.1109/TAFFC.2020.3014842}
  {\path{doi:10.1109/TAFFC.2020.3014842}}.

\bibitem{Akre2023}
S.~Akre, Z.~D. Cohen, A.~Welborn, T.~D. Zbozinek, B.~Balliu, J.~Flint, A.~A.~T.
  Bui, M.~G. Craske, Detection of {{Symptoms}} of {{Depression Using Data
  From}} the {{iPhone}} and {{Apple Watch}}, in: 2023 {{IEEE International
  Conference}} on {{Bioinformatics}} and {{Biomedicine}} ({{BIBM}}), IEEE,
  Istanbul, Turkiye, 2023, pp. 1818--1823.
\newblock \href {https://doi.org/10.1109/BIBM58861.2023.10385797}
  {\path{doi:10.1109/BIBM58861.2023.10385797}}.

\bibitem{Koldijk2014}
S.~Koldijk, M.~Sappelli, S.~Verberne, M.~A. Neerincx, W.~Kraaij,
  \href{https://doi.org/10.1145/2663204.2663257}{The swell knowledge work
  dataset for stress and user modeling research}, in: Proceedings of the 16th
  International Conference on Multimodal Interaction, ICMI '14, Association for
  Computing Machinery, New York, NY, USA, 2014, p. 291–298.
\newblock \href {https://doi.org/10.1145/2663204.2663257}
  {\path{doi:10.1145/2663204.2663257}}.
\newline\urlprefix\url{https://doi.org/10.1145/2663204.2663257}

\bibitem{Siddharth2019}
Siddharth, T.~Jung, T.~J. Sejnowski, Utilizing deep learning towards
  multi-modal bio-sensing and vision-based affective computing, CoRR
  abs/1905.07039 (2019).
\newblock \href {http://arxiv.org/abs/1905.07039} {\path{arXiv:1905.07039}}.

\bibitem{Zeng2024e}
C.~Zeng, Y.~Yu, Z.~Wang, S.~Xia, H.~Cui, X.~Wan, Gsista-net: Generalized
  structure ista networks for image compressed sensing based on optimized
  unrolling algorithm, Multimedia Tools and Applications (Mar. 2024).
\newblock \href {https://doi.org/10.1007/s11042-024-18724-9}
  {\path{doi:10.1007/s11042-024-18724-9}}.

\bibitem{Wang2023a}
Z.~Wang, Z.~Wang, C.~Zeng, Y.~Yu, X.~Wan, High-quality image compressed sensing
  and reconstruction with multi-scale dilated convolutional neural network,
  Circuits, Systems, and Signal Processing 42~(3) (2023) 1593--1616.
\newblock \href {https://doi.org/10.1007/s00034-022-02181-6}
  {\path{doi:10.1007/s00034-022-02181-6}}.

\bibitem{Zeng2023c}
C.~Zeng, S.~Xia, Z.~Wang, X.~Wan, Multi-channel representation learning
  enhanced unfolding multi-scale compressed sensing network for high quality
  image reconstruction, Entropy 25~(12) (2023) 1579.
\newblock \href {https://doi.org/10.3390/e25121579}
  {\path{doi:10.3390/e25121579}}.

\bibitem{Li2023h}
L.~Li, Z.~Wang, T.~Zhang, Gbh-yolov5: Ghost convolution with bottleneckcsp and
  tiny target prediction head incorporating yolov5 for pv panel defect
  detection, Electronics 12~(3) (2023) 1--15.
\newblock \href {https://doi.org/10.3390/electronics12030561}
  {\path{doi:10.3390/electronics12030561}}.

\bibitem{Zeng2022}
C.~Zeng, J.~Ye, Z.~Wang, N.~Zhao, M.~Wu, Cascade neural network-based joint
  sampling and reconstruction for image compressed sensing, Signal, Image and
  Video Processing 16~(1) (2022) 47--54.
\newblock \href {https://doi.org/10.1007/s11760-021-01955-w}
  {\path{doi:10.1007/s11760-021-01955-w}}.

\bibitem{Zheng2024}
Q.~Zheng, Z.~Chen, Z.~Wang, H.~Liu, M.~Lin, Meconformer: Highly representative
  embedding extractor for speaker verification via incorporating selective
  convolution into deep speaker encoder, Expert Systems with Applications 244
  (2024) 123004.
\newblock \href {https://doi.org/10.1016/j.eswa.2023.123004}
  {\path{doi:10.1016/j.eswa.2023.123004}}.

\bibitem{Zeng2024g}
C.~Zeng, Y.~Zhao, Z.~Wang, K.~Li, X.~Wan, M.~Liu, Squeeze-and-excitation
  self-attention mechanism enhanced digital audio source recognition based on
  transfer learning, Circuits, Systems, and Signal Processing (Sep. 2024).
\newblock \href {https://doi.org/10.1007/s00034-024-02850-8}
  {\path{doi:10.1007/s00034-024-02850-8}}.

\bibitem{Wang2023f}
Z.~Wang, J.~Zhan, G.~Zhang, D.~Ouyang, H.~Guo, An end-to-end transfer learning
  framework of source recording device identification for audio sustainable
  security, Sustainability 15~(14) (2023) 11272.
\newblock \href {https://doi.org/10.3390/su151411272}
  {\path{doi:10.3390/su151411272}}.

\bibitem{Zeng2024b}
C.~Zeng, K.~Li, Z.~Wang, Enfformer: Long-short term representation of electric
  network frequency for digital audio tampering detection, Knowledge-Based
  Systems 297 (2024) 111938.
\newblock \href {https://doi.org/10.1016/j.knosys.2024.111938}
  {\path{doi:10.1016/j.knosys.2024.111938}}.

\bibitem{Wang2011a}
Z.-F. Wang, Q.-H. He, X.-Y. Zhang, H.-Y. Luo, Z.-S. Su, Playback attack
  detection based on channel pattern noise, Journal of South China University
  of Technology 39~(10) (2011) 7--12.

\bibitem{Zeng2024f}
C.~Zeng, S.~Kong, Z.~Wang, K.~Li, Y.~Zhao, X.~Wan, Y.~Chen, Discriminative
  component analysis enhanced feature fusion of electrical network frequency
  for digital audio tampering detection, Circuits, Systems, and Signal
  Processing (Jul. 2024).
\newblock \href {https://doi.org/10.1007/s00034-024-02787-y}
  {\path{doi:10.1007/s00034-024-02787-y}}.

\bibitem{Wang2011}
Z.-F. Wang, G.~Wei, Q.-H. He, Channel pattern noise based playback attack
  detection algorithm for speaker recognition, in: 2011 International
  Conference on Machine Learning and Cybernetics, Vol.~4, 2011, pp. 1708--1713.
\newblock \href {https://doi.org/10.1109/ICMLC.2011.6016982}
  {\path{doi:10.1109/ICMLC.2011.6016982}}.

\bibitem{Zeng2024c}
C.~Zeng, S.~Kong, Z.~Wang, K.~Li, Y.~Zhao, X.~Wan, Y.~Chen, Digital audio
  tampering detection based on spatio-temporal representation learning of
  electrical network frequency, Multimedia Tools and Applications (2024)
  1--21\href {https://doi.org/10.1007/s11042-024-18887-5}
  {\path{doi:10.1007/s11042-024-18887-5}}.

\bibitem{Zhu2013}
Z.-Y. Zhu, Q.-H. He, X.-H. Feng, Y.-X. Li, Z.-F. Wang, Liveness detection using
  time drift between lip movement and voice, in: 2013 International Conference
  on Machine Learning and Cybernetics, Vol.~02, 2013, pp. 973--978.
\newblock \href {https://doi.org/10.1109/ICMLC.2013.6890423}
  {\path{doi:10.1109/ICMLC.2013.6890423}}.

\bibitem{Zeng2024d}
C.~Zeng, S.~Kong, Z.~Wang, S.~Feng, N.~Zhao, J.~Wang, Deletion and insertion
  tampering detection for speech authentication based on fluctuating super
  vector of electrical network frequency, Speech Communication 158 (2024)
  103046.
\newblock \href {https://doi.org/10.1016/j.specom.2024.103046}
  {\path{doi:10.1016/j.specom.2024.103046}}.

\bibitem{Wang2015b}
Z.~Wang, Q.~Liu, J.~Chen, H.~Yao, Recording source identification using device
  universal background model, in: 2015 International Conference of Educational
  Innovation through Technology (EITT), IEEE, Wuhan, China, 2015, pp. 19--23.
\newblock \href {https://doi.org/10.1109/EITT.2015.11}
  {\path{doi:10.1109/EITT.2015.11}}.

\bibitem{Zeng2024}
C.~Zeng, S.~Feng, Z.~Wang, X.~Wan, Y.~Chen, N.~Zhao, Spatio-temporal
  representation learning enhanced source cell-phone recognition from speech
  recordings, Journal of Information Security and Applications 80 (2024)
  103672.
\newblock \href {https://doi.org/10.1016/j.jisa.2023.103672}
  {\path{doi:10.1016/j.jisa.2023.103672}}.

\bibitem{Wang2018a}
Z.-F. Wang, J.~Wang, C.-Y. Zeng, Q.-S. Min, Y.~Tian, M.-Z. Zuo, Digital audio
  tampering detection based on enf consistency, in: 2018 International
  Conference on Wavelet Analysis and Pattern Recognition (ICWAPR), IEEE,
  Chengdu, 2018, pp. 209--214.
\newblock \href {https://doi.org/10.1109/ICWAPR.2018.8521378}
  {\path{doi:10.1109/ICWAPR.2018.8521378}}.

\bibitem{Zeng2024a}
C.~Zeng, S.~Feng, Z.~Wang, Y.~Zhao, K.~Li, X.~Wan, Audio source recording
  device recognition based on representation learning of sequential gaussian
  mean matrix, Forensic Science International: Digital Investigation 48 (2024)
  301676.
\newblock \href {https://doi.org/10.1016/j.fsidi.2023.301676}
  {\path{doi:10.1016/j.fsidi.2023.301676}}.

\bibitem{Wang2020h}
Z.~Wang, S.~Duan, C.~Zeng, X.~Yu, Y.~Yang, H.~Wu, Robust speaker identification
  of iot based on stacked sparse denoising auto-encoders, in: 2020
  International Conferences on Internet of Things (iThings), IEEE, Rhodes,
  Greece, 2020, pp. 252--257.
\newblock \href
  {https://doi.org/10.1109/iThings-GreenCom-CPSCom-SmartData-Cybermatics50389.2020.00056}
  {\path{doi:10.1109/iThings-GreenCom-CPSCom-SmartData-Cybermatics50389.2020.00056}}.

\bibitem{Zeng2023a}
C.~Zeng, S.~Kong, Z.~Wang, K.~Li, Y.~Zhao, Digital audio tampering detection
  based on deep temporal--spatial features of electrical network frequency,
  Information 14~(5) (2023) 253.
\newblock \href {https://doi.org/10.3390/info14050253}
  {\path{doi:10.3390/info14050253}}.

\bibitem{Zeng2023}
C.~Zeng, S.~Feng, D.~Zhu, Z.~Wang, Source acquisition device identification
  from recorded audio based on spatiotemporal representation learning with
  multi-attention mechanisms, Entropy 25~(4) (2023) 626.
\newblock \href {https://doi.org/10.3390/e25040626}
  {\path{doi:10.3390/e25040626}}.

\bibitem{Zeng2018}
C.-Y. Zeng, C.-F. Ma, Z.-F. Wang, J.-X. Ye, Stacked autoencoder networks based
  speaker recognition, in: 2018 International Conference on Machine Learning
  and Cybernetics (ICMLC), IEEE, Chengdu, 2018, pp. 294--299.
\newblock \href {https://doi.org/10.1109/ICMLC.2018.8526953}
  {\path{doi:10.1109/ICMLC.2018.8526953}}.

\bibitem{Brunton2016}
S.~L. Brunton, J.~L. Proctor, J.~N. Kutz,
  \href{https://www.pnas.org/doi/abs/10.1073/pnas.1517384113}{Discovering
  governing equations from data by sparse identification of nonlinear dynamical
  systems}, Proceedings of the National Academy of Sciences 113~(15) (2016)
  3932--3937.
\newblock \href {https://doi.org/10.1073/pnas.1517384113}
  {\path{doi:10.1073/pnas.1517384113}}.
\newline\urlprefix\url{https://www.pnas.org/doi/abs/10.1073/pnas.1517384113}

\bibitem{Rogers2024}
A.~W. Rogers, A.~Lane, C.~Mendoza, S.~Watson, A.~Kowalski, P.~Martin, D.~Zhang,
  Integrating knowledge-guided symbolic regression and model-based design of
  experiments to automate process flow diagram development, Chemical
  Engineering Science 300 (2024) 120580.
\newblock \href {https://doi.org/10.1016/j.ces.2024.120580}
  {\path{doi:10.1016/j.ces.2024.120580}}.

\bibitem{Masato2023}
M.~Miyazaki, K.-I. Ishikawa, K.~Nakashima, H.~Shimizu, T.~Takahashi,
  N.~Takahashi,
  \href{https://www.frontiersin.org/journals/artificial-intelligence/articles/10.3389/frai.2023.1039438}{Application
  of the symbolic regression program ai-feynman to psychology}, Frontiers in
  Artificial Intelligence 6 (2023).
\newblock \href {https://doi.org/10.3389/frai.2023.1039438}
  {\path{doi:10.3389/frai.2023.1039438}}.
\newline\urlprefix\url{https://www.frontiersin.org/journals/artificial-intelligence/articles/10.3389/frai.2023.1039438}

\bibitem{Liu2024}
S.~Liu, Q.~Li, X.~Shen, J.~Sun, Z.~Yang,
  \href{https://doi.org/10.1038/s43588-024-00629-0}{Automated discovery of
  symbolic laws governing skill acquisition from naturally occurring data},
  Nature Computational Science 4~(5) (2024) 334--345.
\newblock \href {https://doi.org/10.1038/s43588-024-00629-0}
  {\path{doi:10.1038/s43588-024-00629-0}}.
\newline\urlprefix\url{https://doi.org/10.1038/s43588-024-00629-0}

\bibitem{Zeng2022a}
C.~Zeng, Y.~Yang, Z.~Wang, S.~Kong, S.~Feng, Audio tampering forensics based on
  representation learning of enf phase sequence, International Journal of
  Digital Crime and Forensics 14~(1) (2022) 1--19.
\newblock \href {https://doi.org/10.4018/IJDCF.302894}
  {\path{doi:10.4018/IJDCF.302894}}.

\bibitem{Wang2022t}
Z.~Wang, Y.~Yang, C.~Zeng, S.~Kong, S.~Feng, N.~Zhao, Shallow and deep feature
  fusion for digital audio tampering detection, EURASIP Journal on Advances in
  Signal Processing 2022~(69) (2022) 1--20.
\newblock \href {https://doi.org/10.1186/s13634-022-00900-4}
  {\path{doi:10.1186/s13634-022-00900-4}}.

\bibitem{Zeng2021a}
C.~Zeng, D.~Zhu, Z.~Wang, M.~Wu, W.~Xiong, N.~Zhao, Spatial and temporal
  learning representation for end-to-end recording device identification,
  EURASIP Journal on Advances in Signal Processing 2021~(1) (2021) 41.
\newblock \href {https://doi.org/10.1186/s13634-021-00763-1}
  {\path{doi:10.1186/s13634-021-00763-1}}.

\bibitem{Wang2021m}
Z.~Wang, C.~Zeng, S.~Duan, H.~Ouyang, H.~Xu, Robust speaker recognition based
  on stacked auto-encoders, in: L.~Barolli, K.~F. Li, T.~Enokido, M.~Takizawa
  (Eds.), Advances in Networked-Based Information Systems, Vol. 1264, Springer
  International Publishing, Cham, 2021, pp. 390--399.

\bibitem{Zeng2021b}
C.~Zeng, D.~Zhu, Z.~Wang, Y.~Yang, Deep and shallow feature fusion and
  recognition of recording devices based on attention mechanism, in:
  L.~Barolli, K.~F. Li, H.~Miwa (Eds.), Advances in Intelligent Networking and
  Collaborative Systems, Vol. 1263, Springer International Publishing, Cham,
  2021, pp. 372--381.

\bibitem{Zeng2020}
C.~Zeng, D.~Zhu, Z.~Wang, Z.~Wang, N.~Zhao, L.~He, An end-to-end deep source
  recording device identification system for web media forensics, International
  Journal of Web Information Systems 16~(4) (2020) 413--425.
\newblock \href {https://doi.org/10.1108/IJWIS-06-2020-0038}
  {\path{doi:10.1108/IJWIS-06-2020-0038}}.

\bibitem{Wang2021}
Z.~Wang, C.~Zuo, C.~Zeng, Sae based unified double jpeg compression detection
  system for web image forensics, International Journal of Web Information
  Systems 17~(2) (2021) 84--98.
\newblock \href {https://doi.org/10.1108/IJWIS-11-2020-0073}
  {\path{doi:10.1108/IJWIS-11-2020-0073}}.

\bibitem{Zeng2021c}
C.~Zeng, Z.~Wang, Z.~Wang, K.~Yan, Y.~Yu, Image compressed sensing and
  reconstruction of multi-scale residual network combined with channel
  attention mechanism, Journal of Physics: Conference Series 2010~(1) (2021)
  012134.
\newblock \href {https://doi.org/10.1088/1742-6596/2010/1/012134}
  {\path{doi:10.1088/1742-6596/2010/1/012134}}.

\bibitem{Tian2018}
Y.~Tian, X.~Wang, H.~Yao, J.~Chen, Z.~Wang, L.~Yi, Occlusion handling using
  moving volume and ray casting techniques for augmented reality systems,
  Multimedia Tools and Applications 77~(13) (2018) 16561--16578.
\newblock \href {https://doi.org/10.1007/s11042-017-5228-2}
  {\path{doi:10.1007/s11042-017-5228-2}}.

\bibitem{Wang2015a}
Z.~Wang, Q.~Liu, H.~Yao, J.~Chen, Virtual chime-bells experimental system based
  on multi-modal fusion, in: 2015 International Conference of Educational
  Innovation through Technology (EITT), IEEE, Wuhan, China, 2015, pp. 64--67.
\newblock \href {https://doi.org/10.1109/EITT.2015.20}
  {\path{doi:10.1109/EITT.2015.20}}.

\bibitem{Zeng2020a}
C.~Zeng, Z.~Wang, Z.~Wang, Image reconstruction of iot based on parallel cnn,
  in: 2020 International Conferences on Internet of Things (iThings), IEEE,
  Rhodes, Greece, 2020, pp. 258--263.
\newblock \href
  {https://doi.org/10.1109/iThings-GreenCom-CPSCom-SmartData-Cybermatics50389.2020.00057}
  {\path{doi:10.1109/iThings-GreenCom-CPSCom-SmartData-Cybermatics50389.2020.00057}}.

\bibitem{Min2018}
Q.~Min, Z.~Wang, N.~Liu, An evaluation of html5 and webgl for medical imaging
  applications, Journal of Healthcare Engineering 2018 (2018) e1592821.
\newblock \href {https://doi.org/10.1155/2018/1592821}
  {\path{doi:10.1155/2018/1592821}}.

\bibitem{Wang2017}
Z.-F. Wang, L.~Zhu, Q.-S. Min, C.-Y. Zeng, Double compression detection based
  on feature fusion, in: 2017 International Conference on Machine Learning and
  Cybernetics (ICMLC), IEEE, Ningbo, 2017, pp. 379--384.
\newblock \href {https://doi.org/10.1109/ICMLC.2017.8108951}
  {\path{doi:10.1109/ICMLC.2017.8108951}}.

\bibitem{Rabiner1989}
L.~Rabiner, A tutorial on hidden markov models and selected applications in
  speech recognition, Proceedings of the IEEE 77~(2) (1989) 257--286.
\newblock \href {https://doi.org/10.1109/5.18626} {\path{doi:10.1109/5.18626}}.

\bibitem{Lafferty2001}
J.~D. Lafferty, A.~McCallum, F.~C.~N. Pereira, Conditional random fields:
  Probabilistic models for segmenting and labeling sequence data, in:
  Proceedings of the Eighteenth International Conference on Machine Learning,
  Morgan Kaufmann Publishers Inc., San Francisco, CA, USA, 2001, p. 282–289.

\bibitem{Mikolov2010}
T.~Mikolov, M.~Karafi{\'a}t, L.~Burget, J.~{\v C}ernock{\'y}, S.~Khudanpur,
  Recurrent neural network based language model, in: Interspeech 2010, ISCA,
  2010, pp. 1045--1048.
\newblock \href {https://doi.org/10.21437/Interspeech.2010-343}
  {\path{doi:10.21437/Interspeech.2010-343}}.

\bibitem{Frinken2012}
V.~Frinken, F.~Zamora-Martínez, S.~España-Boquera, M.~J. Castro-Bleda,
  A.~Fischer, H.~Bunke, Long-short term memory neural networks language
  modeling for handwriting recognition, in: Proceedings of the 21st
  International Conference on Pattern Recognition (ICPR2012), 2012, pp.
  701--704.

\bibitem{Gers1999}
F.~Gers, J.~Schmidhuber, F.~Cummins, Learning to forget: continual prediction
  with lstm, in: 1999 Ninth International Conference on Artificial Neural
  Networks ICANN 99. (Conf. Publ. No. 470), Vol.~2, 1999, pp. 850--855 vol.2.
\newblock \href {https://doi.org/10.1049/cp:19991218}
  {\path{doi:10.1049/cp:19991218}}.

\bibitem{Dhavale2022}
M.~Dhavale, P.~Bhandari, Speech emotion recognition using cnn and lstm, in:
  2022 6th International Conference On Computing, Communication, Control And
  Automation (ICCUBEA), 2022, pp. 1--3.
\newblock \href {https://doi.org/10.1109/ICCUBEA54992.2022.10010751}
  {\path{doi:10.1109/ICCUBEA54992.2022.10010751}}.

\bibitem{Shyam2024}
S.~K. Sateesh, S.~BK, U.~D, \href{https://arxiv.org/abs/2408.10328}{Decoding
  human emotions: Analyzing multi-channel eeg data using lstm networks} (2024).
\newblock \href {http://arxiv.org/abs/2408.10328} {\path{arXiv:2408.10328}}.
\newline\urlprefix\url{https://arxiv.org/abs/2408.10328}

\bibitem{Samira2024}
S.~Hazmoune, F.~Bougamouza,
  \href{https://www.sciencedirect.com/science/article/pii/S0952197624004974}{Using
  transformers for multimodal emotion recognition: Taxonomies and state of the
  art review}, Engineering Applications of Artificial Intelligence 133 (2024)
  108339.
\newblock \href
  {https://doi.org/https://doi.org/10.1016/j.engappai.2024.108339}
  {\path{doi:https://doi.org/10.1016/j.engappai.2024.108339}}.
\newline\urlprefix\url{https://www.sciencedirect.com/science/article/pii/S0952197624004974}

\bibitem{Mittal2020}
T.~Mittal, U.~Bhattacharya, R.~Chandra, A.~Bera, D.~Manocha,
  \href{https://ojs.aaai.org/index.php/AAAI/article/view/5492}{M3er:
  Multiplicative multimodal emotion recognition using facial, textual, and
  speech cues}, Proceedings of the AAAI Conference on Artificial Intelligence
  34~(02) (2020) 1359--1367.
\newblock \href {https://doi.org/10.1609/aaai.v34i02.5492}
  {\path{doi:10.1609/aaai.v34i02.5492}}.
\newline\urlprefix\url{https://ojs.aaai.org/index.php/AAAI/article/view/5492}

\bibitem{Kumar2023}
C.~S.~A. Kumar, A.~D. Maharana, S.~M. Krishnan, S.~S.~S. Hanuma, G.~J. Lal,
  V.~Ravi, Speech emotion recognition using cnn-lstm and vision transformer,
  in: A.~Abraham, A.~Bajaj, N.~Gandhi, A.~M. Madureira, C.~Kahraman (Eds.),
  Innovations in Bio-Inspired Computing and Applications, Springer Nature
  Switzerland, Cham, 2023, pp. 86--97.

\bibitem{Wang2025}
Z.~Wang, L.~Li, C.~Zeng, S.~Dong, J.~Sun, Slbdetection-net: Towards closed-set
  and open-set student learning behavior detection in smart classroom of k-12
  education, Expert Systems with Applications 260 (2025) 125392.
\newblock \href {https://doi.org/10.1016/j.eswa.2024.125392}
  {\path{doi:10.1016/j.eswa.2024.125392}}.

\bibitem{Liao2024}
X.~Liao, X.~Zhang, Z.~Wang, H.~Luo, Design and implementation of an ai-enabled
  visual report tool as formative assessment to promote learning achievement
  and self-regulated learning: An experimental study, British Journal of
  Educational Technology 55~(3) (2024) 1253--1276.
\newblock \href {https://doi.org/10.1111/bjet.13424}
  {\path{doi:10.1111/bjet.13424}}.

\bibitem{Wang2024m}
Z.~Wang, M.~Wang, C.~Zeng, L.~Li, Sbd-net: Incorporating multi-level features
  for an efficient detection network of student behavior in smart classrooms,
  Applied Sciences 14~(18) (2024) 8357.
\newblock \href {https://doi.org/10.3390/app14188357}
  {\path{doi:10.3390/app14188357}}.

\bibitem{Dong2024}
S.~Dong, X.~Tao, R.~Zhong, Z.~Wang, M.~Zuo, J.~Sun, Advanced mathematics
  exercise recommendation based on automatic knowledge extraction and
  multilayer knowledge graph, IEEE Transactions on Learning Technologies 17
  (2024) 776--793.
\newblock \href {https://doi.org/10.1109/TLT.2023.3333669}
  {\path{doi:10.1109/TLT.2023.3333669}}.

\bibitem{Wang2023g}
Z.~Wang, J.~Yao, C.~Zeng, L.~Li, C.~Tan, Students' classroom behavior detection
  system incorporating deformable detr with swin transformer and light-weight
  feature pyramid network, Systems 11~(7) (2023) 372.
\newblock \href {https://doi.org/10.3390/systems11070372}
  {\path{doi:10.3390/systems11070372}}.

\bibitem{Wang2023j}
Z.~Wang, W.~Yan, C.~Zeng, Y.~Tian, S.~Dong, A unified interpretable intelligent
  learning diagnosis framework for learning performance prediction in
  intelligent tutoring systems, International Journal of Intelligent Systems
  2023 (2023) e4468025.
\newblock \href {https://doi.org/10.1155/2023/4468025}
  {\path{doi:10.1155/2023/4468025}}.

\bibitem{Wang2023w}
Z.~Wang, M.~Wang, C.~Zeng, J.~Yao, Y.~Yang, H.~Xu, Enhanced convolutional
  neural networks based learner authentication for personalized e-learning
  system, in: 2023 International Conference on Intelligent Education and
  Intelligent Research (IEIR), IEEE, Wuhan, China, 2023, pp. 1--7.
\newblock \href {https://doi.org/10.1109/IEIR59294.2023.10391216}
  {\path{doi:10.1109/IEIR59294.2023.10391216}}.

\bibitem{Wang2023l}
Z.~Wang, L.~Li, C.~Zeng, J.~Yao, Student learning behavior recognition
  incorporating data augmentation with learning feature representation in smart
  classrooms, Sensors 23~(19) (2023) 8190.
\newblock \href {https://doi.org/10.3390/s23198190}
  {\path{doi:10.3390/s23198190}}.

\bibitem{Wang2023d}
Z.~Wang, Y.~Hou, C.~Zeng, S.~Zhang, R.~Ye, Multiple learning features--enhanced
  knowledge tracing based on learner--resource response channels,
  Sustainability 15~(12) (2023) 9427.
\newblock \href {https://doi.org/10.3390/su15129427}
  {\path{doi:10.3390/su15129427}}.

\bibitem{Ma2023b}
L.~Ma, X.~Zhang, Z.~Wang, H.~Luo, Designing effective instructional feedback
  using a diagnostic and visualization system: Evidence from a high school
  biology class, Systems 11~(7) (2023) 364.
\newblock \href {https://doi.org/10.3390/systems11070364}
  {\path{doi:10.3390/systems11070364}}.

\bibitem{Li2023i}
L.~Li, Z.~Wang, Knowledge relation rank enhanced heterogeneous learning
  interaction modeling for neural graph forgetting knowledge tracing, PLOS ONE
  18~(12) (2023) e0295808.
\newblock \href {https://doi.org/10.1371/journal.pone.0295808}
  {\path{doi:10.1371/journal.pone.0295808}}.

\bibitem{Li2023g}
L.~Li, Z.~Wang, Knowledge graph-enhanced intelligent tutoring system based on
  exercise representativeness and informativeness, International Journal of
  Intelligent Systems 2023 (2023) e2578286.
\newblock \href {https://doi.org/10.1155/2023/2578286}
  {\path{doi:10.1155/2023/2578286}}.

\bibitem{Li2023f}
L.~Li, Z.~Wang, Calibrated q-matrix-enhanced deep knowledge tracing with
  relational attention mechanism, Applied Sciences 13~(4) (2023) 1--24.
\newblock \href {https://doi.org/10.3390/app13042541}
  {\path{doi:10.3390/app13042541}}.

\bibitem{Wang2022at}
Z.~Wang, J.~Yao, C.~Zeng, W.~Wu, H.~Xu, Y.~Yang, Yolov5 enhanced learning
  behavior recognition and analysis in smart classroom with multiple students,
  in: 2022 International Conference on Intelligent Education and Intelligent
  Research (IEIR), IEEE, 2022, pp. 23--29.
\newblock \href {https://doi.org/10.1109/IEIR56323.2022.10050042}
  {\path{doi:10.1109/IEIR56323.2022.10050042}}.

\bibitem{Wang2022as}
Z.~Wang, W.~Wu, C.~Zeng, J.~Yao, Y.~Yang, H.~Xu, Smart contract vulnerability
  detection for educational blockchain based on graph neural networks, in: 2022
  International Conference on Intelligent Education and Intelligent Research
  (IEIR), IEEE, 2022, pp. 8--14.
\newblock \href {https://doi.org/10.1109/IEIR56323.2022.10050059}
  {\path{doi:10.1109/IEIR56323.2022.10050059}}.

\bibitem{Lyu2022}
L.~Lyu, Z.~Wang, H.~Yun, Z.~Yang, Y.~Li, Deep knowledge tracing based on
  spatial and temporal representation learning for learning performance
  prediction, Applied Sciences 12~(14) (2022) 1--21.
\newblock \href {https://doi.org/10.3390/app12147188}
  {\path{doi:10.3390/app12147188}}.

\bibitem{Min2019}
Q.~Min, Z.~Wang, N.~Liu, Integrating a cloud learning environment into
  english-medium instruction to enhance non-native english-speaking students'
  learning, Innovations in Education and Teaching International 56~(4) (2019)
  493--504.
\newblock \href {https://doi.org/10.1080/14703297.2018.1483838}
  {\path{doi:10.1080/14703297.2018.1483838}}.

\bibitem{signal_filter}
C.~Setz, B.~Arnrich, J.~Schumm, R.~La~Marca, G.~Tröster, U.~Ehlert,
  Discriminating stress from cognitive load using a wearable eda device, IEEE
  Transactions on Information Technology in Biomedicine 14~(2) (2010) 410--417.
\newblock \href {https://doi.org/10.1109/TITB.2009.2036164}
  {\path{doi:10.1109/TITB.2009.2036164}}.

\bibitem{Bianchi2007}
G.~Bianchi, R.~Sorrentino,
  \href{https://books.google.com.sg/books?id=5S3LCIxnYCcC}{Electronic Filter
  Simulation \& Design}, McGraw-Hill Education, 2007.
\newline\urlprefix\url{https://books.google.com.sg/books?id=5S3LCIxnYCcC}

\bibitem{cvxEDA}
A.~Greco, G.~Valenza, A.~Lanata, E.~Scilingo, L.~Citi, {{cvxEDA}}: A {{Convex
  Optimization Approach}} to {{Electrodermal Activity Processing}}, IEEE
  Transactions on Biomedical Engineering (2016) 1--1\href
  {https://doi.org/10.1109/TBME.2015.2474131}
  {\path{doi:10.1109/TBME.2015.2474131}}.

\bibitem{NeuroKit2}
D.~Makowski, T.~Pham, Z.~J. Lau, J.~C. Brammer, F.~Lespinasse, H.~Pham,
  C.~Schölzel, S.~H.~A. Chen,
  \href{https://doi.org/10.3758/s13428-020-01516-y}{Neurokit2: A python toolbox
  for neurophysiological signal processing}, Behavior Research Methods 53~(4)
  (2021) 1689--1696.
\newblock \href {https://doi.org/10.3758/s13428-020-01516-y}
  {\path{doi:10.3758/s13428-020-01516-y}}.
\newline\urlprefix\url{https://doi.org/10.3758/s13428-020-01516-y}

\bibitem{Non_stationarity}
Y.~Liu, C.~Li, J.~Wang, M.~Long, Koopa: learning non-stationary time series
  dynamics with koopman predictors, NIPS '23, Curran Associates Inc., Red Hook,
  NY, USA, 2024.

\bibitem{grad_importance}
M.~Sundararajan, A.~Taly, Q.~Yan, Axiomatic attribution for deep networks, in:
  D.~Precup, Y.~W. Teh (Eds.), Proceedings of the 34th International Conference
  on Machine Learning, Vol.~70 of Proceedings of Machine Learning Research,
  PMLR, 2017, pp. 3319--3328.

\bibitem{WESAD_intro}
P.~Schmidt, A.~Reiss, R.~Duerichen, C.~Marberger, K.~Van~Laerhoven,
  \href{https://doi.org/10.1145/3242969.3242985}{Introducing wesad, a
  multimodal dataset for wearable stress and affect detection}, ICMI '18,
  Association for Computing Machinery, 2018, p. 400–408.
\newblock \href {https://doi.org/10.1145/3242969.3242985}
  {\path{doi:10.1145/3242969.3242985}}.
\newline\urlprefix\url{https://doi.org/10.1145/3242969.3242985}

\bibitem{Bobade2020}
P.~Bobade, M.~Vani, Stress {{Detection}} with {{Machine Learning}} and {{Deep
  Learning}} using {{Multimodal Physiological Data}}, in: 2020 {{Second
  International Conference}} on {{Inventive Research}} in {{Computing
  Applications}} ({{ICIRCA}}), IEEE, Coimbatore, India, 2020, pp. 51--57.
\newblock \href {https://doi.org/10.1109/ICIRCA48905.2020.9183244}
  {\path{doi:10.1109/ICIRCA48905.2020.9183244}}.

\bibitem{Siirtola2019}
P.~Siirtola, Continuous stress detection using the sensors of commercial
  smartwatch, in: Adjunct {{Proceedings}} of the 2019 {{ACM International Joint
  Conference}} on {{Pervasive}} and {{Ubiquitous Computing}} and
  {{Proceedings}} of the 2019 {{ACM International Symposium}} on {{Wearable
  Computers}}, ACM, London United Kingdom, 2019, pp. 1198--1201.
\newblock \href {https://doi.org/10.1145/3341162.3344831}
  {\path{doi:10.1145/3341162.3344831}}.

\bibitem{Ferdinando2018}
H.~Ferdinando, E.~Alasaarela,
  \href{https://jtec.utem.edu.my/jtec/article/view/4186}{Emotion recognition
  using cvxeda-based features}, Journal of Telecommunication, Electronic and
  Computer Engineering (JTEC) 10~(2-3) (2018) 19–23.
\newline\urlprefix\url{https://jtec.utem.edu.my/jtec/article/view/4186}

\bibitem{Yu2020}
D.~Yu, S.~Sun, \href{https://www.mdpi.com/2078-2489/11/4/212}{A systematic
  exploration of deep neural networks for eda-based emotion recognition},
  Information 11~(4) (2020).
\newblock \href {https://doi.org/10.3390/info11040212}
  {\path{doi:10.3390/info11040212}}.
\newline\urlprefix\url{https://www.mdpi.com/2078-2489/11/4/212}

\end{thebibliography}

\newpage
% \appendix
\section*{Appendix}
\subsection*{A. Features Description}

\begin{figure}[htbp]
  \centering
  \includegraphics[width=0.9\textwidth]{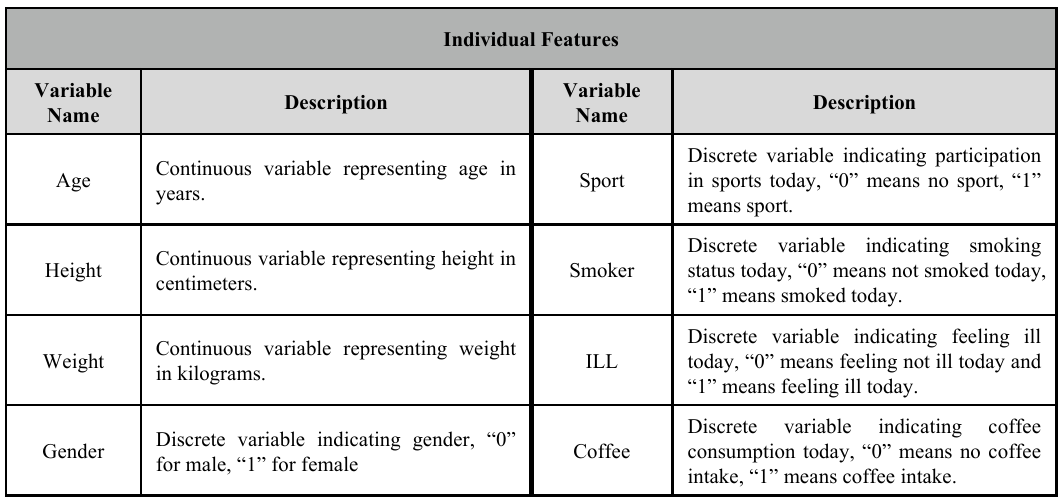}
  \caption{Individual features table, which describes all individual attributes features, their types, and the meanings they represent.}
  \label{A1}
\end{figure}

\begin{figure}[htbp]
  \centering
  \includegraphics[width=0.9\textwidth]{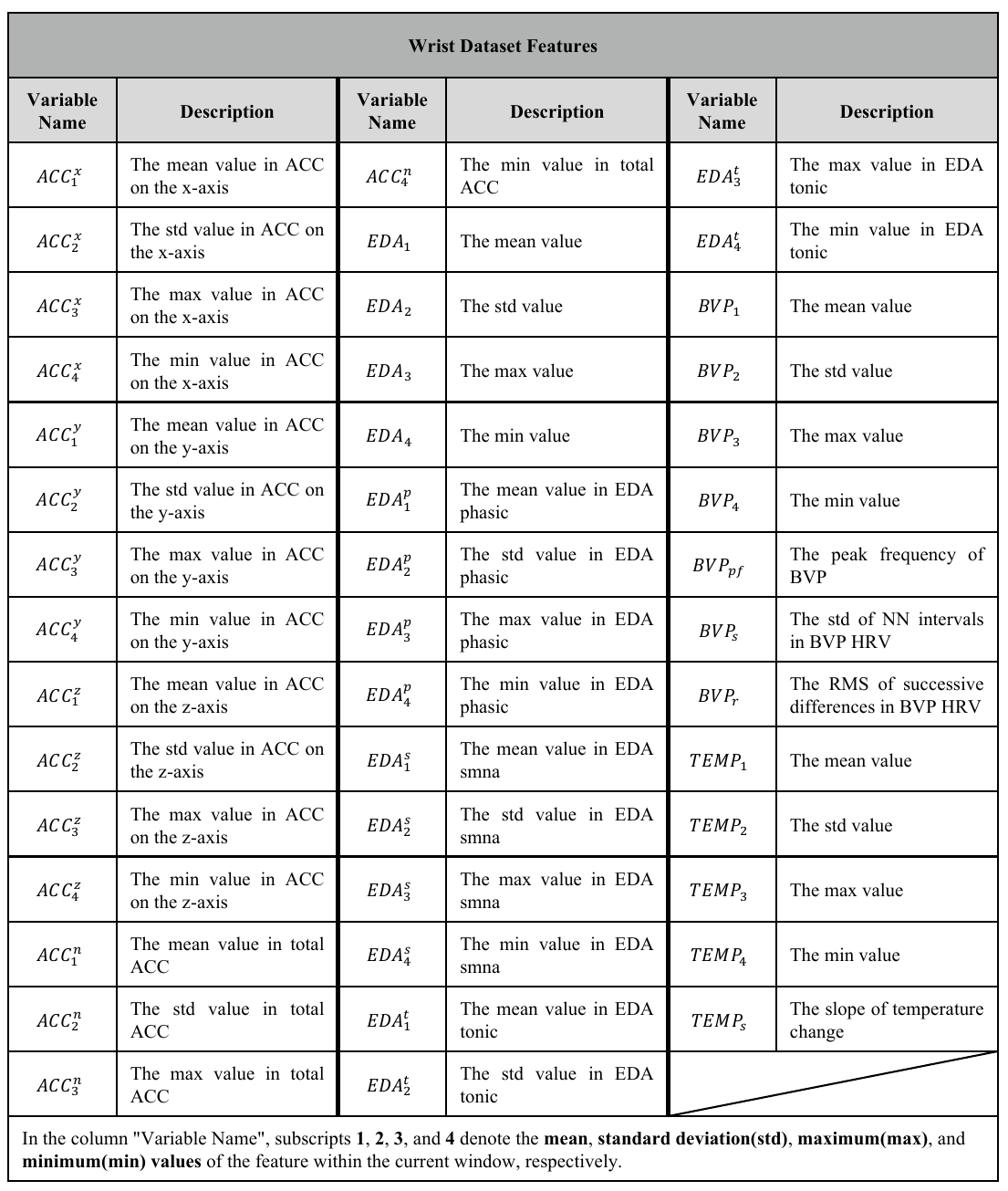}
  \caption{Features table for Wrist dataset, which describes all features and the meanings they represent.}
  \label{A2}
\end{figure}

\begin{figure}[htbp]
  \centering
  \includegraphics[width=0.9\textwidth]{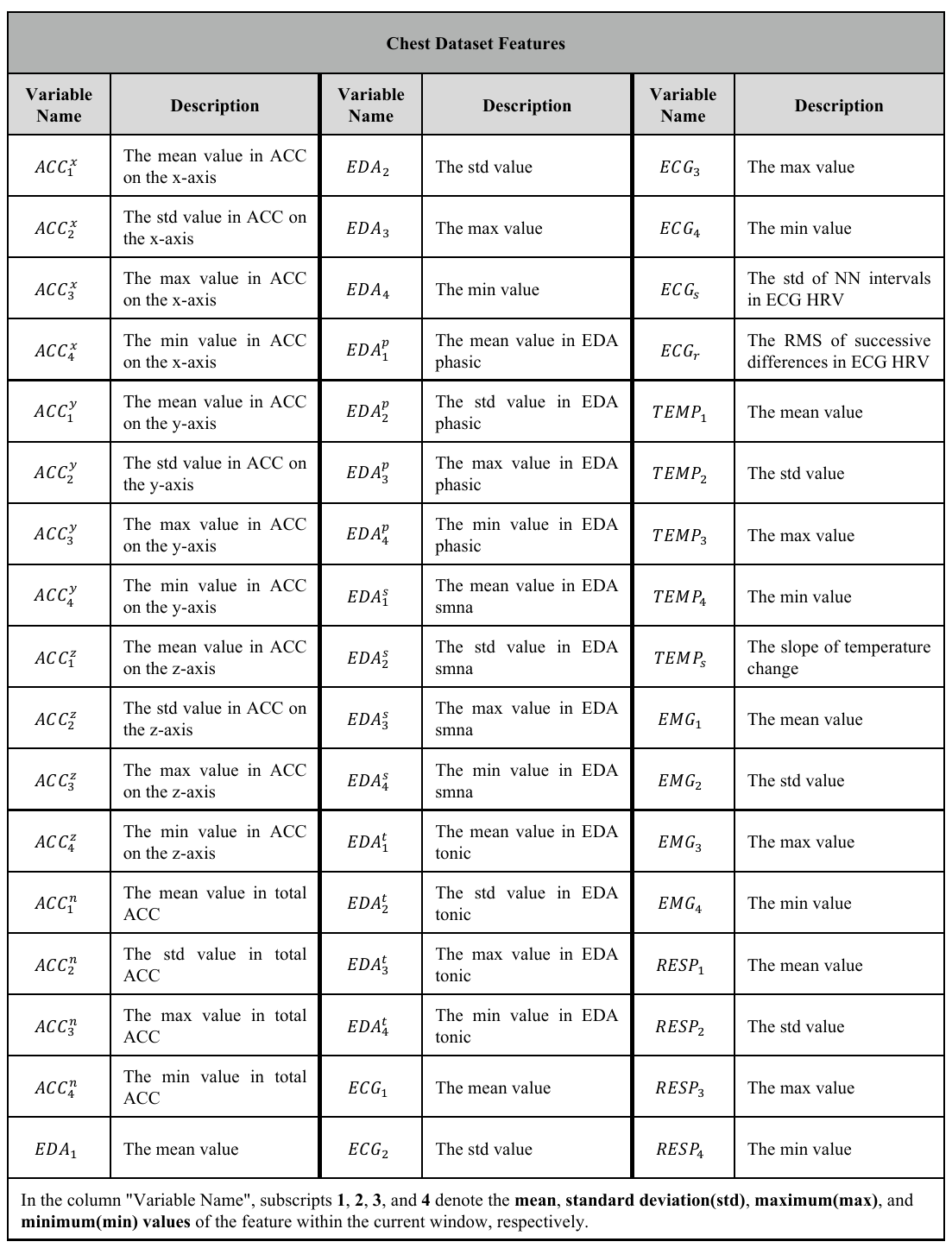}
  \caption{Features table for Chest dataset, which describes all features and the meanings they represent.}
  \label{A3}
\end{figure}

\subsection*{B. Top 10 Features in Terms of Features Importance Tables}

\subsubsection*{Tables of Wrist Dataset}

\begin{figure}[htbp]
  \centering
  \begin{subfigure}{.48\textwidth}
    \centering
    \includegraphics[width=.77\linewidth]{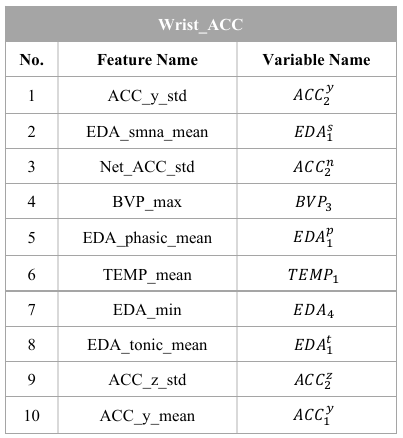}
    \caption{Top 10 features table of ACC in Wrist}
    \label{B1}
  \end{subfigure}
  \begin{subfigure}{.48\textwidth}
    \centering
    \includegraphics[width=.77\linewidth]{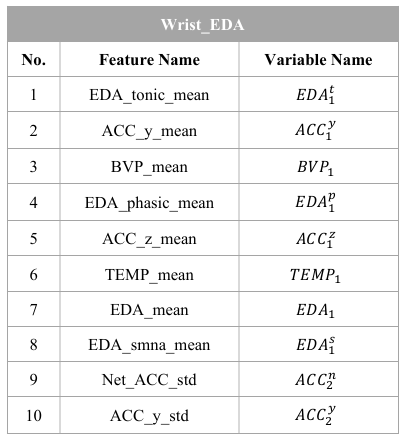}
    \caption{Top 10 features table of EDA in Wrist}
    \label{B2}
  \end{subfigure}
  \begin{subfigure}{.48\textwidth}
    \centering
    \includegraphics[width=.77\linewidth]{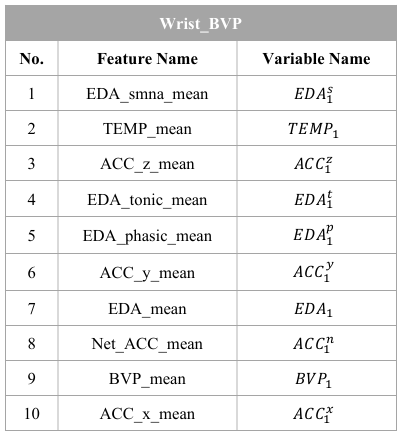}
    \caption{Top 10 features table of BVP in Wrist}
    \label{B3}
  \end{subfigure}
  \begin{subfigure}{.48\textwidth}
    \centering
    \includegraphics[width=.77\linewidth]{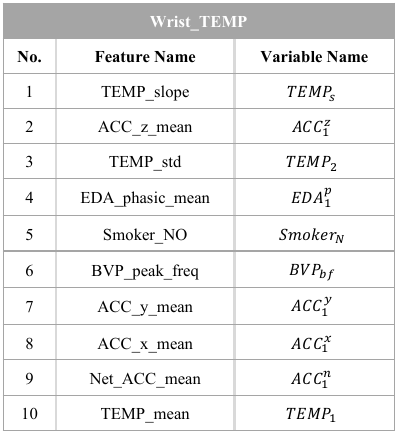}
    \caption{Top 10 features table of TEMP in Wrist}
    \label{B4}
  \end{subfigure}
\end{figure}

\subsubsection*{Tables of Chest Dataset}

\begin{figure}[htbp]
  \centering
  \begin{subfigure}{.48\textwidth}
    \centering
    \includegraphics[width=.77\linewidth]{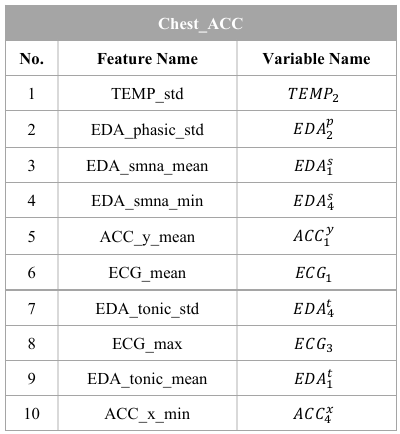}
    \caption{Top 10 features table of ACC in Chest}
    \label{B5}
  \end{subfigure}
  \begin{subfigure}{.48\textwidth}
    \centering
    \includegraphics[width=.77\linewidth]{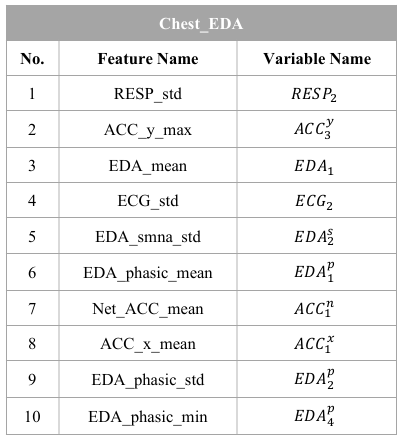}
    \caption{Top 10 features table of EDA in Chest}
    \label{B6}
  \end{subfigure}
  \begin{subfigure}{.48\textwidth}
    \centering
    \includegraphics[width=.77\linewidth]{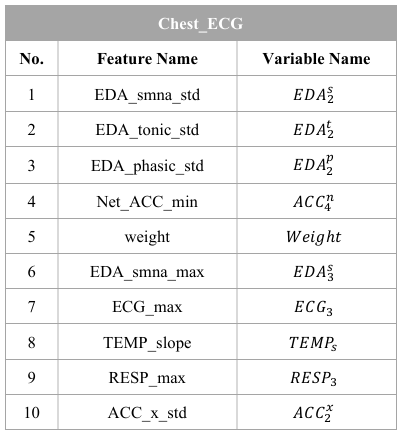}
    \caption{Top 10 features table of ECG in Chest}
    \label{B7}
  \end{subfigure}
  \begin{subfigure}{.48\textwidth}
    \centering
    \includegraphics[width=.77\linewidth]{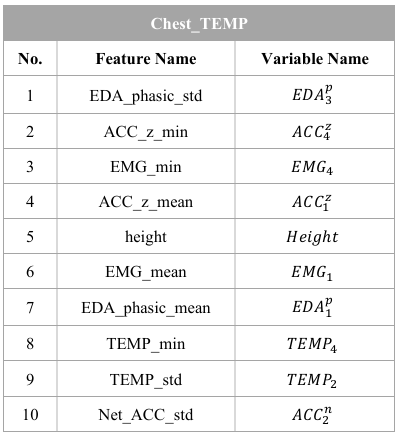}
    \caption{Top 10 features table of TEMP in Chest}
    \label{B8}
  \end{subfigure}
  \begin{subfigure}{.48\textwidth}
    \centering
    \includegraphics[width=.77\linewidth]{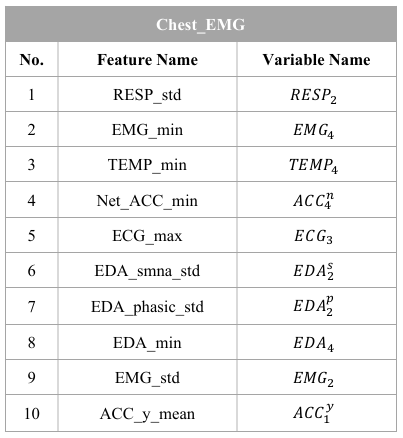}
    \caption{Top 10 features table of EMG in Chest}
    \label{B9}
  \end{subfigure}
  \begin{subfigure}{.48\textwidth}
    \centering
    \includegraphics[width=.77\linewidth]{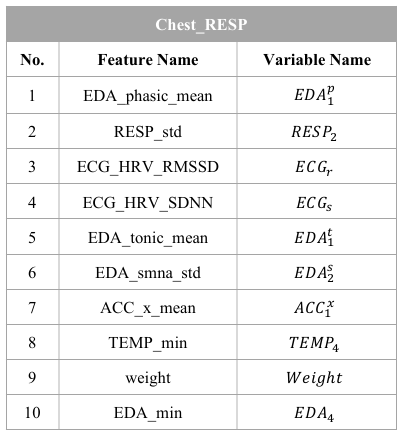}
    \caption{Top 10 features table of RESP in Chest}
    \label{B10}
  \end{subfigure}
\end{figure}

\subsection*{C. Laws Tables}

This part shows the indicators law table of WESAD dataset. The highlight formula indicates the selected formula.

\subsubsection*{Tables of Indicators Law in Wrist Dataset}

\begin{figure}
  \centering
  \includegraphics[width=.75\linewidth]{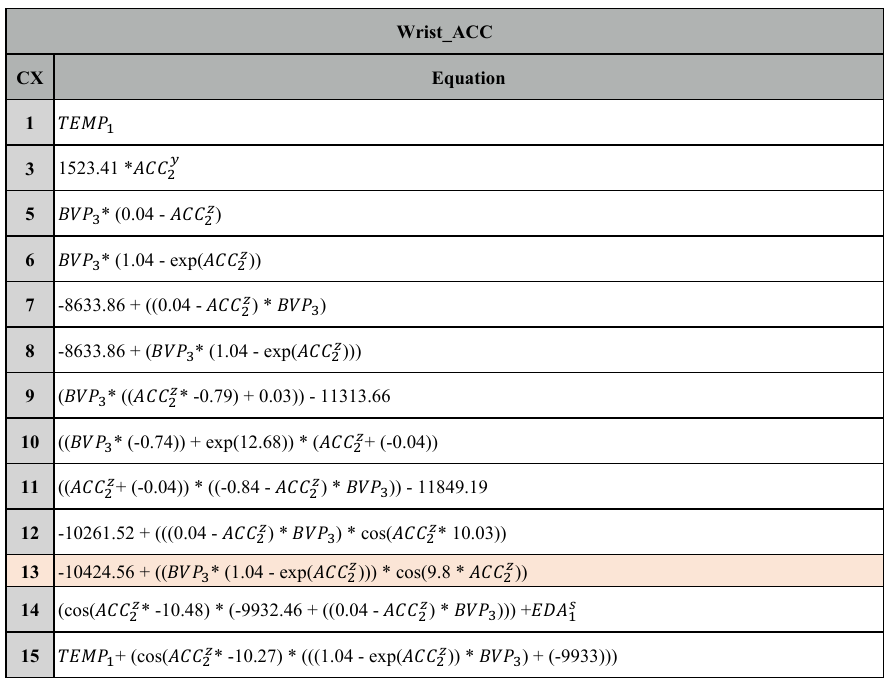}
  \caption{Indicators law table of ACC in Wrist}
\end{figure}

\begin{figure}
  \centering
  \includegraphics[width=.75\linewidth]{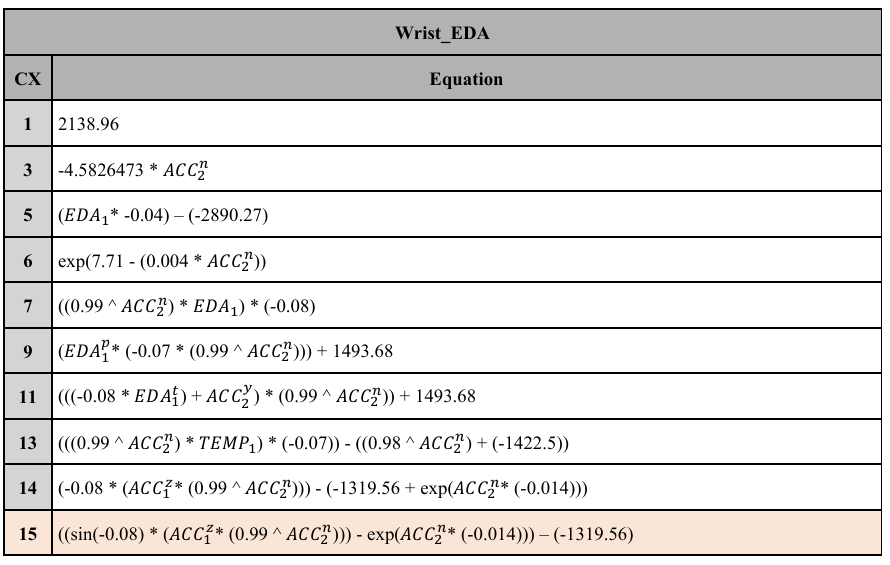}
  \caption{Indicators law table of EDA in Wrist}
\end{figure}

\begin{figure}
  \centering
  \includegraphics[width=.75\linewidth]{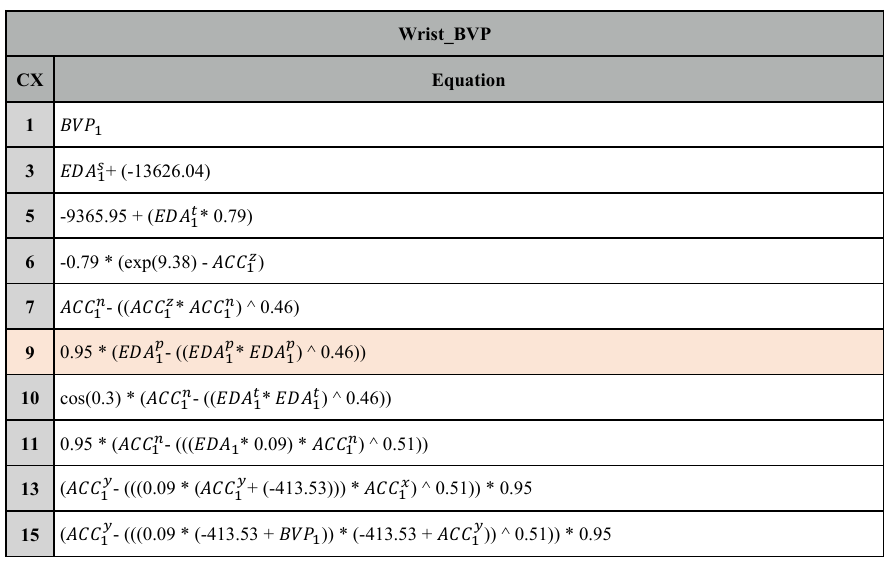}
  \caption{Indicators law table of BVP in Wrist}
\end{figure}

\begin{figure}
  \centering
  \includegraphics[width=.75\linewidth]{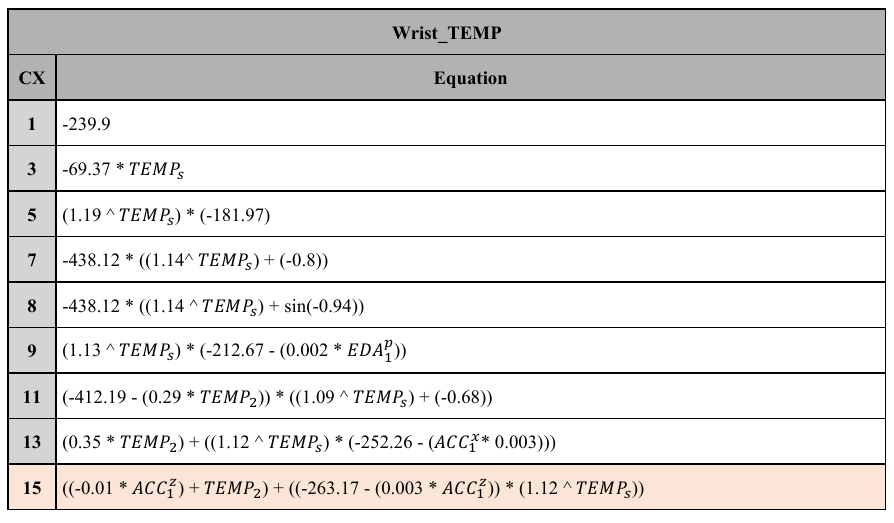}
  \caption{Indicators law table of TEMP in Wrist}
\end{figure}

\subsubsection*{Tables of Indicators Law in Chest Dataset}

\begin{figure}
  \centering
  \includegraphics[width=.75\linewidth]{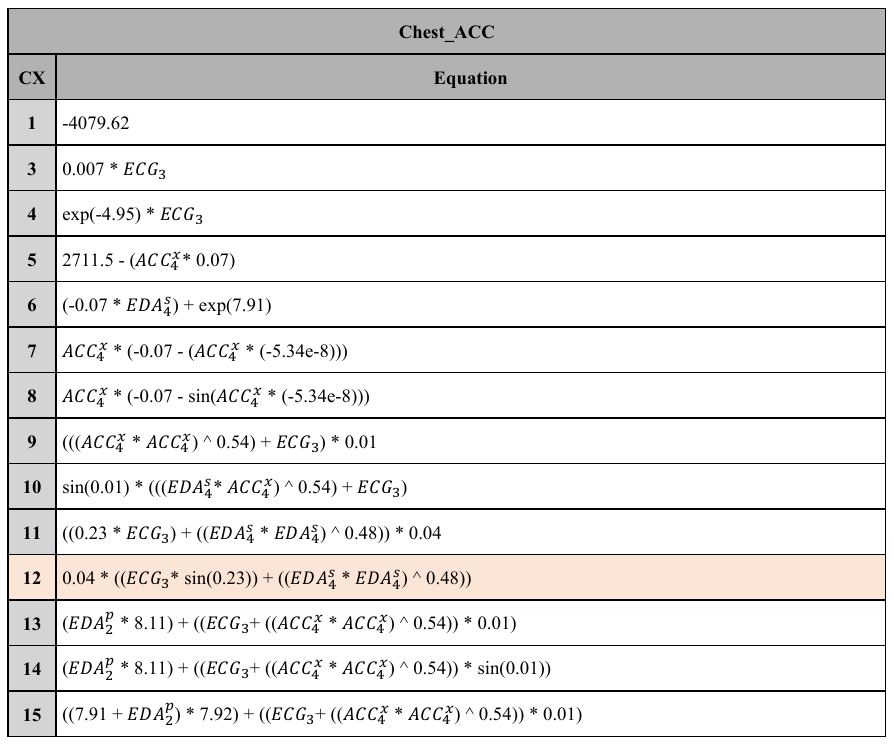}
  \caption{Indicators law table of ACC in Chest}
\end{figure}

\begin{figure}
  \centering
  \includegraphics[width=.75\linewidth]{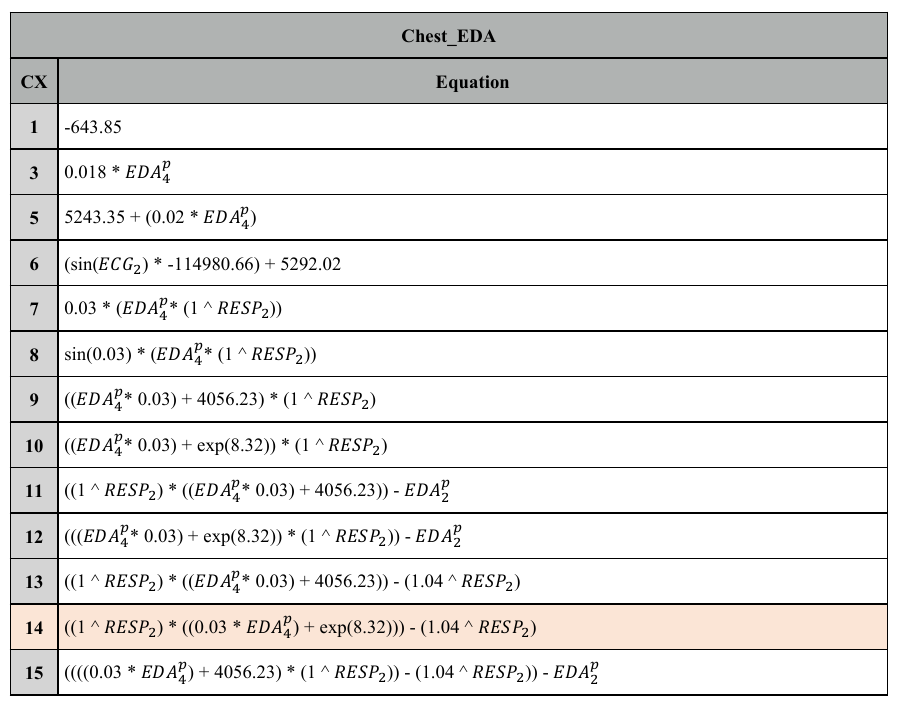}
  \caption{Indicators law table of EDA in Chest}
\end{figure}

\begin{figure}
  \centering
  \includegraphics[width=.75\linewidth]{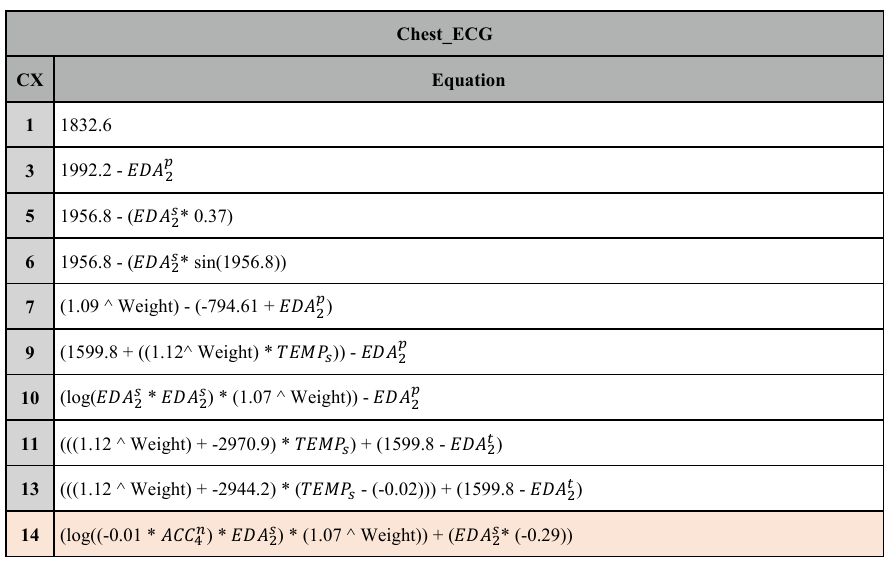}
  \caption{Indicators law table of ECG in Chest}
\end{figure}

\begin{figure}
  \centering
  \includegraphics[width=.75\linewidth]{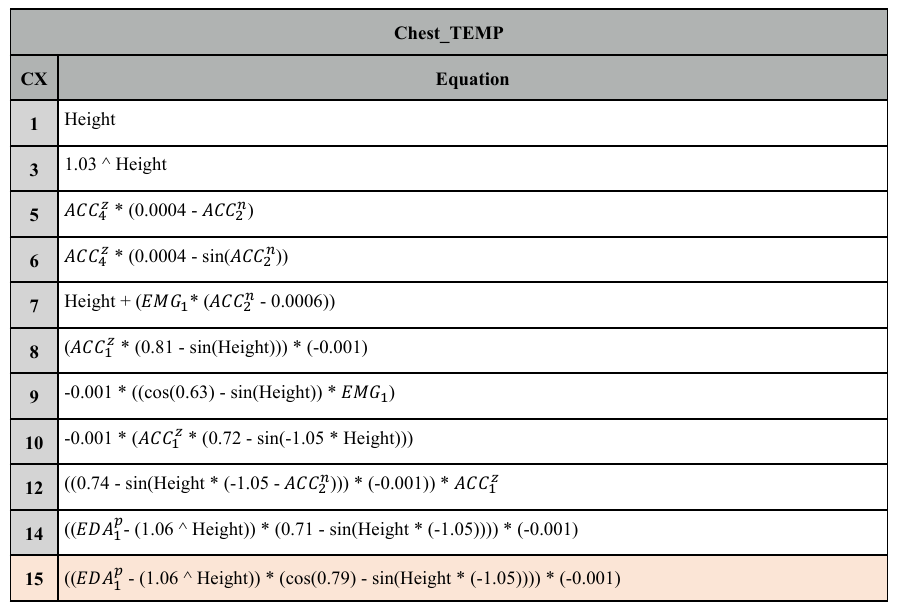}
  \caption{Indicators law table of TEMP in Chest}
\end{figure}

\begin{figure}
  \centering
  \includegraphics[width=.75\linewidth]{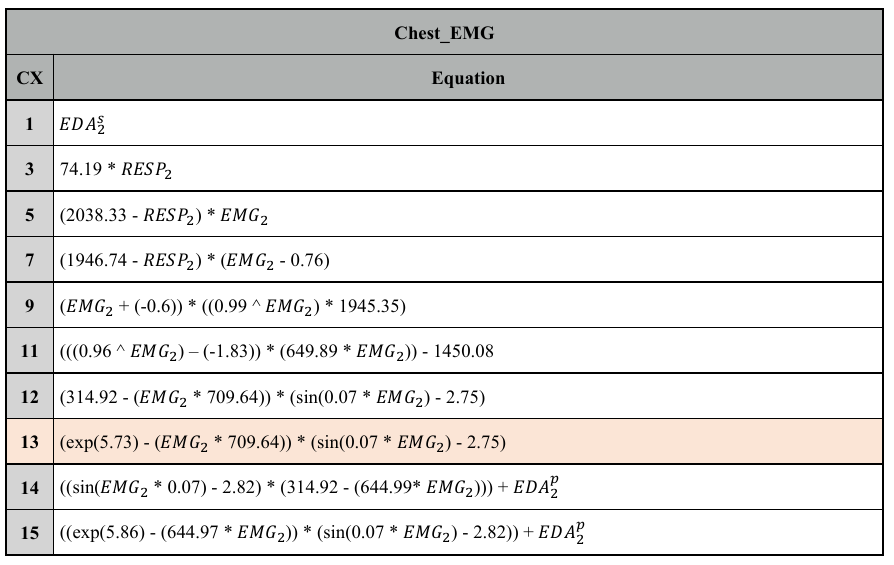}
  \caption{Indicators law table of EMG in Chest}
\end{figure}

\begin{figure}
  \centering
  \includegraphics[width=.75\linewidth]{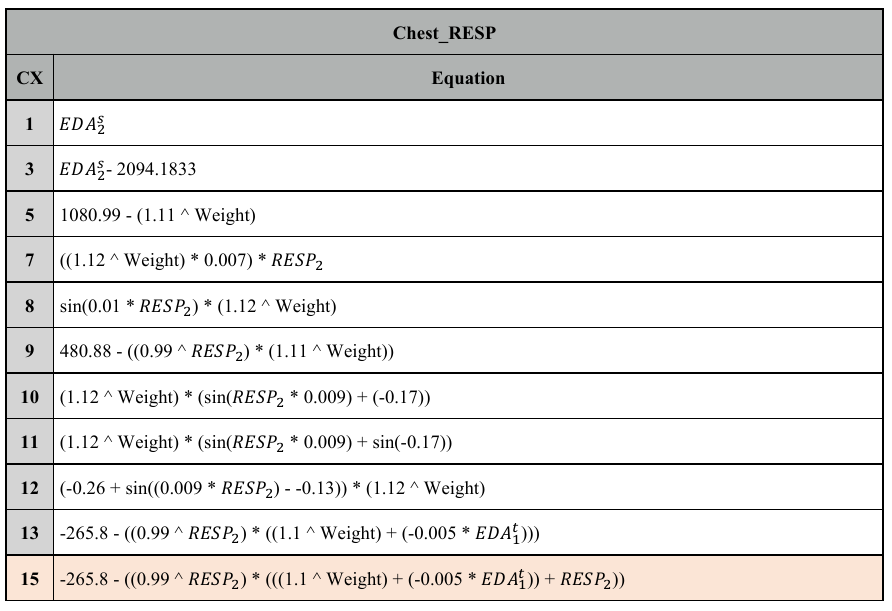}
  \caption{Indicators law table of RESP in Chest}
\end{figure}




% Author biography without author photo.
% Author biography. Author biography. Author biography.
% Author biography. Author biography. Author biography.
% Author biography. Author biography. Author biography.
% Author biography. Author biography. Author biography.
% Author biography. Author biography. Author biography.
% Author biography. Author biography. Author biography.
% Author biography. Author biography. Author biography.
% Author biography. Author biography. Author biography.
% Author biography. Author biography. Author biography.

% \bio{figs/cas-pic1}
% Author biography with author photo.
% Author biography. Author biography. Author biography.
% Author biography. Author biography. Author biography.
% Author biography. Author biography. Author biography.
% Author biography. Author biography. Author biography.
% Author biography. Author biography. Author biography.
% Author biography. Author biography. Author biography.
% Author biography. Author biography. Author biography.
% Author biography. Author biography. Author biography.
% Author biography. Author biography. Author biography.

% \vskip3pc

% \bio{figs/cas-pic1}
% Author biography with author photo.
% Author biography. Author biography. Author biography.
% Author biography. Author biography. Author biography.
% Author biography. Author biography. Author biography.
% Author biography. Author biography. Author biography.

\

\end{document}